\documentclass[11pt,english]{article}
\usepackage[T1]{fontenc}
\usepackage[latin9]{inputenc}
\usepackage{geometry}
\usepackage{color}
\usepackage{float}
\usepackage{amsmath}
\usepackage{amssymb}
\usepackage{graphicx}
\usepackage{esint}

\makeatletter

\providecommand{\tabularnewline}{\\}

\newcommand{\lyxaddress}[1]{
\par {\raggedright #1
\vspace{1.4em}
\noindent\par}
}

\usepackage{multicol}

\makeatother

\usepackage{babel}
\begin{document}

\title{Finite volume form factors in the presence of integrable defects}

\author{Z. Bajnok$^{1}$, F. Buccheri$^{2}$, L. Hollo$^{1}$, J. Konczer$^{1}$
and G. Takacs$^{2,3}$}

\maketitle

\lyxaddress{\begin{center}
$^{1}$MTA Lend\"ulet Holographic QFT Group, Wigner Research Centre
for Physics\\
H-1525 Budapest 114, P.O.B. 49, Hungary\\
~\\
$^{2}$MTA-BME ``Momentum'' Statistical Field Theory Research Group\\
 1111 Budapest, Budafoki út 8, Hungary\\
~\\
$^{3}$Department of Theoretical Physics, \\
 Budapest University of Technology and Economics\\
 1111 Budapest, Budafoki út 8, Hungary\\

\par\end{center}}
\begin{abstract}
We developed the theory of finite volume form factors in the presence
of integrable defects. These finite volume form factors are expressed
in terms of the infinite volume form factors and the finite volume
density of states and incorporate all polynomial corrections in the
inverse of the volume. We tested our results, in the defect Lee-Yang
model, against numerical data obtained by truncated conformal space
approach (TCSA), which we improved by renormalization group methods
adopted to the defect case. To perform these checks we determined
the infinite volume defect form factors in the Lee-Yang model exactly,
including their vacuum expectation values. We used these data to calculate
the two point functions, which we compared, at short distance, to
defect CFT. We also derived explicit expressions for the exact finite
volume one point functions, which we checked numerically. In all of
these comparisons excellent agreement was found.
\end{abstract}

\section{Introduction}

The complete solution of a quantum field theory means the determination
of its spectrum and correlation functions. This ultimate goal is almost
impossible to achieve in general dimensions and for generic interacting
theories. However, in two dimensional integrable models even such
an ambitious plan can be fulfilled.

For many quantities, such as operator matrix elements, a numerical
determination is only possible in finite volumes \cite{Maiani:1990ca}.
Such determination can be performed using the tools of lattice field
theory or, in the case of two-dimensional quantum field theories,
a more specialized method such as the truncated conformal space approach
(TCSA) \cite{Yurov:1989yu}. Finite volume quantities are also interesting
on their own right in statistical field theory, as well as in particle
physics. The general strategy to compute them analytically is to solve
the theory in infinite volume first and then to take into account
the finite size corrections systematically. The infinite volume solution
is carried out in the bootstrap framework, and it consists of the
scattering matrix bootstrap and the form factor bootstrap parts. All
finite size corrections can be expressed purely in terms of these
infinite volume characteristics of the theory in a framework that
was pioneered by Lüscher \cite{Luscher:1985dn,Luscher:1986pf,Luscher:1990ux}. 

The infinite volume solution of an integrable QFT starts with the
S-matrix bootstrap. The scattering matrix satisfies unitarity and
crossing symmetry and all of its poles are located on the imaginary
rapidity axes and correspond to bound-states or some Coleman-Thun
type diagrams. Assuming one single particle in the spectrum with a
self-fusing pole the bootstrap leads to the S-matrix of the scaling
Lee-Yang model. Introducing integrable boundaries or integrable defects
requires additionally to perform this bootstrap program for the reflection
and transmission matrices. The structure of the scattering, reflection
and transmission matrices contain all information about the infinite
volume spectrum of bulk, boundary or defect excitations. Once this
first bootstrap step is completed, the resulting scattering, reflection
and transmission matrices can be used to formulate consistency requirements
for the matrix elements of local operators between asymptotic states
(form factors). Solutions to these requirements compatible with the
analytical structure demanded by physics lead to the determination
of the form factors of all local bulk, boundary and defect operators.
These form factors then can be used to build up all correlation functions
in infinite volume. 

Once all infinite volume characteristics are determined they can be
used to decrease the volume gradually and continue the quantities
for finite volumes. The leading volume dependence is polynomial in
the inverse of the volume, while the sub-leading ones are exponentially
small. The polynomial finite size corrections for the spectrum can
be formulated in terms of the scattering, reflection or transmission
matrices. At this order the dispersion relation is not changed, but
the energy levels are quantized. Momentum is quantized in a finite
volume as, if we move a particle around the 1D 'world', we collect
not only the translational phase, but also the phases of all scatterings
and/or reflections and transmissions: therefore, imposing periodic
boundary conditions restricts the allowed particle momenta. The resulting
equations are called Bethe-Yang equations. Exponentially small corrections
are due to vacuum polarization effects and can be taken into account
by the Thermodynamic Bethe Ansatz method. 

Finite volume form factors are the matrix elements of local operators
between the finite volume eigenstates of the Hamiltonian. Just like
for the spectrum, leading finite size corrections are polynomial in
the inverse of the volume, while sub-leading corrections are exponentially
small. The polynomial corrections are related to the normalization
of the Bethe-Yang eigenvectors and were systematically analyzed in
\cite{Pozsgay:2007kn,Pozsgay:2007gx} for periodic boundary conditions,
while in \cite{Kormos:2007qx} for general integrable boundary conditions.
The aim of the present paper is to generalize this analysis for the
defect case. We mention for completeness that the exponential finite
size corrections have not been described yet, except for some recent
results accounting for the composite structure of bound states \cite{Pozsgay:2008bf,Takacs:2011nb},
and for diagonal form factors in the case of periodic boundary conditions
\cite{Pozsgay:2013jua}. 

Integrable interacting defects are either purely reflective (i.e.
boundaries) or purely transmissive \cite{Delfino:1994nr}. Recently
there was relevant progress in identifying and solving purely transmitting
defect theories \cite{Bowcock:2004my,Bowcock:2005vs,Corrigan:2010ph,Avan:2012rb,Doikou:2012fc}.
In this paper we focus on the simplest integrable defect theory, namely
the defect Lee-Yang model. The T-matrix bootstrap of this model was
performed in \cite{Bajnok:2007jg}, while the form factor bootstrap
program was initiated in \cite{Bajnok:2009hp}.

Our paper is organized as follows: In section 2, we recall the infinite
volume form factor solution of the defect Lee-Yang model. We start
by listing the defect form factor axioms. We then introduce the defect
Lee-Yang model both from the scattering (IR) and from the perturbed
CFT (UV) side. In presenting the defect form factors we go beyond
the results available in the literature, as we determine all form
factor solutions of primary fields. Technical details are relegated
to Appendix A. In order to have properly normalized form factors,
we calculate the vacuum expectation values of all primary fields.
The form factors and normalizations are checked against the defect
CFT two point functions. In section 3 we develop the leading finite
size corrections of defect form factors. We analyze separately the
non-diagonal and diagonal cases as in the latter one disconnected
terms appear which have to be determined carefully. Section 4 contains
the checks of our results against the numerically ``measured'' finite
volume matrix elements calculated by the defect TCSA (DTCSA) method.
In section 5 we derive the exact finite volume vacuum expectation
values of local fields what we also check numerically. Finally we
draw our conclusions in section 6.

\section{Form factors in infinite volume}

In this section we recall the theory of form factors in the presence
of integrable defects following \cite{Bajnok:2009hp}. We focus on
a relativistic integrable theory which contains one particle type
only. Energy and momentum of the particles are parametrized by the
rapidity variable $\theta$:
\begin{equation}
E(\theta)=m\cosh\theta\quad,\quad p(\theta)=m\sinh\theta
\end{equation}
and Lorentz transformation simply shifts the rapidity: $\theta\to\theta+\Lambda$.
Integrability forces the multi-particle scattering matrix to factorize
into pairwise two particle scatterings, which depends on the rapidity
differences $S(\theta_{1}-\theta_{2})$ and satisfies unitarity and
crossing symmetry
\begin{equation}
S(\theta)=S(-\theta)^{-1}\qquad;\qquad S(i\pi-\theta)=S(\theta)\,.
\end{equation}
If the $S$-matrix has a pole at $\theta=i\frac{2\pi}{3}$ (like in
the scaling Lee-Yang model) then it has to satisfy the fusion equation
\begin{equation}
S(\theta)\vert_{\theta\approx i\frac{2\pi}{3}}=-i\frac{\Gamma^{2}}{\theta-i\frac{2\pi}{3}}+\mathrm{reg.\ terms}\quad\longrightarrow\quad S(\theta)=S\left(\theta-i\frac{\pi}{3}\right)S\left(\theta+i\frac{\pi}{3}\right)\,,\label{eq:fusionpole}
\end{equation}
which shows that the particle is a bound-state of itself. 

Introducing an integrable defect means that we cut the space-time
into two halves and associate an amplitude of crossing through the
defect. Particles coming from the left cross with $T_{-}(\theta)$,
while those coming from the right cross with $T_{+}(-\theta)$. The
$T_{+}$ transmission factor is parametrized such that for its physical
domain of rapidities $(\theta<0)$ its argument is always positive.
Unitarity and crossing symmetry relates these two amplitudes as \cite{Bajnok:2004jd}:
\begin{equation}
T_{-}(\theta)=T_{+}(-\theta)^{-1}\qquad;\qquad T_{-}(\theta)=T_{+}(i\pi-\theta)
\end{equation}
A prototype of a parity symmetric defect is a standing particle with
$T_{-}(\theta)=S(\theta)=T_{+}(\theta).$ A non parity symmetric defect
can be realized as an imaginary rapidity ``bound'' particle: $T_{\pm}(\theta)=S(\theta\pm iu).$
In the case of a fusion pole, (\ref{eq:fusionpole}), the defect satisfies
the fusion equation as well: 
\begin{equation}
T_{-}(\theta)=T_{-}\left(\theta-i\frac{\pi}{3}\right)T_{-}\left(\theta+i\frac{\pi}{3}\right)
\end{equation}
and likewise for $T_{+}$. It might happen that the transmission factor
exhibits a pole
\begin{equation}
T_{-}(\theta)\vert_{\theta\approx i\nu}=-i\frac{g^{2}}{\theta-i\nu}+\mathrm{reg.\ terms}
\end{equation}
signaling a defect bound-state. 

Integrable defects are topological in the sense that the location
of the defect can be changed without altering the amplitude of any
multi-particle transmission process unless it crosses an insertion
point of any field. As a consequence multi-particle transmissions
factorize into the product of pairwise scatterings and individual
transmissions. 

In the following we recall the form factor axioms in the presence
of these integrable defects \cite{Bajnok:2009hp}.

\subsection{Summary of the defect form factor bootstrap \label{sub:Summary-of-the}}

Form factors are the matrix elements of local operators between asymptotic
states: 
\begin{equation}
\langle\theta_{m'+n'}^{'},\dots,\theta_{n'+1}^{'};\theta_{n'}^{'},\dots,\theta_{1}^{'}\vert\mathcal{O}(x,t)\vert\theta_{1},\dots,\theta_{n};\theta_{n+1},\dots,\theta_{n+m}\rangle\label{eq:genff}
\end{equation}
 where in the presence of defects we have to distinguish if particles
arrive from the left or from the right. In an initial state the rapidities
are ordered as 
\begin{equation}
\theta_{1}>\dots>\theta_{n}>0>\theta_{n+1}>\dots>\theta_{n+m}
\end{equation}
while in the final state oppositely $\theta_{m'+n'}^{'}>\dots>\theta_{n'+1}^{'}>0>\theta_{n'}^{'}>\dots>\theta_{1}^{'}$.
The form factor is originally defined for initial and final states
and then analytically continued for any orderings of its arguments.
Interestingly, the presence of an integrable defect breaks the translation
invariance by having non-zero momentum (like a bound particle) and
not by destroying the existence of the momentum itself. As a consequence
the space-time dependence of the form factor is 
\begin{eqnarray}
\langle\theta_{m'+n'}^{'},\dots,\theta_{n'+1}^{'};\theta_{n'}^{'},\dots,\theta_{1}^{'}\vert\mathcal{O}(x,t)\vert\theta_{1},\dots,\theta_{n};\theta_{n+1},\dots,\theta_{n+m}\rangle=\qquad\qquad\qquad\qquad\\
\qquad e^{it\Delta E-ix\Delta P}F_{(n',m')(n,m)}^{\mathcal{O}_{\pm}}(\theta_{n'+m'}^{'},\dots,\theta_{n'+1}^{'};\theta_{n'}^{'},\dots,\theta_{1}^{'}\vert\theta_{1},\dots,\theta_{n};\theta_{n+1},\dots,\theta_{n+m})\nonumber 
\end{eqnarray}
with $\Delta E=m(\sum_{j}\cosh\theta_{j}-\sum_{j'}\cosh\theta_{j'}^{'})$
and $\Delta P=m(\sum_{j}\sinh\theta_{j}-\sum_{j'}\sinh\theta_{j'}^{'})$,
and we distinguished if the operator was localized on the left, $\mathcal{O}_{-}$,
or on the right, $\mathcal{O}_{+}$, of the defect as they might not
be continuous there. Same apply for operators localized at the defect
($x=0$). Crossing transformation of any of the form factors 
\begin{eqnarray}
F_{(n',m')(n,m)}^{\mathcal{O}}(\theta_{n'+m'}^{'},\dots,\theta_{n'+1}^{'};\theta_{n'}^{'},\dots,\theta_{1}^{'}\vert\theta_{1},\dots,\theta_{n};\theta_{n+1},\dots,\theta_{n+m})=\qquad\qquad\qquad\qquad\\
\qquad F_{(n',m'+1)(n,m-1)}^{\mathcal{O}}(\theta_{n+m}+i\pi,\theta_{n'+m'}^{'},\dots,\theta_{n'+1}^{'};\theta_{n'}^{'},\dots,\theta_{1}^{'}\vert\theta_{1},\dots,\theta_{n};\theta_{n+1},\dots,\theta_{n+m-1})\nonumber 
\end{eqnarray}
\begin{eqnarray}
F_{(n',m')(n,m)}^{\mathcal{O}}(\theta_{n'+m'}^{'},\dots,\theta_{n'+1}^{'};\theta_{n'}^{'},\dots,\theta_{1}^{'}\vert\theta_{1},\dots,\theta_{n};\theta_{n+1},\dots,\theta_{n+m})=\qquad\qquad\qquad\qquad\\
\qquad F_{(n'+1,m')(n-1,m)}^{\mathcal{O}}(\theta_{n'+m'}^{'},\dots,\theta_{n'+1}^{'};\theta_{n'}^{'},\dots,\theta_{1}^{'},\theta_{1}-i\pi\vert\theta_{2},\dots,\theta_{n};\theta_{n+1},\dots,\theta_{n+m})\nonumber 
\end{eqnarray}
can be used to bring all particles into one side:
\begin{equation}
F_{(n,m)}^{\mathcal{O}}(\theta_{1},\dots,\theta_{n};\theta_{n+1},\dots,\theta_{n+m}):=F_{(0,0)(n,m)}^{\mathcal{O}}(;\vert\theta_{1},\dots,\theta_{n};\theta_{n+1},\dots,\theta_{n+m})
\end{equation}
We can also use transmission 
\begin{equation}
F_{(n,m)}^{\mathcal{O}}(\theta_{1},\dots,\theta_{n};\theta_{n+1},\dots,\theta_{n+m})=T_{-}(\theta_{n})F_{(n-1,m+1)}^{\mathcal{O}}(\theta_{1},\dots,\theta_{n-1};\theta_{n},\theta_{n+1},\dots,\theta_{n+m})\label{eq:axtrans}
\end{equation}
to define the \emph{elementary} form factors
\begin{equation}
F_{n}^{\mathcal{O}}(\theta_{1},\dots,\theta_{n})=F_{(n,0)}^{\mathcal{O}}(\theta_{1},\dots,\theta_{n};)\label{eq:ffdef}
\end{equation}
which satisfies the axioms
\begin{equation}
F_{n}^{\mathcal{O}}(\theta_{1},\dots\theta_{i},\theta_{i+1},\dots,\theta_{n})=S(\theta_{i}-\theta_{i+1})F_{n}^{\mathcal{O}}(\theta_{1},\dots\theta_{i+1},\theta_{i},\dots,\theta_{n})\label{eq:axperm}
\end{equation}
\begin{equation}
F_{n}^{\mathcal{O}}(\theta_{1},\theta_{2},\dots,\theta_{n})=F_{n}^{\mathcal{O}}(\theta_{2},\dots\theta_{n},\dots,\theta_{1}-2i\pi)\label{eq:axper}
\end{equation}
\begin{equation}
-i\mbox{Res}_{\theta=\theta^{\prime}}F_{n+2}^{\mathcal{O}}(\theta+i\pi,\theta^{\prime},\theta_{1},...,\theta_{n})=\bigl(1-\prod_{j=1}^{n}S(\theta-\theta_{j})\bigr)F_{n}^{\mathcal{O}}(\theta_{1},...,\theta_{n})\label{eq:axkin}
\end{equation}

\begin{equation}
-i\mbox{Res}_{\theta=\theta^{\prime}}F_{n+2}^{\mathcal{O}}(\theta+\frac{i\pi}{3},\theta^{\prime}-\frac{i\pi}{3},\theta_{1},\ldots,\theta_{n})=\Gamma F_{n+1}^{\mathcal{O}}(\theta,\theta_{1},\ldots,\theta_{n})\label{eq:axdyn}
\end{equation}

\begin{equation}
-i\mbox{Res}_{\theta=iu}F_{n+1}^{\mathcal{O}}(\theta_{1},\ldots,\theta_{n},\theta)=ig\tilde{F}_{n}^{\mathcal{O}}(\theta_{1},\ldots,\theta_{n})\label{eq:axbnd}
\end{equation}
where $\tilde{F}$ is the form factor on the excited defect state.
Although the axioms (\ref{eq:axperm}-\ref{eq:axdyn}) look the same
as the form factor axioms without the defect they are valid only for
particles coming from the left (see eq. (\ref{eq:ffdef})). For any
particle coming from the right one has to include a transmission factor,
see eq. (\ref{eq:axtrans}). 

The form factor of a bulk operator localized on the left of the defect,
$\mathcal{O}_{-}$, is simply its bulk form factor 
\begin{equation}
F_{n}^{\mathcal{O}_{-}}(\theta_{1},\dots,\theta_{n})=B_{n}^{\mathcal{O}}(\theta_{1},\dots,\theta_{n})
\end{equation}
but when the same operator is localized on the right of the defect,
$\mathcal{O}_{+},$ its form factor is 
\begin{equation}
F_{n}^{\mathcal{O}_{+}}(\theta_{1},\dots,\theta_{n})=\prod_{i}T_{-}(\theta_{i})B_{n}^{\mathcal{O}}(\theta_{1},\dots,\theta_{n})
\end{equation}
These apply for the left/right limits of the bulk fields at the defect
as well. 

As we assume that the bulk form factors are already determined in
\cite{Zamolodchikov:1990bk} we focus on form factors of defect operators.
In general, the solution compatible with the form factor axioms takes
the form 
\begin{equation}
F_{n}^{\mathcal{O}}(\theta_{1},\dots,\theta_{n})=\langle\mathcal{O}\rangle H_{n}\prod_{i}d(\theta_{i})\prod_{i<j}\frac{f(\theta_{i}-\theta_{j})}{x_{i}+x_{j}}Q_{n}(x_{1},\dots,x_{n})
\end{equation}
where $f(\theta)$ is the minimal bulk two particle form factor, which
satisfies: 
\begin{equation}
f(\theta)=S(\theta)f(-\theta)\quad;\qquad f(i\pi-\theta)=f(i\pi+\theta)
\end{equation}
The one particle minimal defect form factor $d(\theta)$ is responsible
for defect bound-states, $H_{n}$ is some normalization constant and
$Q_{n}$ is a symmetric polynomial in its arguments $x_{i}=e^{\theta_{i}}.$

\subsection{Defect form factors in the scaling Lee-Yang model }

The Lee-Yang model is the simplest conformal field theory, the $\mathcal{M}_{2,5}$
minimal model, whose central charge is $c=-\frac{22}{5}$ and which
has only two irreducible Virasoro representations $V_{0}$ and $V_{h}$
with highest weights $0$ and $h=-\frac{1}{5}$, respectively. The
periodic Lee-Yang model carries a representation of $Vir\otimes\overline{Vir}$
and its Hilbert-space is decomposed as
\begin{equation}
\mathcal{H}=V_{0}\otimes\bar{V}_{0}+V_{h}\otimes\bar{V}_{h}
\end{equation}
We can associate a local field for all vector of the Hilbert-space,
and for the highest weight states these fields are the identity $\mathbb{I}$
and $\Phi$, respectively. These local fields form an operator-algebra
with the operator product expansions
\begin{equation}
\Phi\left(z,\bar{z}\right)\Phi\left(0,0\right)=C_{\Phi\Phi}^{\mathbb{I}}\vert z\vert^{-4h}\mathbb{I}+C_{\Phi\Phi}^{\Phi}\vert z\vert^{-2h}\Phi\left(0,0\right)+\dots
\end{equation}
In order to have a real field $\Phi^{\dagger}=\Phi$, we normalized
the field as $C_{\Phi\Phi}^{\mathbb{I}}=-1$. A consistent choice
of the other structure constant is $C_{\Phi\Phi}^{\Phi}=\sqrt{\frac{2}{1+\sqrt{5}}}\frac{\Gamma\left(\frac{1}{5}\right)\Gamma\left(\frac{6}{5}\right)}{\Gamma\left(\frac{3}{5}\right)\Gamma\left(\frac{4}{5}\right)}\approx1.91131\dots$ 

In this subsection we specify the previous defect form factor considerations
for the scaling Lee-Yang model \cite{Bajnok:2007jg}. The scaling
Lee-Yang model is the single relevant perturbation of the Lee-Yang
model 
\begin{equation}
\mathcal{A}=\mathcal{A}_{LY}-\lambda\int d^{2}z\,\Phi(z,\bar{z})\label{eq:SLY}
\end{equation}
and has a single particle in the spectrum with the two-particle scattering
matrix:
\begin{equation}
S(\theta)=\frac{\sinh\theta+i\sin\frac{\pi}{3}}{\sinh\theta-i\sin\frac{\pi}{3}}
\end{equation}
There is a one parameter family of integrable defect perturbation
of the defect Lee-Yang model \cite{Bajnok:2013} which has the transmission
factor 
\begin{equation}
T_{\pm}(\theta)=S\left(\theta\pm i\frac{\pi}{6}(3-b)\right)
\end{equation}
Formally it is like a particle with rapidity $\theta=i\frac{\pi}{6}(3-b)$
and the defect energy and momentum are indeed 
\begin{equation}
e_{d}=m\sin\frac{\pi b}{6}\qquad;\qquad p_{d}=im\cos\frac{\pi b}{6}\label{eq:defectenergyandmomentum}
\end{equation}
This theory can be realized as a unique one parameter family of integrable
perturbation of the defect Lee-Yang model:
\begin{equation}
S^{d}=S_{LY}^{d}-\lambda\int d^{2}z\,\Phi(z,\bar{z})-\mu\int dy\,\varphi(y)-\bar{\mu}\int dy\,\bar{\varphi}(y)
\end{equation}
where integrability forces the constraint: 
\begin{equation}
\lambda=\mu\bar{\mu}\xi^{-2}\quad;\qquad\xi^{-2}=2i\sqrt{\frac{1+\sqrt{5}}{2}}\frac{1-e^{i\pi/5}}{1+e^{i\pi/5}}=0.826608
\end{equation}
 The relation between the Lagrangian and scattering parameters is
\begin{equation}
\lambda=\left(\frac{m}{\kappa}\right)^{\frac{12}{5}}\quad;\qquad\kappa=2^{\frac{19}{12}}\sqrt{\pi}\frac{\left(\Gamma(\frac{3}{5})\Gamma(\frac{4}{5})\right)^{\frac{5}{12}}}{5^{\frac{5}{16}}\Gamma(\frac{2}{3})\Gamma(\frac{5}{6})}=2.642944\label{eq:lambda}
\end{equation}
\begin{equation}
\mu=\left(\frac{m}{\kappa}\right)^{\frac{6}{5}}\xi e^{-\frac{i\pi}{5}(3+b)}\qquad;\qquad\bar{\mu}=\left(\frac{m}{\kappa}\right)^{\frac{6}{5}}\xi e^{\frac{i\pi}{5}(3+b)}\label{eq:mu}
\end{equation}
 Due to the nontrivial defect, the Hilbert space of the defect Lee-Yang
model 
\begin{equation}
\mathcal{H}=V_{0}\otimes\bar{V}_{h}+V_{h}\otimes\bar{V}_{0}+V_{h}\otimes\bar{V}_{h}=:[\bar{d}]+[d]+[D]
\end{equation}
 does not coincide with the operator space localized on the defect
which is 
\begin{equation}
V_{0}\otimes\bar{V}_{0}+V_{h}\otimes\bar{V}_{0}+V_{0}\otimes\bar{V}_{h}+2V_{h}\otimes\bar{V}_{h}=:[\mathbb{I}]+[\varphi]+[\bar{\varphi}]+[\Phi_{\pm}]
\end{equation}
We are going to compute the form factors of these operators.

\subsubsection{Form factor solutions}

The ingredients of the form factor solutions are as follows: the minimal
solution of the bulk two particle form factor is
\begin{equation}
f(\theta)=\frac{x+x^{-1}-2}{x+x^{-1}+1}\, v(i\pi-\theta)\, v(-i\pi+\theta)\quad,\qquad x=e^{\theta}
\end{equation}
where 
\begin{equation}
\log v(\theta)=2\int_{0}^{\infty}\frac{dt}{t}e^{\frac{i\theta t}{\pi}}\frac{\sinh\frac{t}{2}\sinh\frac{t}{3}\sinh\frac{t}{6}}{\sinh^{2}t}\quad,
\end{equation}
which automatically includes the pole of the dynamical singularity.
This minimal solution satisfies the identities 
\begin{eqnarray}
f\left(\theta\right)f\left(\theta+i\pi\right) & = & \frac{\sinh\left(\theta\right)}{\sinh\left(\theta\right)-i\sin\left(\frac{\pi}{3}\right)}\nonumber \\
f\left(\theta+\frac{i\pi}{3}\right)f\left(\theta-\frac{i\pi}{3}\right) & = & \frac{\cosh\left(\theta\right)+\frac{1}{2}}{\cosh\left(\theta\right)+1}f\left(\theta\right).\label{eq:fidentity}
\end{eqnarray}
The normalization factor is the same as in the bulk 
\begin{equation}
H_{n}=\left(\frac{i3^{\frac{1}{4}}}{2^{\frac{1}{2}}v(0)}\right)^{n}\quad.
\end{equation}
The one particle defect form factor, which accommodates the possible
defect bound-state pole is
\begin{equation}
d(\theta)=\frac{1}{\sqrt{3}+x\nu+x^{-1}\bar{\nu}}\label{eq:d}
\end{equation}
where we introduced $\nu=e^{i\frac{\pi b}{6}}$ and $\bar{\nu}=\nu^{-1}$,
satisfying
\begin{eqnarray}
d\left(\theta+i\pi\right)d\left(\theta\right) & = & \frac{1}{1-2\cos\left(\frac{b\pi}{3}-2i\theta\right)}\nonumber \\
d\left(\theta+\frac{i\pi}{3}\right)d\left(\theta-\frac{i\pi}{3}\right) & = & \frac{1}{2\cos\left(\frac{b\pi}{6}-i\theta\right)}d\left(\theta\right).\label{eq:didentity}
\end{eqnarray}
The singularity axioms provide recursion relations between the polynomials
$Q_{n}$ as 
\begin{eqnarray}
Q_{n+2}(-x,x,x_{1},...,x_{n}) & = & K_{n}(x,x_{1},...,x_{n})Q_{n}(x_{1},...,x_{n})\\
Q_{n+1}(x\omega,x\bar{\omega},x_{1},...,x_{n-1}) & = & D_{n}(x,x_{1},...,x_{n-1})Q_{n}(x,x_{1},...,x_{n-1})\nonumber 
\end{eqnarray}
where $\omega=e^{\frac{i\pi}{3}}$, $\bar{\omega}=\omega^{-1}$ and
we explicitly have
\begin{eqnarray}
K_{n}(x,x_{1},...,x_{n}) & = & (-1)^{n}(x^{2}\nu^{2}-1+x^{-2}\nu^{-2})\\
 &  & \frac{x}{2(\omega-\bar{\omega})}\left(\prod_{i=1}^{n}(x\omega+x_{i}\bar{\omega})(x\bar{\omega}-x_{i}\omega)-\prod_{i=1}^{n}(x\omega-x_{i}\bar{\omega})(x\bar{\omega}+x_{i}\omega)\right)\nonumber 
\end{eqnarray}
for the kinematical recursion and
\begin{eqnarray}
D_{n}(x,x_{1},...,x_{n-1}) & = & (\nu x+\nu^{-1}x^{-1})x\prod_{i=1}^{n-1}(x+x_{i})
\end{eqnarray}
for the dynamical one. Since $Q_{n}(x_{1},...,x_{n})$ is a symmetric
polynomial we use the elementary symmetric polynomials $\sigma_{k}^{(n)},\bar{\sigma}_{k}^{(n)}$,
defined by the generating function:
\begin{equation}
\prod_{i=1}^{n}(x+x_{i})=\sum_{k\in\mathbb{Z}}x^{n-k}\sigma_{k}^{(n)}(x_{1},...,x_{n})\quad;\qquad\bar{\sigma}_{k}^{(n)}\left(x_{1},\dots,x_{n}\right)=\sigma_{k}^{(n)}(x_{1}^{-1},\dots,x_{n}^{-1})
\end{equation}
to formulate the results. Note that $\sigma_{k}^{(n)}=0$, if $k>n$
or if $k<0$. We sometimes abbreviate $\sigma_{k}^{(n)}(x_{1},...,x_{n})$
to $\sigma_{k}$ if it does not lead to any confusion. 

The form factors of the left/right limits of the bulk operator $\Phi_{\mp}$
follows from our previous considerations and are trivially related
to the bulk form factors. The low lying form factors of the two chiral
fields living only at the defects have been already calculated%
\footnote{To match with the TCSA calculation we choose a different normalization
for $\varphi$ and $\bar{\varphi}$ than in \cite{Bajnok:2009hp}%
} \cite{Bajnok:2009hp}. The results are summarized in Table \ref{tab:Q}
. 

\begin{table}
\begin{centering}
\begin{tabular}{|c|c|c|}
\hline 
Operator & $Q_{1}$ & $Q_{2}$\tabularnewline
\hline 
\hline 
$\Phi_{-}$ & \emph{$\nu\sigma_{1}+\bar{\nu}\bar{\sigma}_{1}+\sqrt{3}$} & $\sigma_{1}(\nu^{2}\sigma_{2}+\sqrt{3}\nu\sigma_{1}+\sigma_{1}\bar{\sigma}_{1}+1+\sqrt{3}\bar{\nu}\bar{\sigma}_{1}+\bar{\nu}^{2}\bar{\sigma}_{2})$\tabularnewline
\hline 
$\Phi_{+}$ & \emph{$\nu\sigma_{1}+\bar{\nu}\bar{\sigma}_{1}-\sqrt{3}$} & $\sigma_{1}(\nu^{2}\sigma_{2}-\sqrt{3}\nu\sigma_{1}+\sigma_{1}\bar{\sigma}_{1}+1-\sqrt{3}\bar{\nu}\bar{\sigma}_{1}+\bar{\nu}^{2}\bar{\sigma}_{2})$\tabularnewline
\hline 
$\bar{\varphi}$ & $\bar{\nu}\bar{\sigma}_{1}$ & $\bar{\nu}\sigma_{1}(\bar{\nu}\bar{\sigma}_{2}+\nu)$\tabularnewline
\hline 
$\varphi$ & $\nu\sigma_{1}$ & $\nu\sigma_{1}(\nu\sigma_{2}+\bar{\nu})$\tabularnewline
\hline 
\end{tabular}
\par\end{centering}

\caption{The form factor solutions of the primary fields up to level 2}
\label{tab:Q}
\end{table}
In Appendix A we explicitly derive all possible solutions of the form
factor equations and extend these results to any order. Here we list
only the outcome. 

The form factors of the two limits of the bulk fields, $\Phi_{\pm},$
are described in terms of the bulk form factor solutions as 
\begin{equation}
Q_{n}^{\Phi_{\pm}}(x_{1},\dots,x_{n})=\prod_{i=1}^{n}(\nu x_{i}+\bar{\nu}x_{i}^{-1}\mp\sqrt{3})\sigma_{1}\sigma_{n-1}P_{n}\label{eq:QPhipm}
\end{equation}
where
\begin{equation}
P_{n}=\det\Sigma^{(n)}\quad,\qquad\Sigma_{ij}^{(n)}=\sigma_{3i-2j+1}^{(n)}(x_{1},\dots,x_{n})
\end{equation}
The defect form factors have the generic structure:
\begin{equation}
Q_{n}=R(\sigma_{1},\bar{\sigma}_{1})\sigma_{n}P_{n}S_{n}
\end{equation}
where $S_{n}$ does not depend on the operator:
\begin{equation}
S_{n}=\tau_{n-1}+\sum_{m\geq1}\left(-1\right)^{m}\left(\tau_{n+1-6m}+\tau_{n-1-6m}\right)\qquad;\qquad\tau_{k}=\sum_{l\in\mathbb{Z}}\nu^{2l-k}\bar{\sigma}_{k-l}\sigma_{l}\,,
\end{equation}
while the operator dependent parts are
\begin{equation}
R^{\bar{\varphi}}(\sigma_{1},\bar{\sigma}_{1})=\bar{\nu}\bar{\sigma}_{1}\qquad;\qquad R^{\varphi}(\sigma_{1},\bar{\sigma}_{1})=\nu\sigma_{1}\:.
\end{equation}

\subsubsection{Exact vacuum expectation values \label{sub:Exact-vacuum-expectation}}

The form factors are normalized with their vacuum expectation values
(VEVs), so here we determine them. The exact VEV of the bulk field
of the scaling Lee-Yang theory is known from conformal perturbation
theory and from TBA in the ultraviolet limit \cite{Zamolodchikov:1989cf}.

Since the VEV of the bulk field does not depend on its location and
the defect fields $\Phi_{\pm}$ correspond to its limits, we conclude
that
\begin{equation}
\langle\Phi\rangle=\langle\Phi_{+}\rangle=\langle\Phi_{-}\rangle=\frac{1}{\pi\lambda(1-h)}\frac{\pi}{4\sqrt{3}}m^{2}\simeq1.23939m^{2h}\label{eq:PhiVEV}
\end{equation}

In order to obtain the expectation values of the chiral fields $\varphi$,
$\bar{\varphi}$, we adopt an argument similar to \cite{Dorey:2000eh}.
The vacuum energy of the system can be generically parametrized as
\cite{Bajnok:2007jg}

\begin{equation}
E_{0}(L)=e_{d}(b)+LE_{bulk}+E_{0}^{TBA}(L)\label{eq:E0L}
\end{equation}
where the bulk energy constant is $E_{bulk}=-\frac{1}{4\sqrt{3}}m^{2}$,
the defect energy is given in (\ref{eq:defectenergyandmomentum})
and $E_{0}^{TBA}(L)$ is the ground-state energy in the TBA scheme,
given in (\ref{eq:DTBAE-1}) below. 

The presence of an integrable defect does not destroy the existence
of momentum, merely modifies the momentum eigenvalues, yielding a
nonzero value in the ground state. In the same way as for the defect
energy, this defect momentum can be conveniently extracted in the
ultraviolet limit, if one makes use of the expression
\begin{equation}
P_{0}^{TBA}(L)=-\frac{m}{2\pi}\int d\theta\sinh\theta\log\left(1+e^{-\tilde{\varepsilon}(\theta)}\right)=P_{0}(L)-p_{d}(b)\label{eq:momentumdef}
\end{equation}
for the exact finite-volume ground state momentum. Here $\tilde{\epsilon}$
satisfies the ground-state TBA equation (see also Section \ref{sec:Expectation-values-in})
\begin{equation}
\tilde{\epsilon}(\theta)=mL\cosh\theta-\log T_{+}(\frac{i\pi}{2}-\theta)-\varphi(\theta)\star\log(1+e^{-\tilde{\epsilon}(\theta)})\quad;\qquad\varphi(\theta)=-\frac{i}{2\pi}\partial_{\theta}\log S(\theta)\label{eq:DTBA-2}
\end{equation}
 In the $L\to0$ limit, the solution of (\ref{eq:DTBA-2}) develops
two plateau regions, which grow as $\sim\log\frac{2}{mL}$, separated
by a breather region around the origin. The plateaus end in two kink
regions, which do not contribute to the ground state momentum. We
therefore focus onto the central region and make use of the pseudo
energy

\begin{equation}
\varepsilon_{0}(\theta)=\lim_{L\to0}\tilde{\epsilon}(\theta)\label{eq:epsilonzero}
\end{equation}
describing the root distribution in the deep ultraviolet limit. Expanding
the expression (\ref{eq:DTBA-2}) with $L=0$ around $\theta\to\pm\infty$
as in \cite{Zamolodchikov:1990bk,Bajnok:2007jg} produces
\begin{equation}
\varepsilon_{0}(\theta\to\pm\infty)\simeq-A_{\pm}e^{\mp\theta}+\varepsilon_{*}\pm\frac{C}{2\pi}I_{\pm}e^{\mp\theta}\label{eq:asyepsilon}
\end{equation}
where the asymptotic behavior of the kernel and of the source term
of (\ref{eq:epsilonzero}) define $\varphi(\theta\to\pm\infty)\sim Ce^{\mp\theta}$
and $\log T_{+}(i\frac{\pi}{2}-\theta\to\pm\infty)\sim A_{\pm}e^{\mp\theta}$.
For the model at hand, $C=-2\sqrt{3}$ and $A_{\pm}=\mp2i(e^{\pm i\pi\frac{b+1}{6}}+e^{\pm i\pi\frac{b-1}{6}})$.
Finally, $\varepsilon_{*}=\log\frac{1+\sqrt{5}}{2}$ is the plateau
value of the limit solution for the pseudo energy and we introduce
the notation:
\begin{equation}
I_{\pm}=\int_{-\infty}^{\infty}d\theta e^{\pm\theta}\frac{d}{d\theta}\log(1+e^{-\varepsilon_{0}(\theta)})
\end{equation}

We define $Y(\theta)=e^{-\varepsilon_{0}(\theta)}$, which satisfies
the 

\begin{equation}
Y\left(\theta+i\frac{\pi}{3}\right)Y\left(\theta-i\frac{\pi}{3}\right)=1+Y(\theta),
\end{equation}
functional relation, i.e. the Lee-Yang $Y$ system \cite{Bajnok:2007jg}.
This function has period $\frac{5i\pi}{3}$ in the imaginary direction,
hence its asymptotic expansion can only contain powers of $e^{\pm\frac{6}{5}\theta}$:
cancellation of the terms proportional to $e^{\pm\theta}$ in (\ref{eq:asyepsilon})
allows to compute $I_{\pm}$ exactly. From the expression (\ref{eq:momentumdef}),
one then has 
\begin{equation}
\lim_{L\to0}\frac{P_{0}^{TBA}}{m}=-\frac{A_{+}-A_{-}}{2C}
\end{equation}
so that the defect momentum is (\ref{eq:defectenergyandmomentum}).

As a last step, we take the vacuum expectation values of the Hamiltonian
and the momentum operator, (\ref{eq:HTCSA}) (\ref{eq:PTCSA}), and
differentiate them with respect to the defect parameter $b$
\begin{equation}
\frac{\partial}{\partial b}\langle H\rangle=i\frac{\pi}{5}(\mu\langle\varphi\rangle-\bar{\mu}\langle\bar{\varphi}\rangle)\qquad\qquad\frac{\partial}{\partial b}\langle P\rangle=i\frac{\pi}{5}(\mu\langle\varphi\rangle+\bar{\mu}\langle\bar{\varphi}\rangle)\label{eq:dbHP}
\end{equation}
 By virtue of (\ref{eq:mu}), (\ref{eq:defectenergyandmomentum}),
we obtain the vacuum expectation values of the fields $\varphi$ and
$\bar{\varphi}$, 
\begin{equation}
\langle\varphi\rangle=-\frac{5i}{12\mu}e^{-i\frac{\pi b}{6}}\qquad\qquad\langle\bar{\varphi}\rangle=\frac{5i}{12\bar{\mu}}e^{i\frac{\pi b}{6}}\label{eq:varphiVEV}
\end{equation}
which we compare with the numerical values in Section \ref{sec:Numerical-comparison}.
Observe that differentiating in eq. (\ref{eq:dbHP}) the exact finite
volume ground state energy (\ref{eq:E0L}) and ground state momentum
(\ref{eq:momentumdef}), instead of their asymptotic values (\ref{eq:defectenergyandmomentum}),
we could exactly derive the complete finite size one point function
of the defect operators $\varphi$ and $\bar{\varphi}.$

\subsubsection{Spectral expansion of two-point functions \label{sub:Spectral-expansion-of}}

The form factor expansion for correlation functions has proven to
be extremely rapidly convergent in the Lee-Yang theory \cite{Zamolodchikov:1990bk},
providing a good estimate for the correlation function 
\begin{equation}
\langle\Phi(x,t)\Phi(0,0)\rangle
\end{equation}
up to very small values of the separation between the fields. It is
therefore a legitimate check of the form factors expressions, as well
as a due comparison of the convergence properties of the spectral
expansion in the case of operators living on the defect, to repeat
this analysis for the various two-point functions of $\varphi$, $\bar{\varphi}$,
$\Phi_{\pm}$.

In the following we analyze the two point functions of local operators
$\langle\hat{O}_{1}(r)\hat{O}_{2}(0)\rangle$ by inserting a resolution
of the identity 
\begin{equation}
\langle\hat{O}_{1}(r)\hat{O}_{2}(0)\rangle=\sum_{n=0}^{\infty}\langle0\vert\hat{O}_{1}(0)\vert n\rangle\langle n\vert\hat{O}_{2}(0)\vert0\rangle e^{-E_{n}r}
\end{equation}
The various matrix elements are the generic form factors (\ref{eq:genff}),
which all can be expressed in terms of the elementary ones. Truncation
of the series up to two particle terms gives a good approximation
valid even for very small separations, which can be compared to the
short distance CFT predictions. In so doing we assume that the local
fields in the perturbed theory are in one-to-one correspondence with
the operator content of the CFT, apart from additive renormalization
constants \cite{Zamolodchikov:1990bk}. The products of fields living
on the defect $\hat{O}_{1}(r),\hat{O}_{2}(0)$, for small Euclidean
time separation, $r$, can be treated by exploiting their operator
product expansion in the short-distance CFT

\begin{equation}
\hat{O}_{1}(r)\hat{O}_{2}(0)\sim\sum_{j}\frac{C_{12}^{j}\hat{O}_{j}}{|r|^{h_{1}+h_{2}-h_{j}}}\label{eq:ope}
\end{equation}
where $h_{i}$ denote the scaling dimension of the operators and the
structure constants $C_{12}^{j}$ were given in \cite{Bajnok:2013}.
Knowledge of the exact vacuum expectation values (\ref{eq:PhiVEV}),
(\ref{eq:varphiVEV}), allows one to extract the behavior of the correlation
function as $r\to0$. 

We found that the spectral series reproduce the correlation functions
to very good accuracy, even for small values of $r$, by a restricted
number of terms only. In the following we focus on the one- and two-particle
contributions.

\begin{figure}[H]
\centering{}\includegraphics[width=0.7\textwidth]{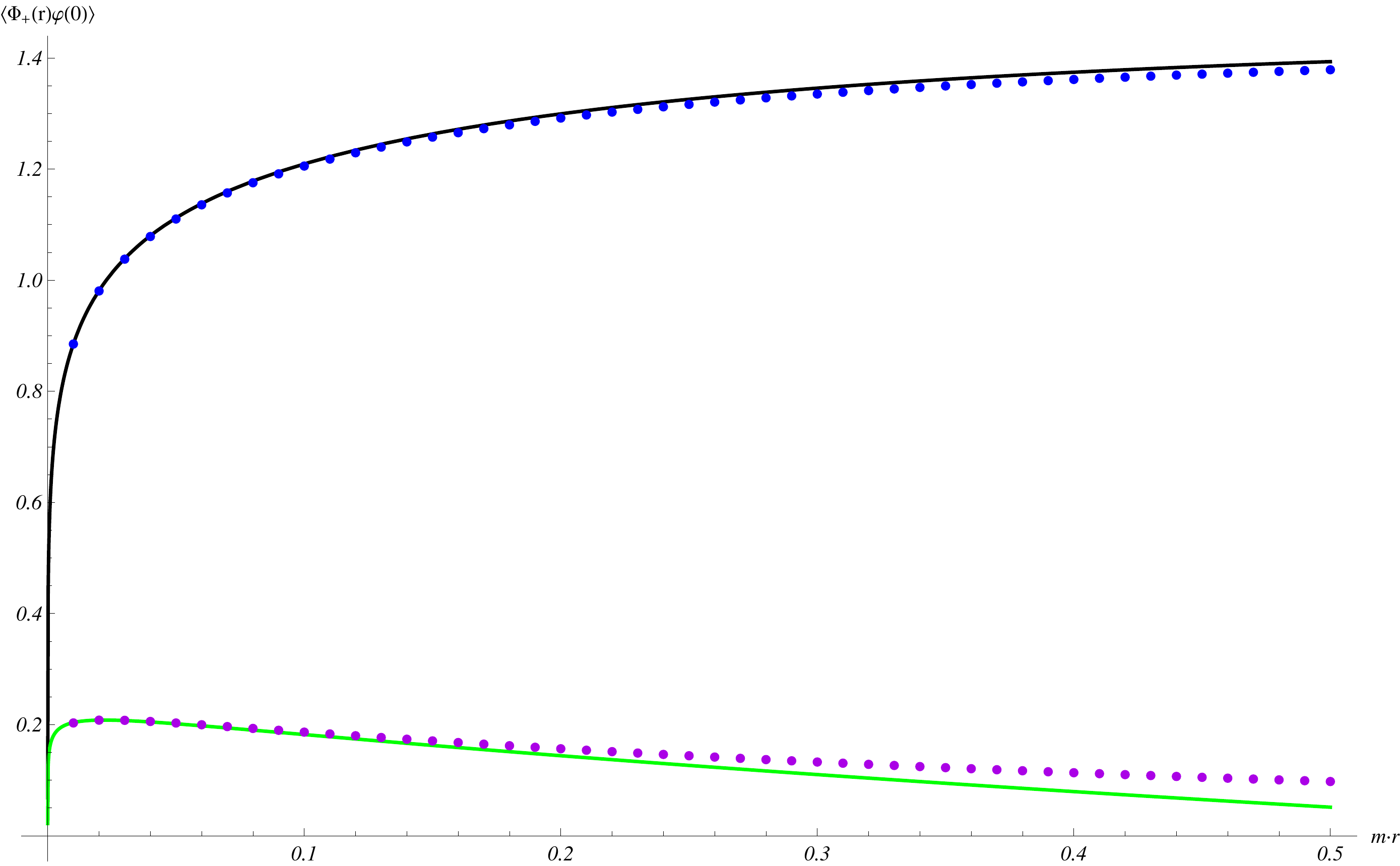}\caption{$\Phi_{+}\varphi$ correlation function: blue and purple dots show
the real and imaginary part calculated from the form factor expansion
up to two particle terms, while black and green curves the first order
CFT perturbation results\label{fig:Phippertphi}}
\end{figure}

The relevant structure constants are written in terms of the constants
$\beta=\sqrt{\frac{2}{1+\sqrt{5}}}$, $\alpha=\sqrt{\frac{\Gamma(1/5)\Gamma(6/5)}{\Gamma(3/5)\Gamma(4/5)}}$
and $\eta=e^{i\frac{\pi}{5}}$. In particular, we show in Figure \ref{fig:Phippertphi}
the real and imaginary part of the $\Phi_{+}\varphi$ correlation
function, which for short distance is expanded as 
\begin{equation}
\langle\Phi_{+}(r)\varphi(0)\rangle\sim C_{\Phi_{+}\varphi}^{\Phi_{+}}\langle\Phi_{+}\rangle r^{1/5}+C_{\Phi_{+}\varphi}^{\Phi_{-}}\langle\Phi_{-}\rangle r^{1/5}+C_{\Phi_{+}\varphi}^{\bar{\varphi}}\langle\bar{\varphi}\rangle r^{2/5}
\end{equation}
with $C_{\Phi_{+}\varphi}^{\Phi_{+}}=\frac{\alpha}{2}(\beta+\beta^{-1}+\frac{i}{\sqrt[4]{5}})$,
$C_{\Phi_{+}\varphi}^{\Phi_{-}}=\frac{\alpha\beta}{2}(1-\frac{i(\beta-\beta^{-1})}{\sqrt[4]{5}})$,
$C_{\Phi_{+}\varphi}^{\bar{\varphi}}=-\frac{\eta}{\beta}$. The short
distance CFT expansion is shown with continuous lines, while the form
factor expansion with dots. 
The two point functions of other fields are analyzed in Appendix B.

\section{Finite volume form factors: theoretical framework }

To verify the predictions made by the defect form factor bootstrap
one can use the finite volume form factor formalism developed by Pozsgay
and Takacs \cite{Pozsgay:2007kn,Pozsgay:2007gx}. Based on the description
of the finite volume spectrum provided by the Bethe-Yang equations,
the formalism gives all finite volume corrections that decay as a
power in the inverse volume. The remaining corrections are suppressed
exponentially as the volume increases, and at present only a partial
description is available for them \cite{Takacs:2011nb}. In this paper
we confine ourselves to the power corrections, as these are sufficient
to verify the validity of the defect form factor bootstrap.

\subsection{Finite volume energy levels}

The finite volume energy levels can be identified with multi-particle
states $|\{I_{1},\dots,I_{n}\}\rangle_{L}$ containing $n$ particles,
labeled by quantum numbers $I_{1},\dots,I_{n}$ which parametrize
the quantization of particle momenta. The quantization conditions
satisfied by the particle rapidities $\theta_{k}$ are the Bethe-Yang
equations 
\begin{equation}
e^{imL\sinh\theta_{k}}T_{-}(\theta_{k})\prod_{j\neq k}S(\theta_{k}-\theta_{j})=1\qquad k=1,\dots,n\label{eq:expdybe}
\end{equation}
Taking the logarithm, these equations can be rewritten as
\begin{equation}
Q_{k}(\theta_{1},\dots,\theta_{n})_{L}=2\pi I_{k}\qquad k=1,\dots,n\label{eq:dby1}
\end{equation}
where
\begin{equation}
Q_{k}(\theta_{1},\dots,\theta_{n})_{L}=mL\sinh\theta_{k}-i\log T_{-}(\theta_{k})-\sum_{j\neq k}i\log S(\theta_{k}-\theta_{j})\label{eq:dby2}
\end{equation}
and the energy is given by
\begin{equation}
E(L)=E_{0}(L)+\sum_{k=1}^{n}m\cosh\theta_{k}+O(e^{-\mu L})
\end{equation}
where $\mu$ is some characteristic scale, and $E_{0}(L)$ is the
ground state (vacuum) energy.

The $k$-th equation in (\ref{eq:dby1}) characterizes the monodromy
of the wave functions under moving the $k$-th particle to the right
and around the circle; in doing so, one picks up the phase from crossing
the defect, the scattering with the other particles (note the order
of the rapidity difference inside $S$, which corresponds to particle
$k$ entering the scattering from the left!). The reason why only
$T_{-}(\theta)$ enters is that it is the phase-shift suffered by
a particle of $\theta>0$ when crossing the defect from the left;
on the other hand, if a given particle has $\theta<0$ its monodromy
when crossing from the right to the left would be given by $T_{+}(-\theta)$;
therefore, when crossing from the left to the right the phase-shift
is given by
\begin{equation}
T_{+}(-\theta)^{-1}
\end{equation}
which is equal to $T_{-}(\theta)$ by defect unitarity. As a result,
equations (\ref{eq:dby1}), (\ref{eq:dby2}) describe all states in
the spectrum by letting the sign of the rapidities free, i.e. by letting
$I_{k}$ to take any integer values. However, due to $S(0)=-1$ the
multi-particle wave-functions are non-vanishing only if all the rapidities,
and therefore all the quantum numbers, take distinct values. 

The density of states in finite volume can be obtained from the Jacobi
determinant of the mapping from rapidity space to the space of the
quantum numbers:
\begin{equation}
\rho_{n}(\theta_{1},\dots,\theta_{n})_{L}=\det\left\{ \frac{\partial Q_{k}(\theta_{1},\dots,\theta_{n})_{L}}{\partial\theta_{j}}\right\} _{k,j=1,\dots,n}
\end{equation}

\subsection{Non-diagonal matrix elements}

Using the arguments in the work \cite{Pozsgay:2007kn}, the finite
volume matrix elements of a defect operator can be obtained as 
\begin{eqnarray}
 &  & \left\langle \left\{ J_{1},\dots,J_{n}\right\} \right|\mathcal{O}\left(t=0\right)\left|\left\{ I_{1},\dots,I_{k}\right\} \right\rangle _{L}=\label{eq:Finite_volume_FF}\\
 &  & \pm\frac{F^{\mathcal{O}}\left(\tilde{\theta}_{n}'+i\pi,\dots,\tilde{\theta}_{1}'+i\pi,\tilde{\theta}_{1},\dots,\tilde{\theta}_{k}\right)}{\sqrt{\rho_{n}\left(\tilde{\theta}_{1}',\dots,\tilde{\theta}_{n}'\right)_{L}\rho_{k}\left(\tilde{\theta}_{1},\dots,\tilde{\theta}_{k}\right)}_{L}}\Phi\left(\tilde{\theta}_{1}',\dots,\tilde{\theta}_{n}'\right)^{*}\Phi\left(\tilde{\theta}_{1},\dots,\tilde{\theta}_{k}\right)+O\left(e^{-\mu L}\right)\nonumber 
\end{eqnarray}
where $F^{\mathcal{O}}$ is the infinite volume form factor, $\tilde{\theta}_{1},\dots,\tilde{\theta}_{k}$
and $\tilde{\theta}_{1}',\dots,\tilde{\theta}_{n}'$ are the solutions
to the Bethe-Yang equations (\ref{eq:dby1}), (\ref{eq:dby2}) with
quantum numbers $I_{1},\dots,I_{k}$ and $J_{1},\dots,J_{n}$, respectively,
and the $\Phi$ are phase factors ensuring that the finite volume
scattering state is symmetric in its arguments and is invariant under
crossing any particle of the defect. Clearly in finite volume the
particles are not ordered and are on the left and on the right of
the defect in the same time. The correct phase factor with this properties
have the form
\begin{equation}
\Phi\left(\theta_{1},\dots,\theta_{n}\right)=\pm\left(\prod_{{j,l=1\atop j<l}}^{n}S(\theta_{j}-\theta_{l})\prod_{j=1}^{n}T_{-}(\theta_{j})\right)^{-\frac{1}{2}}
\end{equation}
We remark that the phases $\Phi$ have no physical significance as
they cancel in any correlation function computed in the theory, whether
in finite or infinite volume; it is only necessary to keep track of
them when analytically continuing (\ref{eq:Finite_volume_FF}) to
complex values of the rapidities, as in \cite{Pozsgay:2008bf,Takacs:2011nb}.
Finally, the $\pm$ sign results from the ambiguity in choosing the
branch of the square root functions; the same ambiguity is present
in TCSA as the eigenvectors still have a residual sign ambiguity even
after imposing a suitable reality condition.

In the case of the Lee-Yang model we have a physical normalization
of the TCSA eigenvectors. We chose the normalization of the UV primary
defect creating fields as 
\begin{equation}
C_{dd}^{\mathbb{I}}=C_{\bar{d}\bar{d}}^{\mathbb{I}}=1\quad;\qquad C_{DD}^{\mathbb{I}}=-1,
\end{equation}
the same way as in \cite{Bajnok:2013}. The ground state of the perturbed
CFT flows to the lowest energy state in the UV limit which is the
field $D$ with negative squared norm. This is therefore natural to
chose the ground state TCSA vectors at any volume to be purely imaginary.

The second and the third lowest energy states in the TCSA (which correspond
to the slowest left- and right-moving one particle states in the scattering
theory point of view) never cross any other lines at any volume, therefore,
they flow to the second and the third lowest energy states in the
conformal theory, namely $\vert d\rangle$ and $\vert\bar{d}\rangle$.
Consequently we normalized the corresponding TCSA vectors to be real.

In the UV limit the states of the Verma modules built over the highest
weight states $d$ and $\bar{d}$ are related by parity transformation
while the Verma module of $D$ is parity symmetric. In case of a parity
symmetric defect the only parity symmetric states in the scattering
theory are the even particle states composed by particles with opposite
rapidities, so these states flow to some parity symmetric state in
the $D$-module. These arguments suggest us that the even particle
TCSA states flow to the $D$-module while the odd particle TCSA states
to either to the $d$-module or to the $\bar{d}$-module in the conformal
limit. For this reason we normalized the even particle TCSA states
to be purely imaginary while the odd particle states to be real. This
assumption is tested \textit{\textcolor{black}{a posteriori}}, measuring
the phases (or equivalently both the real and imaginary part) of the
form factors from the TCSA and comparing them to the theoretically
computed finite volume form factors allow us a non-trivial check,
and we found a perfect agreement for all the studied states.

For bulk operators, the extension of (\ref{eq:Finite_volume_FF})
is simple. As noted in \cite{Bajnok:2009hp}, the presence of an integrable
defect only breaks translational invariance by having a defect momentum
$p_{D}$ (in addition to a defect energy $E_{D}$). The total momentum,
which includes the defect momentum as well, is conserved; therefore
the space-time dependence of a bulk operator can be computed by multiplying
(\ref{eq:Finite_volume_FF}) by the phase factor
\begin{equation}
e^{-it\Delta E+ix\Delta P}
\end{equation}
where
\begin{eqnarray}
\Delta E & = & \sum_{i=1}^{k}m\cosh\tilde{\theta}_{i}-\sum_{j=1}^{n}m\cosh\tilde{\theta}_{j}'\nonumber \\
\Delta P & = & \sum_{i=1}^{k}m\sinh\tilde{\theta}_{i}-\sum_{j=1}^{n}m\sinh\tilde{\theta}_{j}'
\end{eqnarray}

\subsection{Diagonal matrix elements}

Formula (\ref{eq:Finite_volume_FF}) is valid if there are no disconnected
terms in the matrix element. Disconnected terms arise when there is
at least one rapidity value among the $\tilde{\theta}_{1},\dots,\tilde{\theta}_{m}$
which coincides with a value occurring in $\tilde{\theta}_{1}',\dots,\tilde{\theta}_{n}'$.
Following two energy levels as the volume $L$ varies, this can occur
at particular isolated values of $L$, but these cases are not interesting
as the matrix element can be evaluated by taking the limit of (\ref{eq:Finite_volume_FF})
in the volume. Therefore the only interesting cases are when disconnected
terms are present for a continuous range of the volume $L$. Due to
the presence of interactions (the $S$ terms) in (\ref{eq:dby1}),
(\ref{eq:dby2}) this can only occur in very specific situations;
the only generic class is when the matrix element is diagonal, i.e.
the two states are eventually identical, in which case disconnected
terms are present for all values of $L$.

In this case, we can proceed by analogy to the bulk and boundary cases.
For diagonal matrix elements
\begin{equation}
\langle\{I_{1},\dots,I_{n}\}\vert\mathcal{O}\vert\{I_{1},\dots,I_{n}\}\rangle_{L}
\end{equation}
equation (\ref{eq:Finite_volume_FF}) shows that the relevant form
factor expression is 
\begin{equation}
F^{\mathcal{O}}(\theta_{n}+i\pi,...,\theta_{1}+i\pi,\theta_{1},...,\theta_{n})
\end{equation}
Due to the existence of kinematical poles this must be regularized;
however the end result depends on the direction of the limit. The
terms that are relevant in the limit can be written in the following
general form: 
\begin{eqnarray}
F^{\mathcal{O}}(\theta_{n}+i\pi+\epsilon_{n},...,\theta_{1}+i\pi+\epsilon_{1},\theta_{1},...,\theta_{n})=\label{mostgeneral}\\
\prod_{i=1}^{n}\frac{1}{\epsilon_{i}}\cdot\sum_{i_{1}=1}^{n}...\sum_{i_{n}=1}^{n}\mathcal{A}_{i_{1}...i_{n}}(\theta_{1},\dots,\theta_{n})\epsilon_{i_{1}}\epsilon_{i_{2}}...\epsilon_{i_{n}}+\dots\nonumber 
\end{eqnarray}
 where $\mathcal{A}_{i_{1}...i_{n}}^{a_{1}\dots a_{n}}$ is a completely
symmetric tensor of rank $n$ in the indices $i_{1},\dots,i_{n}$,
and the ellipsis denote terms that vanish when taking $\epsilon_{i}\rightarrow0$
simultaneously. 

The connected matrix element can be identified as the $\epsilon_{i}$
independent part of equation (\ref{mostgeneral}), i.e. the part which
does not diverge whenever any of the $\epsilon_{i}$ is taken to zero:
\begin{equation}
F_{conn}^{\mathcal{O}}(\theta_{1},\theta_{2},...,\theta_{n})=n!\,\mathcal{A}_{1\dots n}(\theta_{1},\dots,\theta_{n})\label{eq:connected}
\end{equation}
where the appearance of the factor $n!$ is simply due to the permutations
of the $\epsilon_{i}$. 

Following \cite{Pozsgay:2007gx,Kormos:2007qx} , we are lead to the
following expression
\begin{eqnarray}
 &  & \,\langle\{I_{1}\dots I_{n}\}|\mathcal{O}|\{I_{1}\dots I_{n}\}\rangle_{L}=\label{eq:diaggenrulesaleur}\\
 &  & \frac{1}{\rho_{n}(\tilde{\theta}_{1},\dots,\tilde{\theta}_{n})_{L}}\,\sum_{A\subset\{1,2,\dots n\}}F_{conn}^{\mathcal{O}}(\{\tilde{\theta}_{k}\}_{k\in A})\tilde{\rho_{n}}(\tilde{\theta}_{1},\dots,\tilde{\theta}_{n}|A)_{L}+O(\mathrm{e}^{-\mu L})\nonumber 
\end{eqnarray}
where the summation runs over all subsets $A$ of $\{1,2,\dots n\}$,
and $\left\{ \tilde{\theta}_{1},\dots,\tilde{\theta}_{n}\right\} $
are the Bethe-Yang rapidities corresponding to the set of quantum
numbers $\left\{ I_{1},\dots,I_{n}\right\} $. For any such subset,
we define the appropriate sub-determinant
\begin{equation}
\tilde{\rho}_{n}(\tilde{\theta}_{1},\dots,\tilde{\theta}_{n}|A)=\det\mathcal{J}_{A}(\tilde{\theta}_{1},\dots,\tilde{\theta}_{n})
\end{equation}
of the $n\times n$ Bethe-Yang Jacobi matrix 
\begin{equation}
\mathcal{J}(\tilde{\theta}_{1},\dots,\tilde{\theta}_{n})_{kl}=\frac{\partial Q_{k}(\theta_{1},\dots,\theta_{n})_{L}}{\partial\theta_{l}}\label{eq:jacsubmat}
\end{equation}
where $\mathcal{J}_{A}$ is obtained by deleting the rows and columns
corresponding to the subset of indices $A$. The determinant of the
empty sub-matrix (i.e. when $A=\{1,2,\dots n\}$) is defined to equal
to $1$ by convention. Note also that diagonal matrix elements have
no space-time dependence at all, therefore (\ref{eq:diaggenrulesaleur})
is true both for operators located on the defect and in the bulk. 

We note for bulk theories on a spatial circle without defects, that
there exists another class of matrix elements with disconnected contributions
when there is a particle of exactly zero momentum in both states.
However, due to the presence of the defect this class is absent here.
For example, from (\ref{eq:expdybe}) it follows that the existence
of a state with a single stationary particle would require
\begin{equation}
T_{-}(\theta=0)=1
\end{equation}
but this is not satisfied for any finite value of the defect parameter
$b$. The only class of matrix elements that has disconnected pieces
at more than isolated values of $L$ is the diagonal one treated above.

\section{Numerical comparison\label{sec:Numerical-comparison}}

A valuable tool for investigating statistical field theories in the
vicinity of the critical point was devised in \cite{Yurov:1989yu}
and successfully tested on the scaling Lee-Yang theory. Being based
on the knowledge of the Hilbert space at the conformal point, where
the spectrum is discrete, and on the truncation of the constituting
Verma modules at a certain level $m$, it has been dubbed truncated
conformal space approach (TCSA).

In the present case, the Hilbert space is spanned by the defect-creating
operators, whose corresponding hw. states are denoted by $|D\rangle$,
$|d\rangle$, $|\bar{d}\rangle$ in \cite{Bajnok:2013}, with conformal
dimensions $(h,\bar{h})=(-\frac{1}{5},-\frac{1}{5}),(-\frac{1}{5},0),(0,-\frac{1}{5})$.

In order to evaluate matrix elements of the Hamiltonian, the theory
is mapped on the plane by the transformation

\begin{equation}
z=e^{-i\frac{2\pi}{L}\zeta},\quad\bar{z}=e^{i\frac{2\pi}{L}\bar{\zeta}}\label{eq:expmap}
\end{equation}
where $\zeta=x+iy$ and $\bar{\zeta}=x-iy$ are the Euclidean coordinates
on the cylinder. Operators will appear as finite-size matrices on
the states $\vert j\rangle$ of the conformal Hilbert space on the
plane and the Hamiltonian and momentum operators read
\begin{eqnarray}
\frac{H}{m} & = & \frac{2\pi}{mL}\biggl(L_{0}+\bar{L}_{0}+\frac{11}{30}+\xi\Bigr(\frac{L}{2\pi\kappa}\Bigr)^{1+\frac{1}{5}}\biggl(a\left(G^{-1}\hat{\varphi}\right){}_{jk}+\bar{a}\left(G^{-1}\hat{\bar{\varphi}}\right){}_{jk}\biggr)\label{eq:HTCSA}\\
 &  & \,\,\,\,\,\,\,\,\,\,\,\,\,\,\,\,\,\,\,\,\,\,\,\,\,\,\,\,\,\,\,\,\,\,\,\,\,\,\,\,\,\,\,\,\,\,\,\,\,\,\,\,\,\,\,\,+\Bigr(\frac{L}{2\pi\kappa}\Bigr)^{2+\frac{2}{5}}\left(G^{-1}\Phi\right){}_{jk}\biggr)\nonumber 
\end{eqnarray}

\begin{equation}
\frac{P}{m}=\frac{2\pi}{mL}\left(L_{0}-\bar{L}_{0}+\xi\left(\frac{L}{2\pi\kappa}\right)^{1+\frac{1}{5}}\left(a(G^{-1}\hat{\varphi})_{jk}-\bar{a}(G^{-1}\hat{\bar{\varphi}})_{jk}\right)\right)\label{eq:PTCSA}
\end{equation}
where $G_{jk}=\langle j|k\rangle$, $\hat{\varphi}_{jk}=\langle j|\varphi(1)|k\rangle$,
$\hat{\bar{\varphi}}_{jk}=\langle j|\bar{\varphi}(1)|k\rangle$, $\hat{\Phi}_{jk}=\langle j|\Phi(1,1)|k\rangle P_{jk}$
and the matrix 
\begin{equation}
P_{jk}=\begin{cases}
\begin{array}{c}
-2\pi\\
-2e^{-i\pi(h_{k}-\bar{h}_{k}-h_{j}+\bar{h}_{j})}\frac{\sin\pi(h_{k}-\bar{h}_{k}-h_{j}+\bar{h}_{j})}{(h_{k}-\bar{h}_{k}-h_{j}+\bar{h}_{j})}
\end{array} & \begin{array}{c}
\mbox{ if }h_{k}-\bar{h}_{k}-h_{j}+\bar{h}_{j}=0\\
\mbox{otherwise}
\end{array}\end{cases}
\end{equation}
is obtained by performing the integration on the spatial coordinate
of the bulk perturbing field. Also, there appear the parameters $\mbox{\ensuremath{\kappa}=\ensuremath{\frac{2^{19/12}\sqrt{\pi}\left(\Gamma\left(\frac{3}{5}\right)\Gamma\left(\frac{4}{5}\right)\right)^{5/12}}{5^{5/16}\Gamma\left(\frac{2}{3}\right)\Gamma\left(\frac{5}{6}\right)}}}$
, $\xi=\sqrt[4]{\frac{5}{8}+\frac{3\sqrt{5}}{8}}$, $a=e^{-i\frac{\pi}{5}(b-2)}$
and $\bar{a}=e^{i\frac{\pi}{5}(b-2)}$.

Diagonalization of the truncated Hamiltonian yields the energy levels,
which can be compared with the corresponding quantities obtained from
Bethe-Yang equations (\ref{eq:expdybe}), finding excellent agreement
in all intermediate regimes \cite{Bajnok:2013}.

Truncation of the Hilbert space introduces an error in the computation
of energy levels and matrix elements. However, at least for an operator
with weight $h,\bar{h}<\frac{1}{2}$, such an error is expected to
be smaller and smaller as the truncation level is increased. We found
that, due to the rapid growth of the Hilbert space and the consequent
processor memory usage, it was impossible for us to go beyond truncation
level $n=18$, where the effects due to the finiteness of the space
are still noticeable. This forced us to improve the precision by renormalization
group (RG) methods. 

A renormalization group for the truncation level dependence was introduced
in \cite{Feverati:2006ni,Konik:2007cb,Giokas:2011ix} and extended
to VEVs in \cite{Szecsenyi:2013gna}. Here we briefly repeat the basic
steps of the derivation in \cite{Szecsenyi:2013gna} for matrix elements
with the perturbed action (\ref{eq:SLY}), and refer to the original
derivation for more extensive explanation of the procedure.

The matrix element of a given local operator $\mathcal{O}$, represented
on a finite-dimensional space, will be denoted by 

\begin{equation}
O_{ab}(n)=\langle a|P_{n}\mathcal{O}(0,0)P_{n}e^{-\lambda\int d^{2}rV(\vec{r})-\eta\int dtW(t)}P_{n}|b\rangle
\end{equation}
where the last factor in the expectation value represents a combined
bulk and defect perturbation, as in (\ref{eq:SLY}), the operator
$P_{n}$ is a projector onto the states up to level $n$, and the
operator products are assumed to be time ordered.

The idea is to study the difference $O_{ab}(n+1)-O_{ab}(n)$ by perturbation
theory using that the couplings $\lambda$, $\eta$ are relevant,
and so when the descendant states at the cutoff level $n$ have energy
much higher than the mass gap $\frac{4\pi n}{L}\gg m$ their effect
can be taken into account perturbatively. The bulk perturbation has
already been analyzed in the original literature and the method can
be straightforwardly extended to a combined bulk and defect perturbation;
we concentrate here on the defect part, and only present the main
differences with respect to the original derivation in \cite{Szecsenyi:2013gna}.
It is possible to consider separately the two components of the perturbation
because the (projected) parts of the evolution operators associated
to the two perturbations commute to first order in $\lambda,\eta$.
As a first step, we write the matrix element in the form
\begin{equation}
Q_{ab}(n)=\langle a|U_{+}^{(n)}(\eta)P_{n}O(0)P_{n}U_{-}^{(n)}(\eta)|b\rangle
\end{equation}
where $U_{\pm}^{(n)}$ are past and future evolution operators at
a given cutoff. We evaluate the difference 

\begin{eqnarray}
Q_{ab}(n+1)-Q_{ab}(n) & = & -2\eta\left(\frac{L}{2\pi}\right)^{1-h_{W}-\overline{h}_{W}}\intop_{0}^{1}d\rho\rho^{-1+h_{W}+\overline{h}_{W}}\\
 &  & \qquad\langle a|U_{+}^{(n)}(\eta)O(1)(P_{n+1}-P_{n})W(\rho)U_{-}^{(n)}(\eta)|b\rangle+O(\eta^{2})\nonumber 
\end{eqnarray}
where we applied the exponential map (\ref{eq:expmap}), with $z=\rho e^{i\gamma}$
and put the defect at the radius $\gamma=0$. We then use the operator
product expansion

\begin{equation}
O(0)W(z)=\sum_{A}\frac{C_{OW}^{A}A(0)}{|1-\rho|^{h_{O}+h_{W}-h_{A}+\overline{h}_{O}+\overline{h}_{W}-\overline{h}_{A}}}
\end{equation}
where the summation runs over the scaling fields of the Hilbert space
at the conformal point, while the $C_{OW}^{A}$ are the associated
structure constants. In the above expression, the projector on the
$n$-th level singles out the corresponding descendants in the sum,
and after further manipulations in the limit $n\gg1$, one obtains

\begin{equation}
\frac{d}{dn}Q_{ab}(n)\propto\sum_{A}K_{A}n^{h_{O}+h_{W}-h_{A}+\overline{h}_{O}+\overline{h}_{W}-\overline{h}_{A}-2}
\end{equation}
which implies that truncation errors, in the case of a defect perturbation,
decay as

\begin{equation}
Q_{ab}(n)=Q_{ab}(\infty)+\sum_{A}\tilde{K}_{A}n^{h_{O}+h_{W}-h_{A}+\overline{h}_{O}+\overline{h}_{W}-\overline{h}_{A}-1}
\end{equation}
Conversely, in the case in which the action contains a bulk perturbation
only, it was found that

\begin{equation}
Q(n)=Q(\infty)+\sum_{A}\tilde{K}_{A}n^{h_{O}+h_{W}-h_{A}+\overline{h}_{O}+\overline{h}_{W}-\overline{h}_{A}-2}\label{eq:cutcorr}
\end{equation}
which shows why truncation effects from the defect perturbation are
generically more relevant than those resulting from the bulk perturbation.

For practical reasons, since the sub-leading terms which are not taken
into account in the formula above are $O(1/n)$, one retains in the
sum only the most relevant defect fields, which correspond to the
leading terms for large $n$.

Understanding the behavior of the matrix elements allows to extrapolate
their value when the cutoff in the state number tends to infinity
and allows comparison with the quantities computed in section 3. Explicit
examples can be obtained by using the OPE given in \cite{Bajnok:2013}:
in the case of $\varphi$ and $\bar{\varphi}$, the slowest correction
arises from the presence of the $\Phi_{\pm}$ channel in the OPE of
$\varphi$ and $\bar{\varphi}$ and gives a power of $n^{-1}$; conversely,
from the fact that the structure constants $C_{\Phi_{\pm}\varphi}^{\bar{\varphi}},\: C_{\Phi_{\pm}\bar{\varphi}}^{\varphi}$
are different from zero, one obtains that the cut-off dependence vanishes
as $n^{-6/5}$ for $\Phi_{\pm}$.

\subsection{Vacuum expectation values \label{sub:Form-factors-and}}

We would like to compare the expectation values of the fields $\Phi_{\pm}$,
$\varphi$, $\bar{\varphi}$ derived in section \ref{sub:Exact-vacuum-expectation},
to the data extracted from TCSA. Throughout this section, we work
at a fixed value of $b=-3+0.5i$. As explained in \cite{Bajnok:2013},
such a value ensures a real spectrum, thereby facilitating the comparison
between the energy and momentum data computed from the Bethe-Yang
equations and from TCSA. We chose a value of the defect parameter
with a small imaginary part in order to avoid the occurrence of degenerate
subspaces (corresponding, in the infrared, to particles traveling
with opposite rapidity). In the form factor comparison we are interested
not only in the eigenvalues but also in the eigenvectors of the TCSA
energy and momentum. As energy levels might be degenerate for specific
volumes the identification and systematic tracking of states can be
problematic. To avoid this we combine the self-adjoint $H$ and $P$
into $H+i\, P$, which has nondegenrate complex spectrum, and by following
its eigenstates we could identify $127$ states (up to four-particle
ones) from TCSA.

In the following we present the real and imaginary parts of measured,
extrapolated and theoretically calculated form factors. All of them
have the same consistent legend: green color shows the real part of
measured form factors at cut level 12, 14, 16, 18 $(\bullet,{\scriptstyle \blacksquare},{\scriptstyle \blacklozenge},\blacktriangle)$;
red squares show the real part of extrapolated data with confidence
intervals; and black line shows the real part of theoretical values.
The imaginary parts have the same structure, with colors purple, blue
and orange, respectively.

Concerning the confidence intervals we used Mathematica 9 to fit the
leading cut dependence via (\ref{eq:cutcorr}). We found that the
real and imaginary parts are basically not correlated thus we fit
them separately. On all the figures the confidence level is 95\%.
The theoretical data can be outside the confidence interval for two
reasons. For small volumes the TCSA data are reliable but the exponentially
supressed vacuum polarization effects of the form factor are no longer
negligible. For large volume the finite volume form factors are reliable
but the sub-leading TCSA cut dependence becomes relevant. 

\begin{figure}
\begin{centering}
\includegraphics[width=0.75\textwidth]{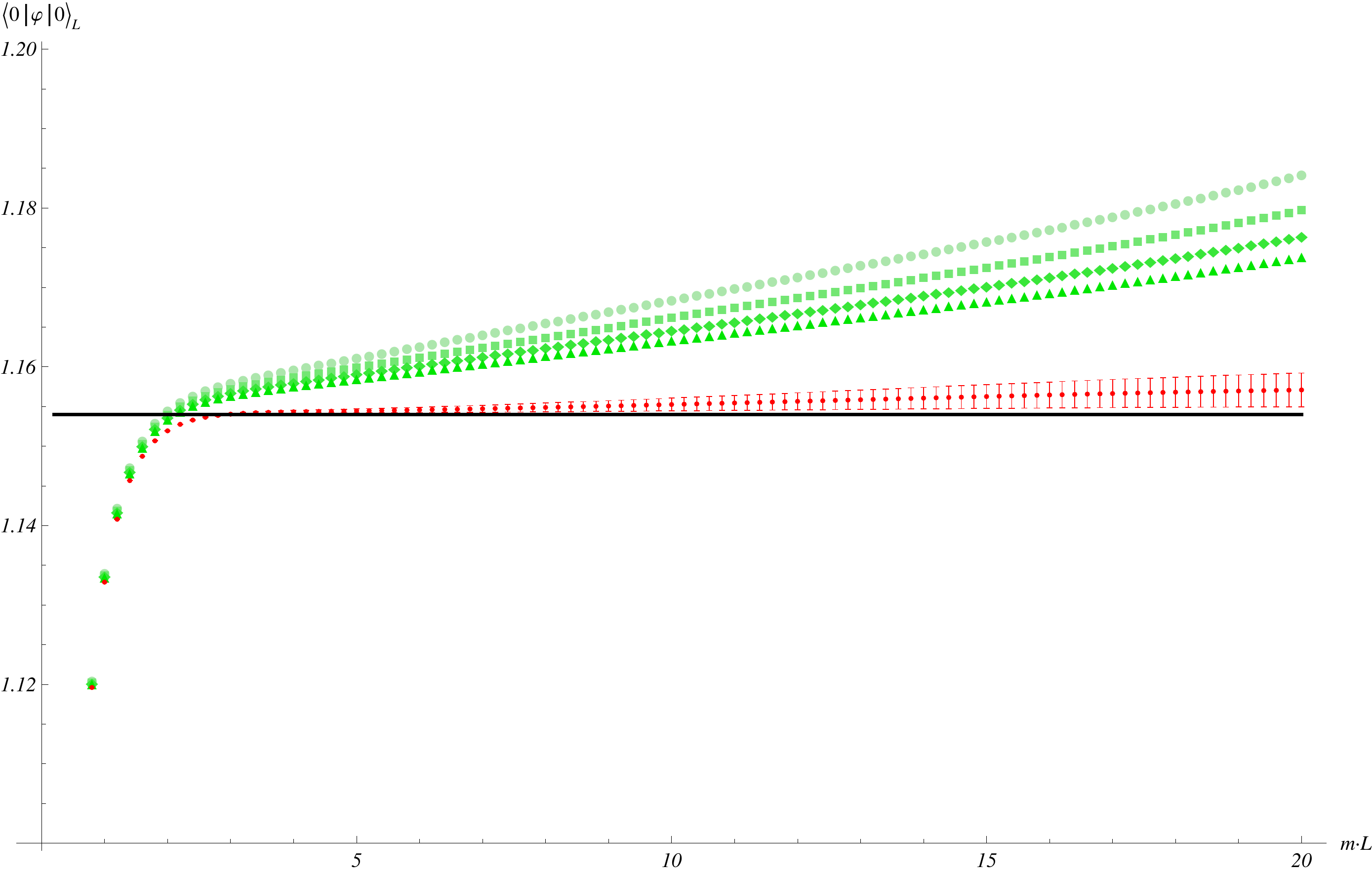}
\par\end{centering}

\caption{Comparison between the TCSA points at different cuts, the extrapolated
values and the theoretical vacuum expectation value of $\varphi$.}
\label{cutdep}
\end{figure}

First, we present an example of how the extrapolation procedure works
in the case of the defect field $\varphi$. In figure \ref{cutdep},
we show the TCSA data and the extrapolated value together with the
theoretical value for cylinder sizes between $0.2$ and $20$.\textbf{\textcolor{red}{{} }}

We report below the numerical comparison between the exact expectation
values of the operators living on the defect and the ones extrapolated
at a large enough volume $mL=15$, such that the exponential finite-size
corrections can be safely neglected.

\begin{table}[h]
\begin{centering}
\begin{tabular}{|c|c|c|c|}
\hline 
 & $\varphi$ & $\bar{\varphi}$ & $\Phi_{+}$\tabularnewline
\hline 
\hline 
TCSA extrapolated VEV & $1.2795$ & $1.1529$ & $1.2368$\tabularnewline
\hline 
Theoretical VEV & $1.2814$ & $1.154$ & $1.2394$\tabularnewline
\hline 
\end{tabular}
\par\end{centering}

\caption{Numerical comparison of the TCSA extrapolated and the theoretical
values of the vacuum expectation values of different operators at
volume $mL=15$. }
\end{table}

\subsection{One particle form factors }

We now turn to one-particle form factors, which, in the presence of
a defect, already carry a nontrivial dependence on the rapidity of
the particle due to the transmission factors from Section \ref{sub:Summary-of-the}
above.

We can collect extrapolated data from states labeled by different
quantization numbers and from different volumes (from $4$ to $20$).
The particle rapidity in a given state is determined from the BY equations
(\ref{eq:expdybe}). From this, it is known how to relate the finite
volume matrix elements and the infinite volume form factors, through
(\ref{eq:Finite_volume_FF}).

We now examine the one-particle form factors of the fields $\varphi$
and $\overline{\varphi}$. Following the procedure outlined above,
we obtain confirmation of the expected $\theta-$dependence.

\begin{figure}[H]
\centering{}\includegraphics[width=0.45\textwidth]{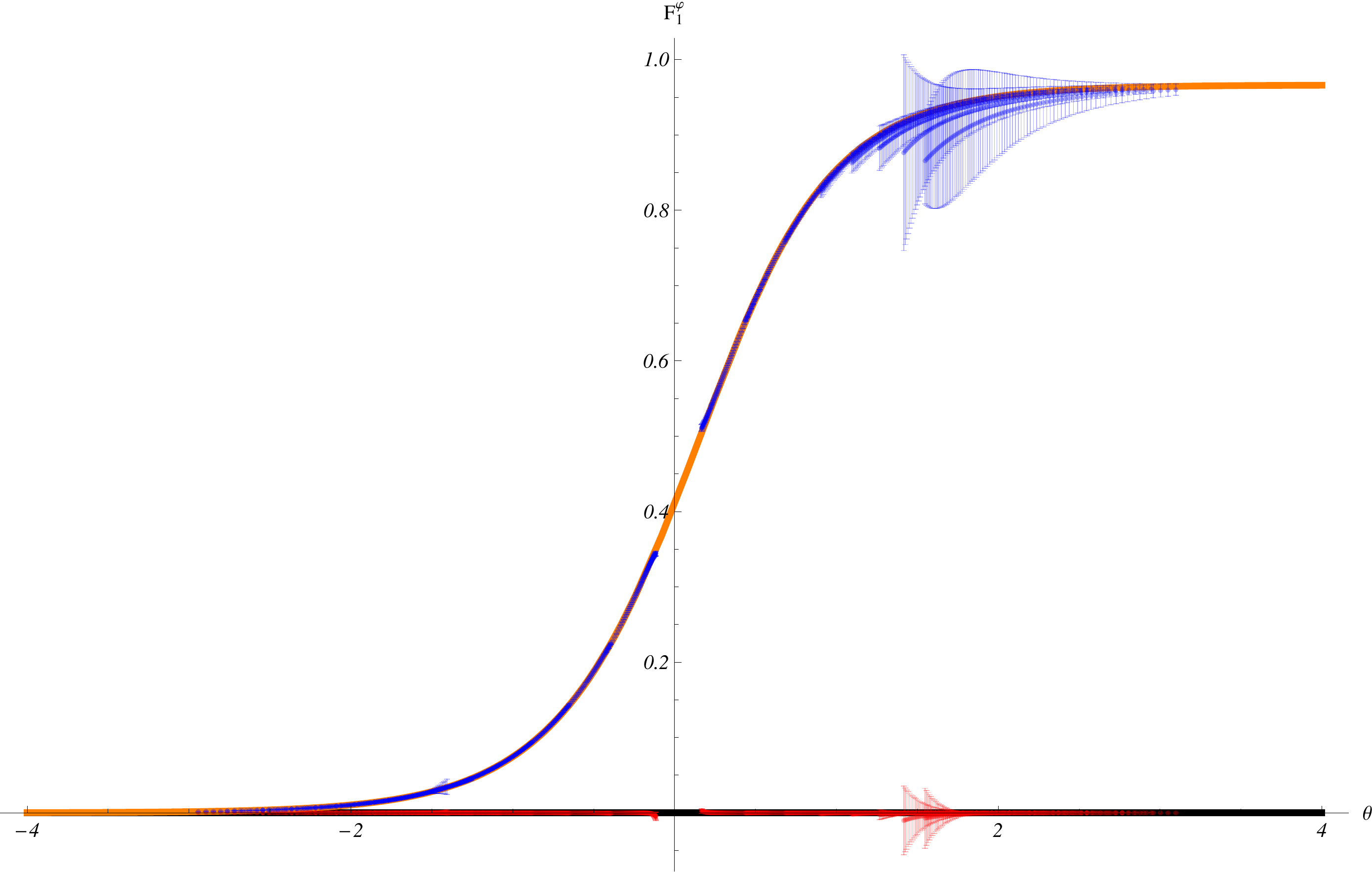}\includegraphics[width=0.45\textwidth]{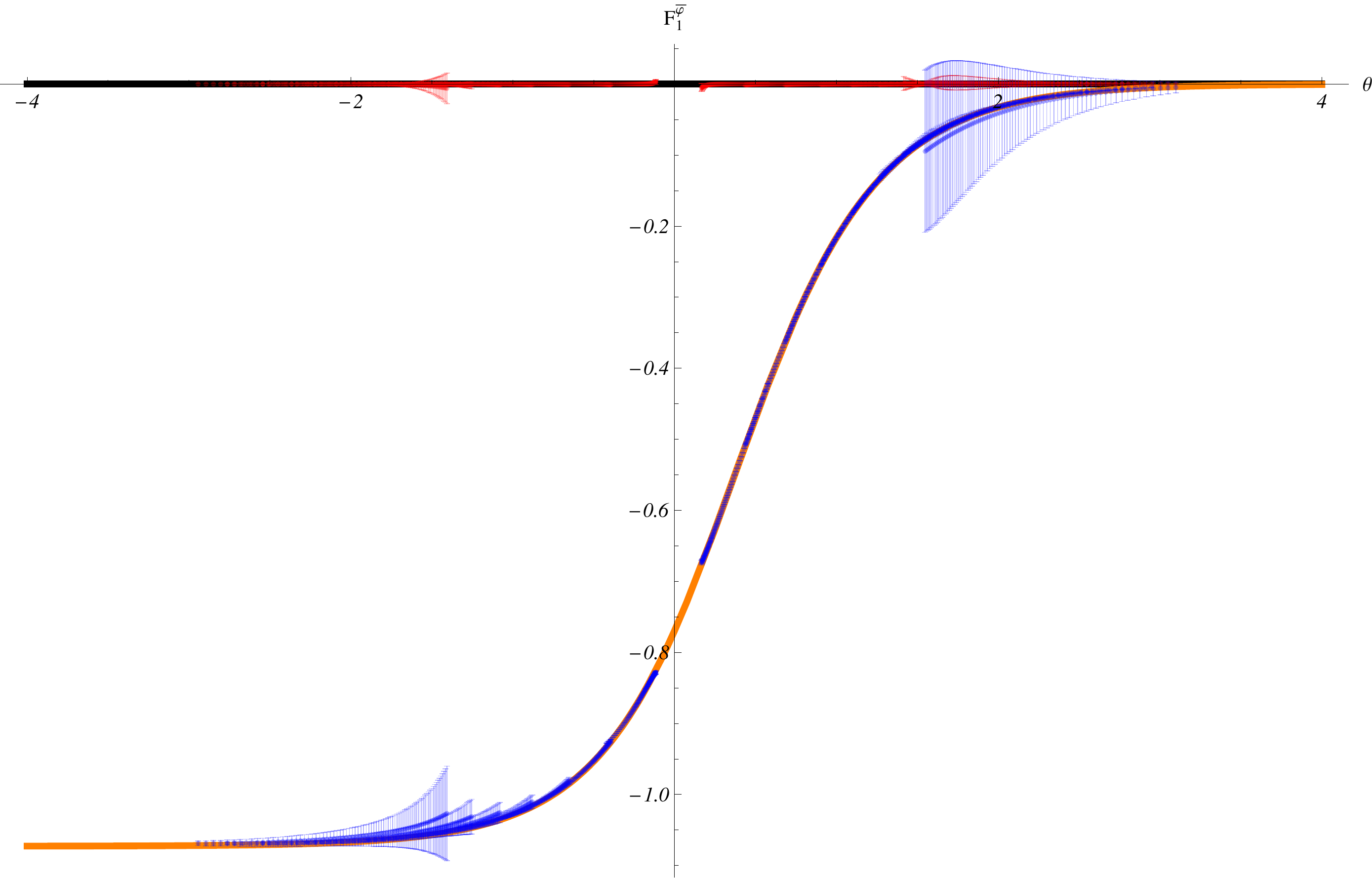}\caption{Comparison between the extrapolated TCSA data (dots with confidence
bars) and the theoretical prediction (solid line) for the one-particle
form factor of the operators $\varphi$ and $\bar{\varphi}$ .}
\end{figure}

Tho similar analysis for the one particle form factors of the operators
$\Phi_{\pm}$ can be found in Appendix B.

\subsection{Multiparticle form factors}

Analogously to the one particle form factors above we can check the
two or more particle form factors. However, they will generally depend
on more than one rapidities separately. For this reason it is easier
to analyze the volume dependence of a form factor for a given state,
identified by the quantization numbers of its rapidities in the Bethe-Yang
equations (\ref{eq:expdybe}). Such lines are identified from the
corresponding energy and momentum levels. Here below, we present some
examples of different states. A more exhaustive list of data can be
found in Appendix B. 

\begin{figure}[H]
\begin{centering}
\includegraphics[width=0.45\textwidth]{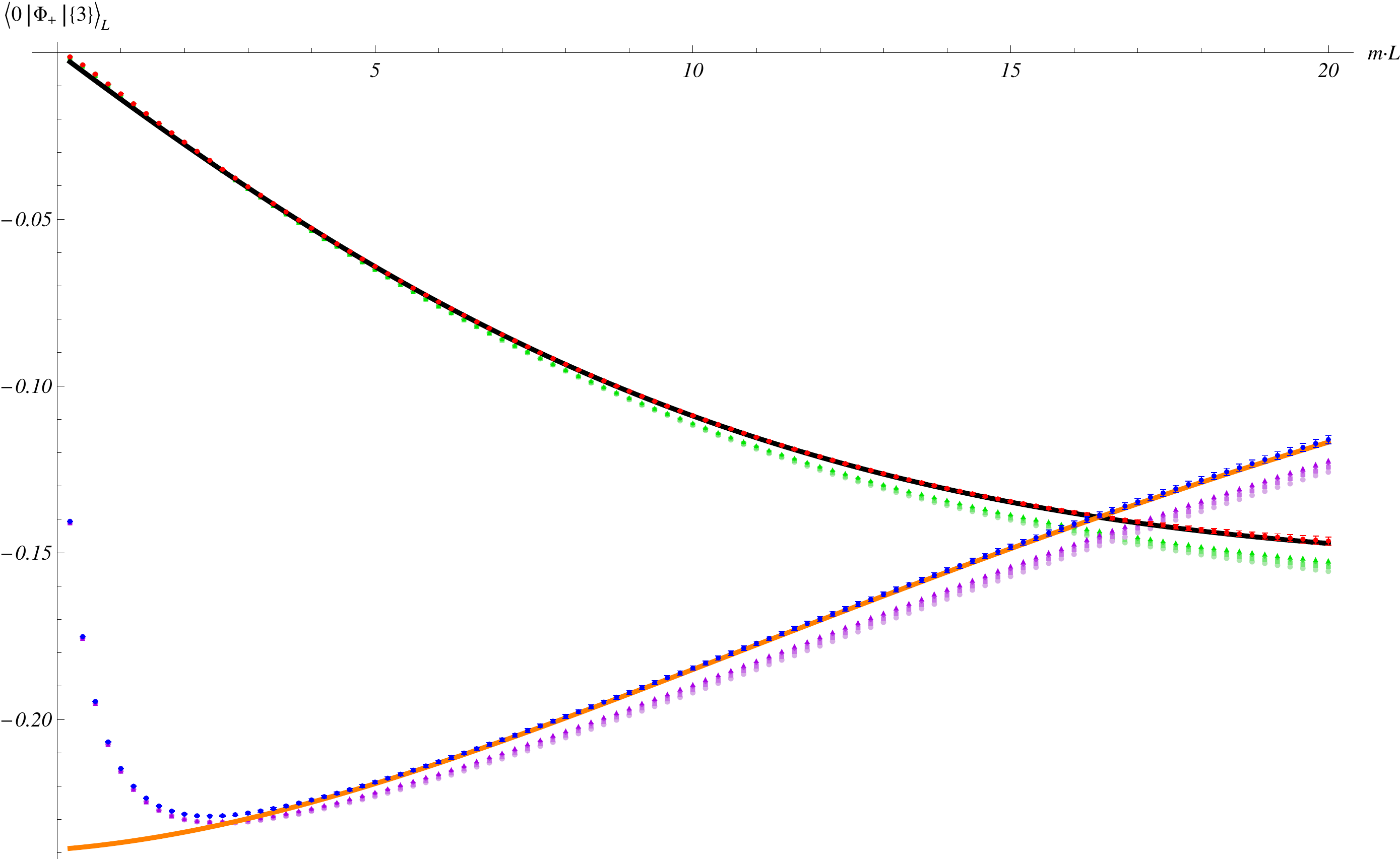}\includegraphics[width=0.45\textwidth]{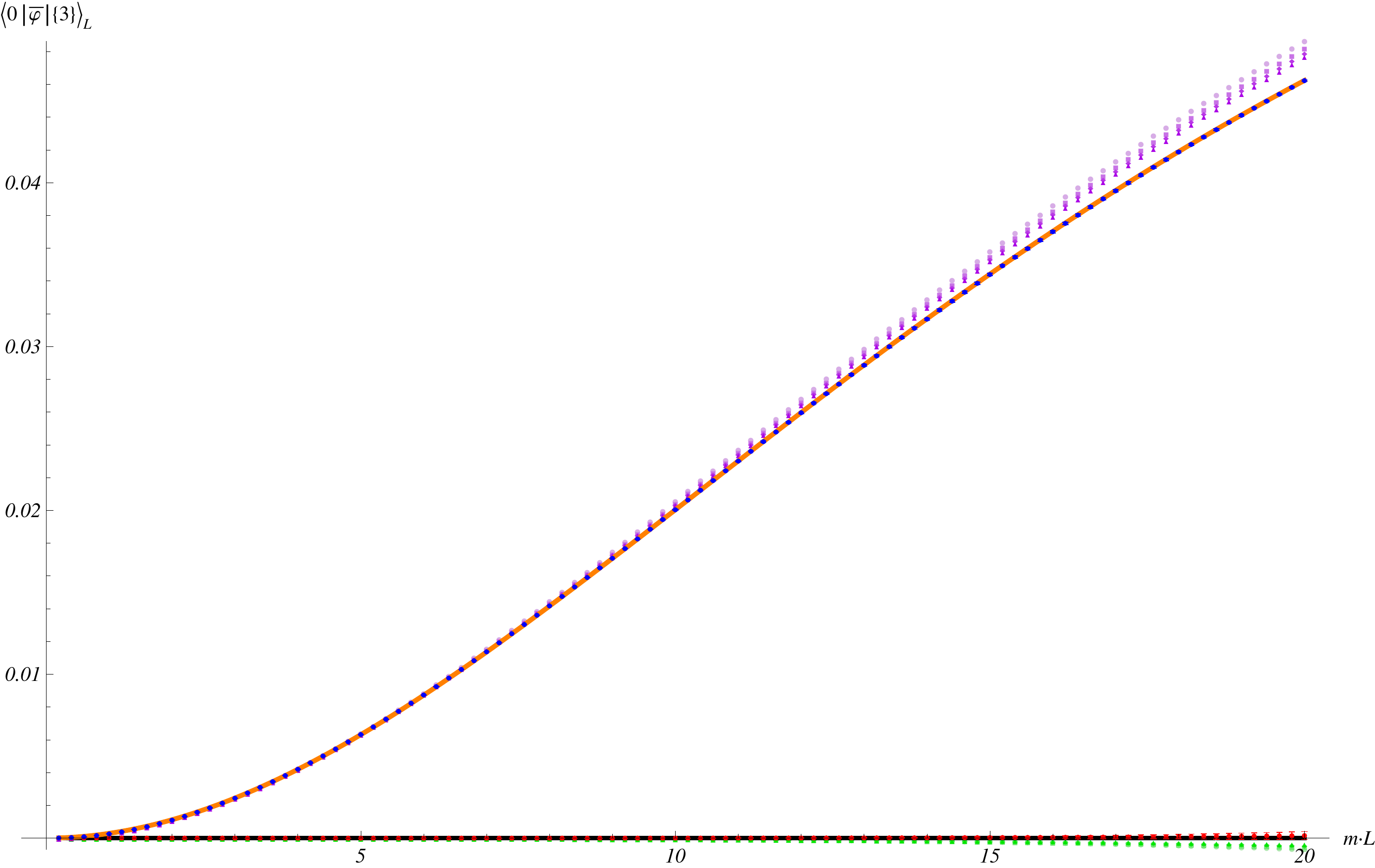}
\par\end{centering}

\caption{Left: one-particle form factor of the operator $\Phi_{+}$ on the
state labeled by quantum number $n_{1}=3$. Right: one-particle form
factor of the operator $\bar{\varphi}$ on the state $n_{1}=3$. The
solid lines are computed from formula (\ref{eq:Finite_volume_FF}),
while the dots with confidence bars are obtained by extrapolated TCSA
data. (The consistent legend is described in the beginning of subsection
\ref{sub:Form-factors-and})}
\end{figure}

We remark that the extrapolation fails for very large values of the
dimensionless volume $mL$. The reason is that for relevant perturbing
fields the dimensionless couplings in the Hamiltonian depends on a
positive power of the coupling, and therefore perturbation theory
is only applicable when the condition 
\begin{equation}
\frac{4\pi n}{mL}\gg1
\end{equation}
is satisfied \cite{Szecsenyi:2013gna}; if this is not the case, it
is necessary to evaluate the cutoff dependence to higher orders in
perturbation theory, which is a very complicated task and out of the
scope of the present work.\emph{ }These extra terms appear during
the extrapolation procedure as systematic errors, and this is why
the confidence intervals calculated during data fittings do not always
contain the theoretical data (mainly for larger volumes). However
these confidence intervals were kept, because usually they correctly
reflect the reliability of TCSA method.

\subsection{Diagonal form factors}

To calculate the diagonal form factors in finite volume we need the
connected form factors defined in equation (\ref{eq:connected}).
We can make use of the identities (\ref{eq:fidentity}) and (\ref{eq:didentity})
to simplify these expressions. For operators $\Phi_{\pm}$ we get
\begin{equation}
F_{2n,c}^{\Phi_{\pm}}\left(\theta_{1},\ldots,\theta_{n}\right)=\langle\Phi\rangle\left(\frac{\sqrt[4]{3}i}{\sqrt{2}v(0)}\right)^{2n}f\left(i\pi\right)^{n}\frac{Q_{n}^{\Phi_{\pm}}(\theta_{1},\ldots,\theta_{n})}{\prod_{j<k}^{n}\left(\sinh^{2}(\text{\ensuremath{\theta_{j}}}-\text{\ensuremath{\theta_{k}}})+\sin^{2}\left(\frac{\pi}{3}\right)\right)}
\end{equation}
with $Q_{n}^{\Phi_{\pm}}$ given in equation (\ref{eq:QPhipm}).

Here we show the one particle diagonal matrix elements. The multiparticle
matrix elements up to four particle number can be found in Appendix
B. 

\begin{figure}[h]
\begin{centering}
\includegraphics[width=0.48\textwidth]{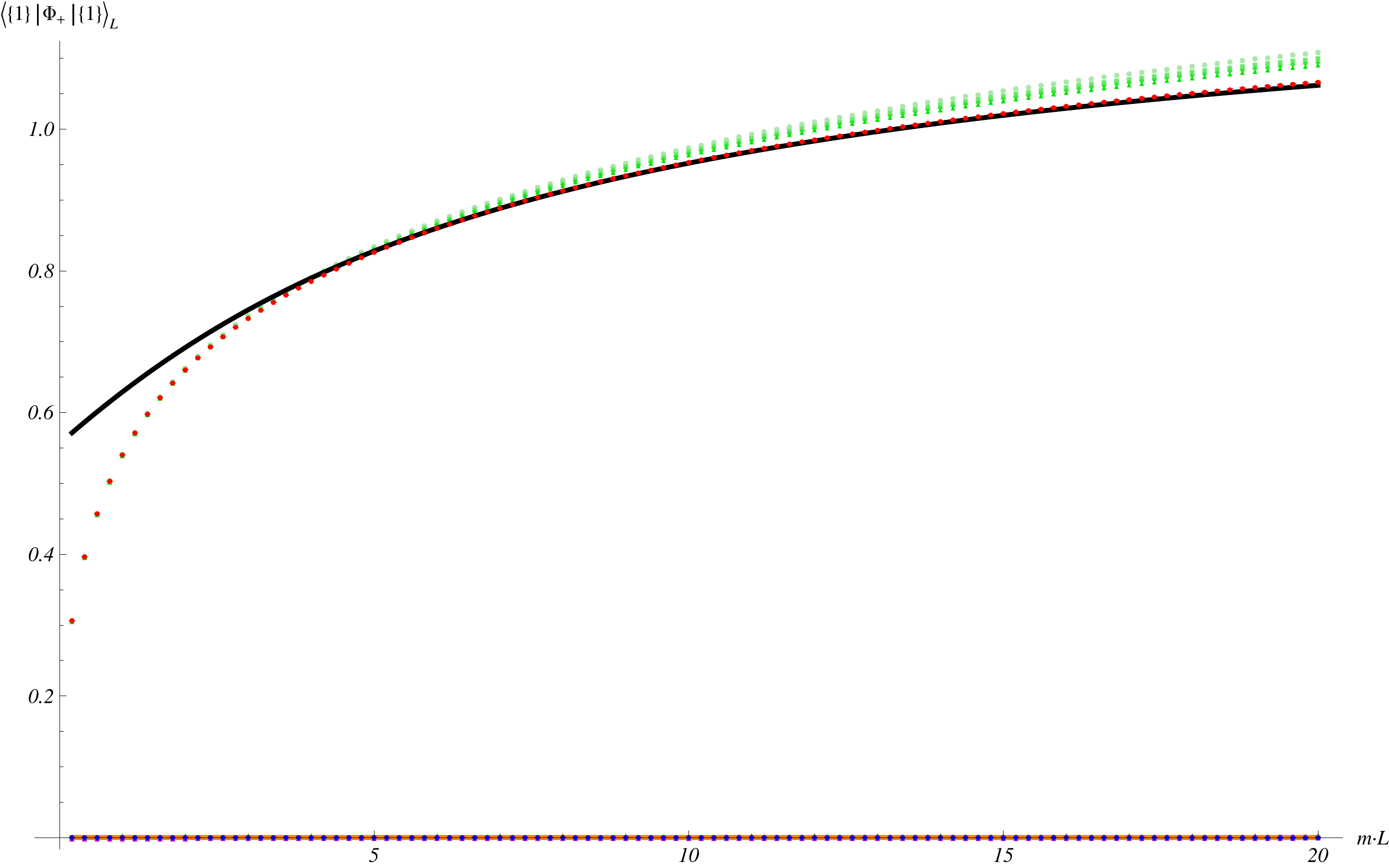}\includegraphics[width=0.48\textwidth]{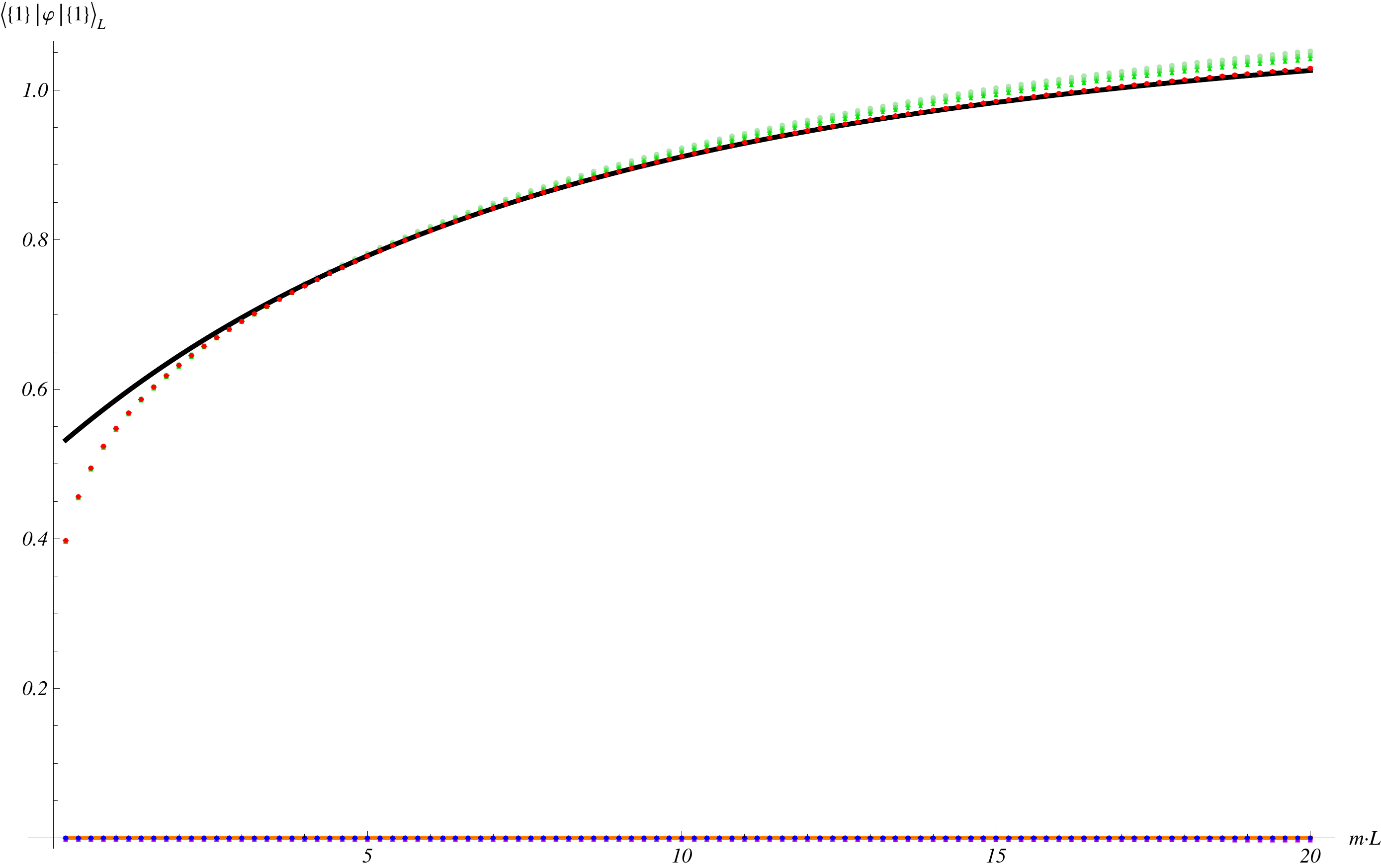}
\par\end{centering}

\caption{Left: one-particle diagonal form factor of the operator $\Phi_{+}$
on the state labeled by quantum number $n_{1}=1$. Right: one-particle
diagonal form factor of the operator $\varphi$ on the state $n_{1}=1$.
The solid lines are computed from formula (\ref{eq:diaggenrulesaleur}),
while the dots with confidence bars are obtained by extrapolated TCSA
data. (The consistent legend is described in the beginning of subsection
\ref{sub:Form-factors-and})}
\end{figure}

\section{Expectation values in finite volume/temperature \label{sec:Expectation-values-in}}

In this section we derive the exact finite volume vacuum expectation
value of our fields. We follow the derivation in \cite{Pozsgay:2010xd}
and indicate the slight modifications only. We start with an operator,
$\mathcal{O}$, localized in the bulk at $x=-1$ and $y=0$. The defect
is localized at $x=0$ and the size of our system is $L$. As usual
we do the calculation on the torus by exchanging the role of space
and time: 
\begin{equation}
_{L}\langle0\vert\mathcal{O}(-1,0)\vert0\rangle_{L}=\lim_{R\to\infty}\frac{\mbox{Tr}(\mathcal{O}(0,-1)e^{-H(R)L}D)}{\mbox{Tr}(e^{-H(R)L}D)}
\end{equation}
In the mirror (exchanged) theory the defect acts like an operator,
which we denote by $D$. As the location of the defect operator is
irrelevant we can follow the derivation of the defect TBA equation
\cite{Bajnok:2007jg} to redefine the Hamiltonian to be 
\begin{equation}
\tilde{H}(R)=H(R)-\frac{1}{L}\log D.
\end{equation}
This will have no other effect then to change the dispersion relation
as
\begin{equation}
m\cosh\theta\to m\cosh\theta-\frac{1}{L}\log T_{+}\left(\frac{i\pi}{2}-\theta\right)
\end{equation}
With these changes the TBA equation takes the form
\begin{equation}
\tilde{\epsilon}(\theta)=mL\cosh\theta-\log T_{+}\left(\frac{i\pi}{2}-\theta\right)-\int_{-\infty}^{\infty}\frac{d\theta'}{2I\pi}\varphi(\theta-\theta')\log(1+e^{-\tilde{\epsilon}(\theta')})\label{eq:DTBA-1}
\end{equation}
giving the ground state energy as 
\begin{equation}
E_{0}^{TBA}(L)=-m\int_{-\infty}^{\infty}\frac{d\theta}{2\pi}\cosh(\theta)\log(1+e^{-\tilde{\epsilon}(\theta)})\label{eq:DTBAE-1}
\end{equation}
The only deference compared to the periodic situation is that the
pseudo energy has changed. Thus the derivation of \cite{Pozsgay:2010xd}
will lead to the result 
\begin{equation}
_{L}\langle0\vert\mathcal{O}\vert0\rangle_{L}=\sum_{n=0}^{\infty}\frac{1}{n!}\prod_{j=1}^{n}\int\frac{d\theta_{j}}{2\pi}\frac{1}{1+e^{\tilde{\epsilon}(\theta_{j})}}F_{2n}^{C}(\theta_{1},\dots,\theta_{n})\label{eq:DLM-1}
\end{equation}
for the exact finite volume vacuum expectation value of any bulk operator.
Clearly this results holds for the two limits $\Phi_{\pm}$ of the
bulk field. The calculation of the exact finite volume vacuum expectation
values of the defect fields $\varphi$ and $\bar{\varphi}$ follows
from the exact ground state energy and momentum as we explained at
the end of section 2.2.2.

As far as the fields $\Phi_{\pm}$ are concerned, the knowledge of
the form factors of the trace of the bulk stress tensor \cite{Zamolodchikov:1990bk}
is sufficient to exploit the connected multi-particle form factors. 

\begin{figure}
\begin{centering}
\includegraphics[width=0.5\textwidth]{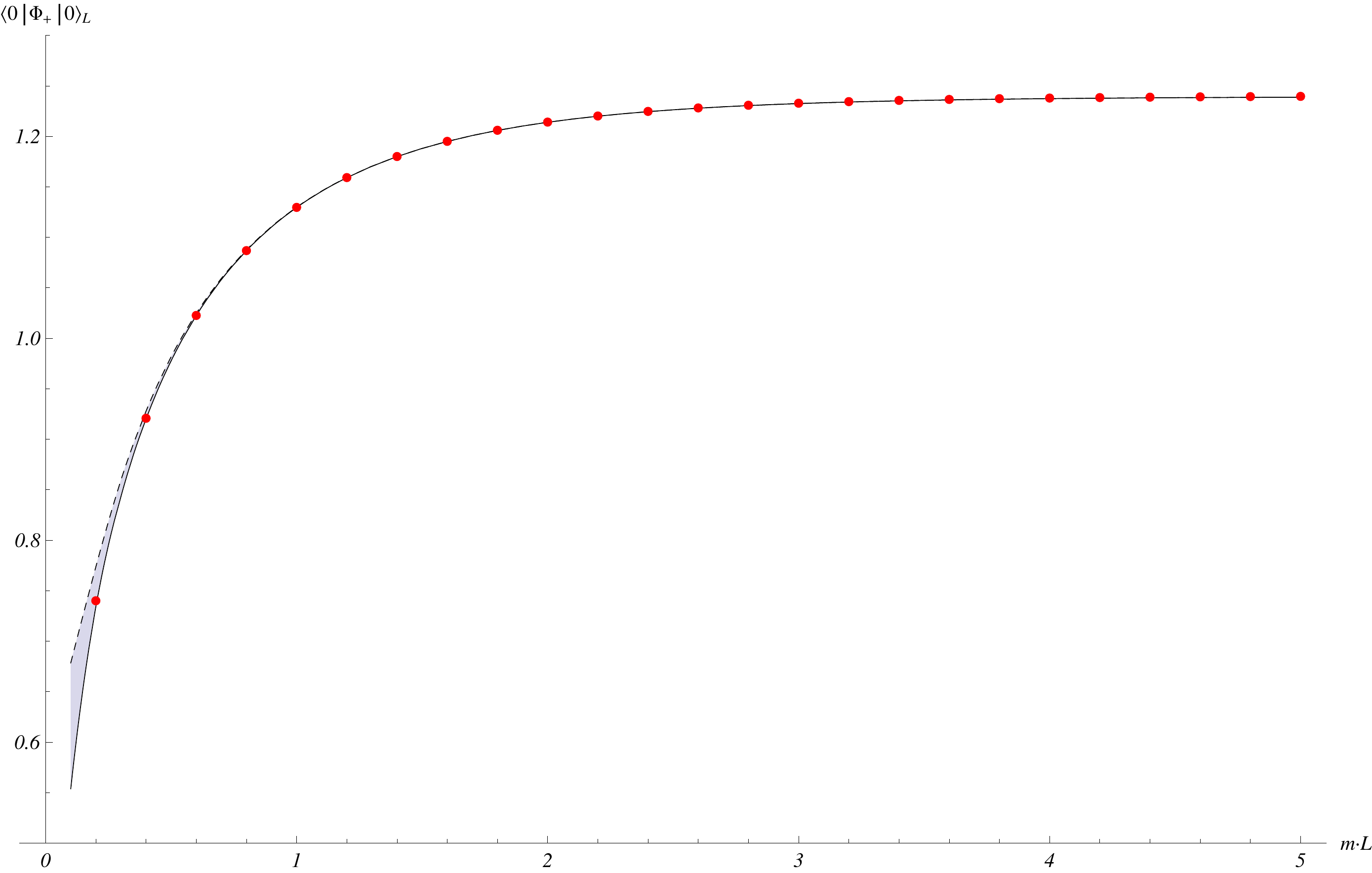}
\par\end{centering}

\caption{Comparison of the TCSA data (red dots) to the prediction from the
series (\ref{eq:DLM-1}) up to 3 terms (black curve) and up to 2 terms
(black dashed curve) for the finite-volume vacuum expectation value
of the fields $\Phi_{\pm}$\emph{\label{fig:DLM}.}}
\end{figure}

The numerical comparison goes in two steps: first, we solve eq. (\ref{eq:DTBA-1})
iteratively for the pseudo energy, starting from a trial solution
containing only the hyperbolic term in large volume and decreasing
the volume gradually. Then, we compute the series (\ref{eq:DLM-1})
with the form factors given above. In Figure \ref{fig:DLM}, we show
a comparison between the extrapolated one-point function of the field
$\Phi_{\pm}$ and the corresponding defect LeClair-Mussardo series,
up to three-particle contributions. Note that the outcome of the procedure
is different from the bulk case.

The numerical values of these points, for more precise comparison,
are reported in Table \ref{tab:DLM}. 

\begin{figure}
\begin{centering}
\emph{}%
\begin{tabular}{|c|c|c|}
\hline 
$mL$ & DLM & DTCSA\tabularnewline
\hline 
\hline 
$1$ & $1.12971$ & $1.12944$\tabularnewline
\hline 
$2$ & $1.21394$ & $1.21368$\tabularnewline
\hline 
$3$ & $1.23245$ & $1.23216$\tabularnewline
\hline 
$4$ & $1.23733$ & $1.23705$\tabularnewline
\hline 
$5$ & $1.23875$ & $1.23851$\tabularnewline
\hline 
$6$ & $1.23919$ & $1.23901$\tabularnewline
\hline 
$7$ & $1.23933$ & $1.23925$\tabularnewline
\hline 
$8$ & $1.23937$ & $1.23942$\tabularnewline
\hline 
$9$ & $1.23939$ & $1.23957$\tabularnewline
\hline 
$10$ & $1.23939$ & $1.23972$\tabularnewline
\hline 
\end{tabular}
\par\end{centering}

\caption{Numerical values for Figure \ref{fig:DLM}. \label{tab:DLM} }
\end{figure}

\section{Conclusions}

We developed the theory of finite volume form factors in the presence
of integrable defects. Our framework is valid for large volumes and
takes into account all polynomial finite size corrections but neglects
the exponentially small effects. We expressed these finite volume
form factors in terms of the infinite volume form factors and the
finite volume density of states, which depends on the scattering and
transmission matrices. 

We tested these ideas in the Lee-Yang model against the data of the
truncated conformal space approach. Within this framework, we numerically
diagonalized the Hamiltonian of the finite volume system on a truncated
Hilbert space and evaluated the matrix elements of local operators.
We performed a systematic comparison: first we compared the vacuum
expectation values of all local fields. In so doing we derived exact
explicit expressions for the vacuum expectation values of all of the
defects fields. We then determined all form factors of the defects
fields. We used these results to calculate the two point functions
of defect fields, which we compared to the short distance expansion,
which contains information on the conformal structure constants and
the vacuum expectation values. We also compared the finite volume
form factors against the TCSA data. Finally, we derived an explicit
expression for the exact finite volume vacuum expectation value of
any defect operator in terms of the multi-particle form factors and
the defect thermodynamic Bethe Ansatz pseudo energy, which we also
checked numerically. In all of these comparisons we used a renormalization
group-improved version of TCSA, which we adapted to the present case,
and found excellent agreement.

Our methods developed for the Lee-Yang model have a much wider application
and can be directly generalized for any diagonal scattering theories.
Especially, the parametrization of the defect form factors in terms
of the bulk form factors and extra polynomials should be applied to
other models. The generalization of our approach for non-diagonal
theories is a non-trivial and rather interesting problem. As the transmission
matrix bootstrap program was completed for many purely transmitting
defect theories \cite{Bowcock:2005vs,Corrigan:2007gt,Corrigan:2010ph},
it would be nice to formulate and solve their form factor bootstrap
program, too. 

Here we analyzed only the polynomial corrections in the inverse of
the volume for the finite volume form factors. It is a challenging
problem to calculate systematically the exponentially small finite
size correction.

\subsection*{Acknowledgments}

We thank OTKA 81461 and Lendulet grants LP2012-18/2012 and LP2012-50/2012
for support. 

\appendix

\section{Exact Form factor solutions\label{sec:AppendixA}}

In this appendix we present all form factors of primary operators.
The two limits of the bulk fields $\Phi_{\pm}$ are the simplest as
they can be expressed in terms of the bulk form factors. For this
reason we recall the bulk form factors first.

\subsection{Bulk form factors}

Bulk form factors of the operator $\Phi$ are parametrized as 

\begin{equation}
B_{n}\left(\theta_{1},\dots,\theta_{n}\right)=\langle\Phi\rangle H_{n}\prod_{i<j}\frac{f(\theta_{i}-\theta_{j})}{x_{i}+x_{j}}Q_{n}^{bulk}(x_{1},\dots,x_{n})
\end{equation}
In the Lee-Yang model $\Phi$ is the perturbing operator itself, thus
it is proportional to the trace of the energy momentum tensor. As
a consequence the form factor must have the form
\begin{eqnarray}
Q_{1}^{bulk}(x_{1}) & = & 1\\
Q_{2}^{bulk}(x_{1},x_{2}) & = & \sigma_{1}^{(2)}(x_{1},x_{2})\\
Q_{n}^{bulk}(x_{1},\dots,x_{n}) & = & \sigma_{1}^{(n)}(x_{1},\dots,x_{n})\sigma_{n-1}^{(n)}(x_{1},\dots,x_{n})P_{n}(x_{1},\dots,x_{n})\qquad\mathrm{if}\ n\geq3
\end{eqnarray}
where the $P_{n}$ polynomials satisfy the following recurrence relations
\begin{eqnarray}
P_{n+2}\left(x,-x,x_{1},\dots,x_{n}\right) & = & \frac{\left(\prod_{i=1}^{n}\left(x+\omega x_{i}\right)\left(x-\bar{\omega}x_{i}\right)-\prod_{i=1}^{n}\left(x-\omega x_{i}\right)\left(x+\bar{\omega}x_{i}\right)\right)}{2x\left(\omega-\bar{\omega}\right)}\times\nonumber \\
 &  & \left(-1\right)^{n+1}P_{n}(x_{1},\dots,x_{n})\\
P_{n+1}\left(\omega x,\bar{\omega}x,x_{1},\dots,x_{n}\right) & = & \prod_{i=1}^{n-1}\left(x+x_{i}\right)P_{n}\left(x,x_{1},\dots,x_{n}\right)
\end{eqnarray}
Te solution of this recursion can be written in terms of a determinant:
\begin{equation}
P_{n}\left(x_{1},\dots,x_{n}\right)=\det\Sigma^{(n)}\quad,\qquad\Sigma_{ij}^{(n)}\left(x_{1},\dots,x_{n}\right)=\sigma_{3i-2j+1}^{(n)}(x_{1},\dots,x_{n})
\end{equation}

\subsection{Form factors of $\Phi_{\pm}$}

Based on our previous discussion the form factor of $\Phi_{-}$ is
the bulk form factor: 
\begin{equation}
F_{n}^{\Phi_{-}}(\theta_{1},\dots,\theta_{n})=B_{n}(\theta_{1},\dots,\theta_{n})
\end{equation}
while the form factor is $\Phi_{+}$ is of the form
\begin{equation}
F_{n}^{\Phi_{+}}(\theta_{1},\dots,\theta_{n})=\prod_{n}T_{-}(\theta_{i})B_{n}(\theta_{1},\dots,\theta_{n})
\end{equation}
Comparing the parametrization of the bulk and defect form factors
we can conclude that 
\begin{eqnarray}
Q_{n}^{\Phi_{\pm}}(x_{1},\dots,x_{n}) & = & \prod_{i=1}^{n}(\nu x_{i}+\bar{\nu}x_{i}^{-1}\mp\sqrt{3})Q_{n}^{bulk}(x_{1},\dots,x_{n})\\
 &  & \prod_{i=1}^{n}(\nu x_{i}+\bar{\nu}x_{i}^{-1}\mp\sqrt{3})\sigma_{1}^{(n)}(x_{1},\dots,x_{n})\sigma_{n-1}^{(n)}(x_{1},\dots,x_{n})P_{n}(x_{1},\dots,x_{n})\nonumber 
\end{eqnarray}

\subsection{Form factors of $\bar{\varphi}$ and $\varphi$}

Let's focus on the form factor solutions for a generic defect field.
Calculating explicitly the first few $Q$ polynomials we found for
$\varphi$: 
\begin{eqnarray}
Q_{1}^{\varphi} & = & \nu\sigma_{1}\\
Q_{2}^{\varphi} & = & \sigma_{1}+\nu^{2}\sigma_{1}\sigma_{2}\nonumber \\
Q_{3}^{\varphi} & = & \bar{\nu}\sigma_{1}^{2}+\nu\sigma_{1}^{2}\sigma_{2}+\nu^{3}\sigma_{1}\sigma_{2}\sigma_{3}\nonumber \\
Q_{4}^{\varphi} & = & \bar{\nu}^{2}\sigma_{1}^{2}\sigma_{2}+\sigma_{1}^{2}\sigma_{2}^{2}+\nu^{2}\sigma_{1}\sigma_{2}^{2}\sigma_{3}+\nu^{4}\sigma_{1}\sigma_{2}\sigma_{3}\sigma_{4}\nonumber \\
Q_{5}^{\varphi} & = & \bar{\nu}^{3}\left(\sigma_{1}^{2}\sigma_{2}\sigma_{3}-\sigma_{1}^{2}\sigma_{5}\right)+\bar{\nu}\left(\sigma_{1}^{2}\sigma_{2}^{2}\sigma_{3}-\sigma_{1}^{2}\sigma_{2}\sigma_{5}\right)+\nu\left(\sigma_{1}\sigma_{5}^{2}+\sigma_{1}\sigma_{2}^{2}\sigma_{3}^{2}-2\sigma_{1}\sigma_{2}\sigma_{3}\sigma_{5}\right)+\nonumber \\
 &  & +\nu^{3}\left(\sigma_{1}\sigma_{2}\sigma_{3}^{2}\sigma_{4}-\sigma_{1}\sigma_{3}\sigma_{4}\sigma_{5}\right)+\nu^{5}\left(\sigma_{1}\sigma_{2}\sigma_{3}\sigma_{4}\sigma_{5}-\sigma_{1}\sigma_{4}\sigma_{5}^{2}\right)\nonumber 
\end{eqnarray}
where at each level $n=1,\dots,5$ we abbreviated $\sigma_{k}^{(n)}$
by $\sigma_{k}$. 

These explicit solutions suggest to parametrize the general form factors
as
\begin{equation}
Q_{n}\left(n\right)=R\left(\sigma_{1}^{\left(n\right)},\bar{\sigma}_{1}^{\left(n\right)}\right)\sigma_{n}^{\left(n\right)}P_{n}^{\left(n\right)}S_{n}^{\left(n\right)}\label{eq:Qansatz}
\end{equation}
for $n\geq3$, where $R$ is some polynomial which depends on the
operator. Plugging this expressions into the recursive relations we
obtain the recursions for $S_{n}$: 
\begin{eqnarray}
S_{n+1}\left(\omega x,\bar{\omega}x,x_{1},\dots,x_{n-1}\right) & = & \left(\nu x+\bar{\nu}\bar{x}\right)S_{n}\left(x,x_{1},\dots,x_{n-1}\right)\nonumber \\
S_{n+2}\left(x,-x,x_{1},\dots,x_{n}\right) & = & -\left(x^{2}\nu^{2}-1+x^{-2}\nu^{-2}\right)S_{n}\left(x_{1},\dots,x_{n}\right)
\end{eqnarray}
The first few solutions are: 
\begin{eqnarray}
S_{3} & = & \left(\nu^{-2}\bar{\sigma}_{2}+\bar{\sigma}_{1}\sigma_{1}+\nu^{2}\sigma_{2}\right)\\
S_{4} & = & \left(\nu^{-3}\bar{\sigma}_{3}+\nu^{-1}\bar{\sigma}_{2}\sigma_{1}+\nu\bar{\sigma}_{1}\sigma_{2}+\nu^{3}\sigma_{3}\right)\nonumber \\
S_{5} & = & \left(\nu^{-4}\bar{\sigma}_{4}+\nu^{2}\bar{\sigma}_{3}\sigma_{1}+\bar{\sigma}_{2}\sigma_{2}+\nu^{2}\bar{\sigma}_{1}\sigma_{3}+\nu^{4}\sigma_{4}\right)-1\nonumber \\
S_{6} & = & \left(\nu^{-5}\bar{\sigma}_{5}+\nu^{-3}\bar{\sigma}_{4}\sigma_{1}+\nu^{-1}\bar{\sigma}_{3}\sigma_{2}+\nu\bar{\sigma}_{2}\sigma_{3}+\nu^{3}\bar{\sigma}_{1}\sigma_{4}+\nu^{5}\sigma_{5}\right)-\left(\nu^{-1}\bar{\sigma}_{1}+\nu\sigma_{1}\right)\nonumber \\
S_{7} & = & \left(\nu^{-6}\bar{\sigma}_{6}+\nu^{-4}\bar{\sigma}_{5}\sigma_{1}+\nu^{-2}\bar{\sigma}_{4}\sigma_{2}+\bar{\sigma}_{3}\sigma_{3}+\nu^{2}\bar{\sigma}_{2}\sigma_{4}+\nu^{4}\bar{\sigma}_{1}\sigma_{5}+\nu^{6}\sigma_{6}\right)-\nonumber \\
 &  & -\left(\nu^{-2}\bar{\sigma}_{2}+\bar{\sigma}_{1}\sigma_{1}+\nu^{2}\sigma_{2}\right)-1\nonumber 
\end{eqnarray}
These suggest that $S_{n}$'s are Laurent-polynomials, which are invariant
under the simultaneous exchanges of $x\leftrightarrow x^{-1}$ and
$\nu\leftrightarrow\nu^{-1}$. In $S_{n}$ the coefficient of $\nu^{k}$
is a homogeneous symmetric Laurent-polynomial of degree $k$, for
$k\in\left\{ -n+1,-n+3,\dots,n-1\right\} .$ Let us define then 
\begin{equation}
\tau_{k}\left(x_{1},\dots,x_{n}\right)=\sum_{l\in\mathbb{Z}}\nu^{2l-k}\bar{\sigma}_{k-l}\left(x_{1},\dots,x_{n}\right)\sigma_{l}\left(x_{1},\dots,x_{n}\right)
\end{equation}
Here we sum over all integers $l$, but note that this sum is always
finite. These Laurent-polynomials satisfy the following useful identities:
\begin{eqnarray}
\tau_{n+1}\left(x_{1},\dots,x_{n}\right) & = & \tau_{n-1}\left(x_{1},\dots,x_{n}\right)\\
\tau_{k+2}\left(x,-x,x_{1},\dots,x_{n}\right) & = & \tau_{k+2}\left(x_{1},\dots,x_{n}\right)+\tau_{k-2}\left(x_{1},\dots,x_{n}\right)\nonumber \\
 &  & -\left(x^{2}\nu^{2}+x^{-2}\nu^{-2}\right)\tau_{k}\left(x_{1},\dots,x_{n}\right)\\
\tau_{k}\left(x,x_{1},\dots,x_{n}\right) & = & (x\nu+x^{-1}\nu^{-1})\tau_{k-1}\left(x_{1},\dots,x_{n}\right)\nonumber \\
 &  & +\tau_{k-2}\left(x_{1},\dots,x_{n}\right)+\tau_{k}(x_{1},\dots,x_{n})\\
\tau_{k+1}\left(x\omega,x\omega^{-1},x_{1},\dots,x_{n-1}\right) & = & \left(x^{-1}\nu^{-1}+x\nu\right)\tau_{k}\left(x,x_{1},\dots,x_{n-1}\right)-\tau_{k-1}\left(x_{1},\dots,x_{n-1}\right)\nonumber \\
 &  & +\tau_{k-3}\left(x_{1},\dots,x_{n-1}\right)+\tau_{k+1}\left(x_{1},\dots,x_{n-1}\right)
\end{eqnarray}
Using these properties it is easy to show by induction that the form
factor solution is 
\begin{eqnarray}
S_{n}\left(x_{1},\dots,x_{n}\right) & = & \tau_{n-1}\left(x_{1},\dots,x_{n}\right)-\left(\tau_{n-5}\left(x_{1},\dots,x_{n}\right)+\tau_{n-7}\left(x_{1},\dots,x_{n}\right)\right)\\
 &  & +\left(\tau_{n-11}\left(x_{1},\dots,x_{n}\right)+\tau_{n-13}\left(x_{1},\dots,x_{n}\right)\right)-\dots\nonumber \\
 & = & \tau_{n-1}\left(x_{1},\dots,x_{n}\right)+\sum_{m\geq1}\left(-1\right)^{m}\left(\tau_{n+1-6m}\left(x_{1},\dots,x_{n}\right)+\tau_{n-1-6m}\left(x_{1},\dots,x_{n}\right)\right)\nonumber 
\end{eqnarray}
The operator-dependent prefactors for $\varphi$ and $\bar{\varphi}$
in (\ref{eq:Qansatz}) are
\begin{equation}
R^{\bar{\varphi}}(\sigma_{1},\bar{\sigma}_{1})=\bar{\nu}\bar{\sigma}_{1}\qquad;\qquad R^{\varphi}(\sigma_{1},\bar{\sigma}_{1})=\nu\sigma_{1}
\end{equation}

\subsection{Form factors of descendant operators}

Descendant operators correspond to the kernels of the recursion equations.
These kernels are the same as for the bulk theory: at level $n$ they
are
\begin{equation}
K_{n}^{kin}(x_{1},\dots,x_{n})=\prod_{1\leq i<j\leq n}(x_{i}+x_{j})\quad;\qquad K_{n}^{dyn}(x_{1},\dots,x_{n})=\prod_{1\leq i<j\leq n}(x_{i}^{2}+x_{j}^{2}+x_{i}x_{j})
\end{equation}
The common kernel is 
\begin{equation}
K_{n}(x_{1},\dots,x_{n})=K_{n}^{kin}(x_{1},\dots,x_{n})K_{n}^{dyn}(x_{1},\dots,x_{n})
\end{equation}
Adding formally $K_{1}=1$ all solutions of the form factor equations
originates form a top representative at level $n$ of the form
\begin{equation}
\sigma_{1}^{a_{1}}\sigma_{2}^{a_{2}}\dots\sigma_{n}^{a_{n}}K_{n}\qquad;\qquad a_{1},\dots,a_{n-1}\in\mathbb{N}\quad,\quad a_{n}\in\mathbb{Z}
\end{equation}
but it is a highly nontrivial task to relate them to the space of
local defect operators.

\section{Numerical Data}

In this appendix we extensively present our numerical data.

\subsection{Short distance expansion of the two point function}

\begin{figure}[H]
\centering{}\includegraphics[width=0.7\textwidth]{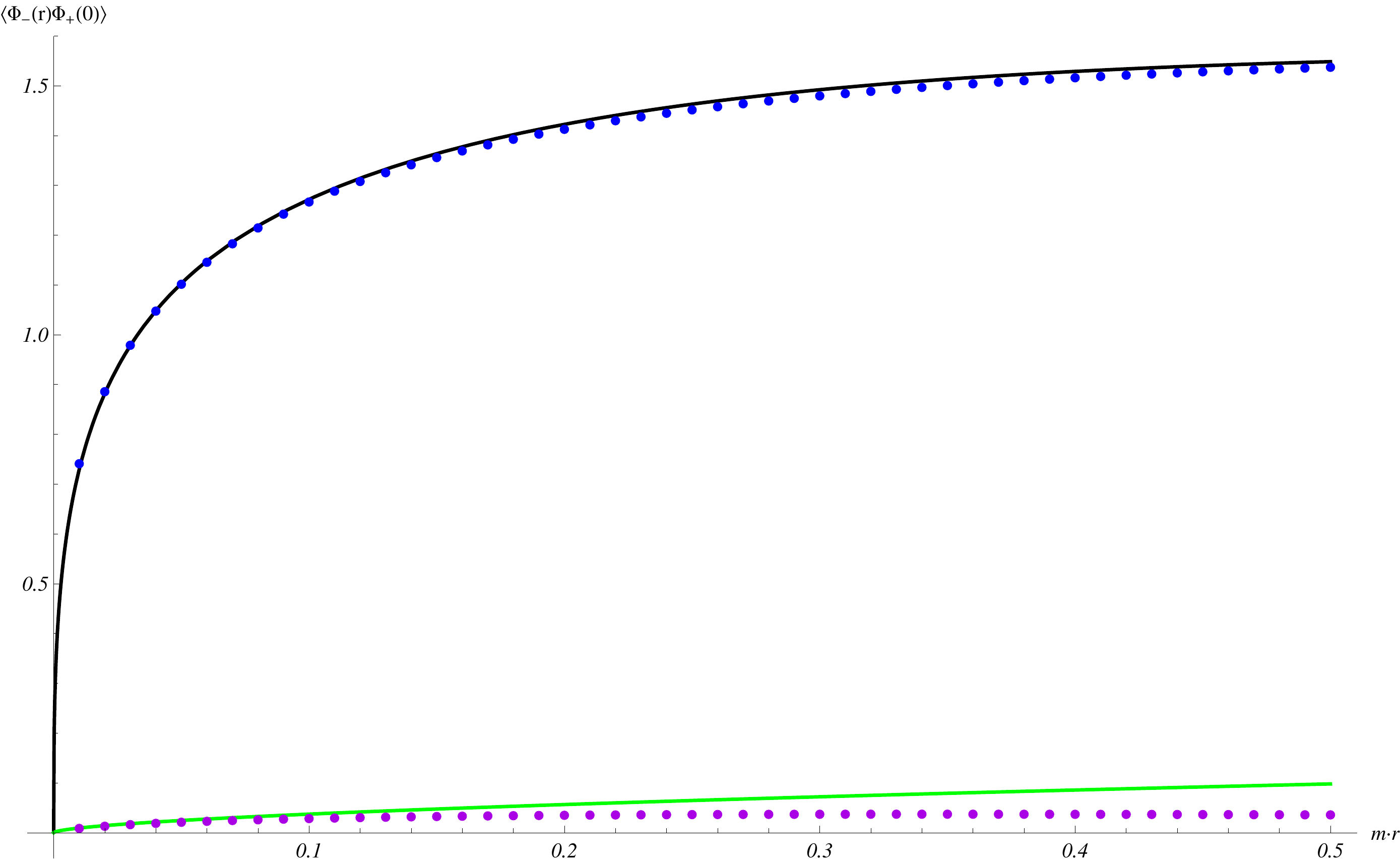}\caption{$\Phi_{-}\Phi_{+}$ correlation function: blue and purple dots show
the real and imaginary part calculated from the form factor expansion
up to two particle terms, while black and green curves the first order
CFT perturbation results\label{fig:PhipPhim}}
\end{figure}

In this subsection we show the real and imaginary parts of various
correlation function. The short distance CFT expansion is shown with
continuous lines, while the form factor expansion with dots. In Figure
\ref{fig:PhipPhim} we compare
\begin{equation}
\langle\Phi_{-}(r)\Phi_{+}(0)\rangle\sim C_{\Phi_{-}\Phi_{+}}^{I}r^{4/5}+C_{\Phi_{-}\Phi_{+}}^{\Phi_{+}}\langle\Phi_{+}\rangle r^{2/5}+C_{\Phi_{-}\Phi_{+}}^{\Phi_{-}}\langle\Phi_{-}\rangle r^{2/5}+C_{\Phi_{-}\Phi_{+}}^{\varphi}\langle\varphi\rangle r^{3/5}+C_{\Phi_{-}\Phi_{+}}^{\bar{\varphi}}\langle\bar{\varphi}\rangle r^{3/5}
\end{equation}

where $C_{\Phi_{-}\Phi_{+}}^{I}=(1+\beta^{-2})$, $C_{\Phi_{-}\Phi_{+}}^{\varphi}=-i\alpha\eta^{-2}\sqrt{\sqrt{5}(1+\beta^{-2})}$,
$C_{\Phi_{-}\Phi_{+}}^{\bar{\varphi}}=i\alpha\eta^{2}\sqrt{\sqrt{5}(1+\beta^{-2})}$,
$C_{\Phi_{-}\Phi_{+}}^{\Phi_{+}}=C_{\Phi_{-}\Phi_{+}}^{\Phi_{-}}=\alpha^{2}/\beta$

\begin{figure}[H]
\centering{}\includegraphics[width=0.7\textwidth]{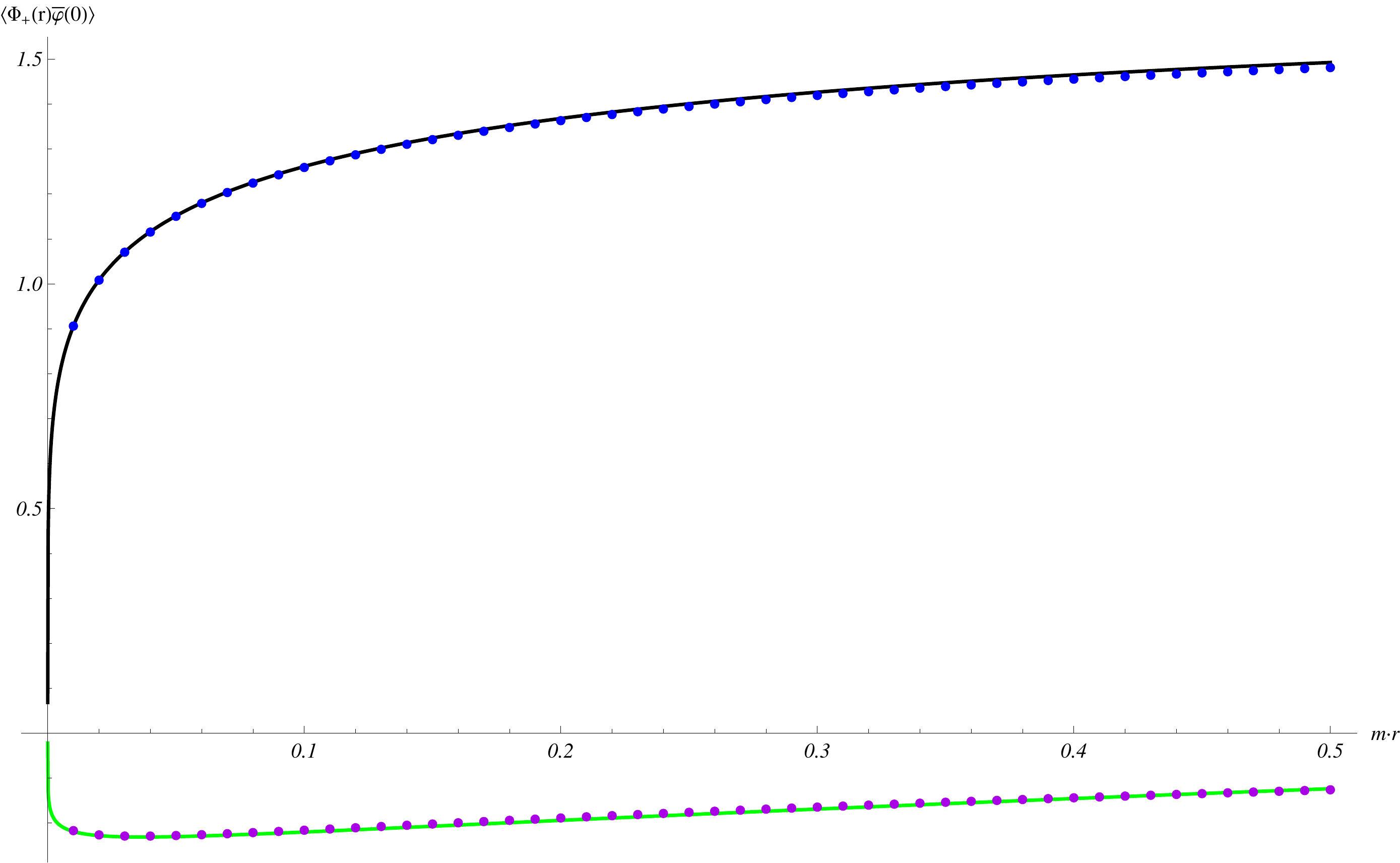}\caption{$\Phi_{+}\bar{\varphi}\equiv\Phi_{-}\varphi$ correlation function:
blue and purple dots show the real and imaginary part calculated from
the form factor expansion up to two particle terms, while black and
green curves the first order CFT perturbation results\label{fig:Phimpertphi}}
\end{figure}

Also, in Figure \ref{fig:Phimpertphi}, we present the correlator
\begin{equation}
\langle\Phi_{-}(r)\varphi(0)\rangle\sim C_{\Phi_{-}\varphi}^{\Phi_{+}}\langle\Phi_{+}\rangle r^{1/5}+C_{\Phi_{-}\varphi}^{\Phi_{-}}\langle\Phi_{-}\rangle r^{1/5}+C_{\Phi_{-}\varphi}^{\bar{\varphi}}\langle\bar{\varphi}\rangle r^{2/5}
\end{equation}
with $C_{\Phi_{-}\varphi}^{\Phi_{-}}=\frac{\alpha}{2}(\beta+\beta^{-1}-\frac{i}{\sqrt[4]{5}})$,
$C_{\Phi_{-}\varphi}^{\Phi_{+}}=\frac{\alpha\beta}{2}(1+(\beta-\beta^{-1})\frac{i}{\sqrt[4]{5}})$,
$C_{\Phi_{-}\varphi}^{\bar{\varphi}}=-\frac{1}{\eta\beta}$

\begin{figure}[H]
\centering{}\includegraphics[width=0.7\textwidth]{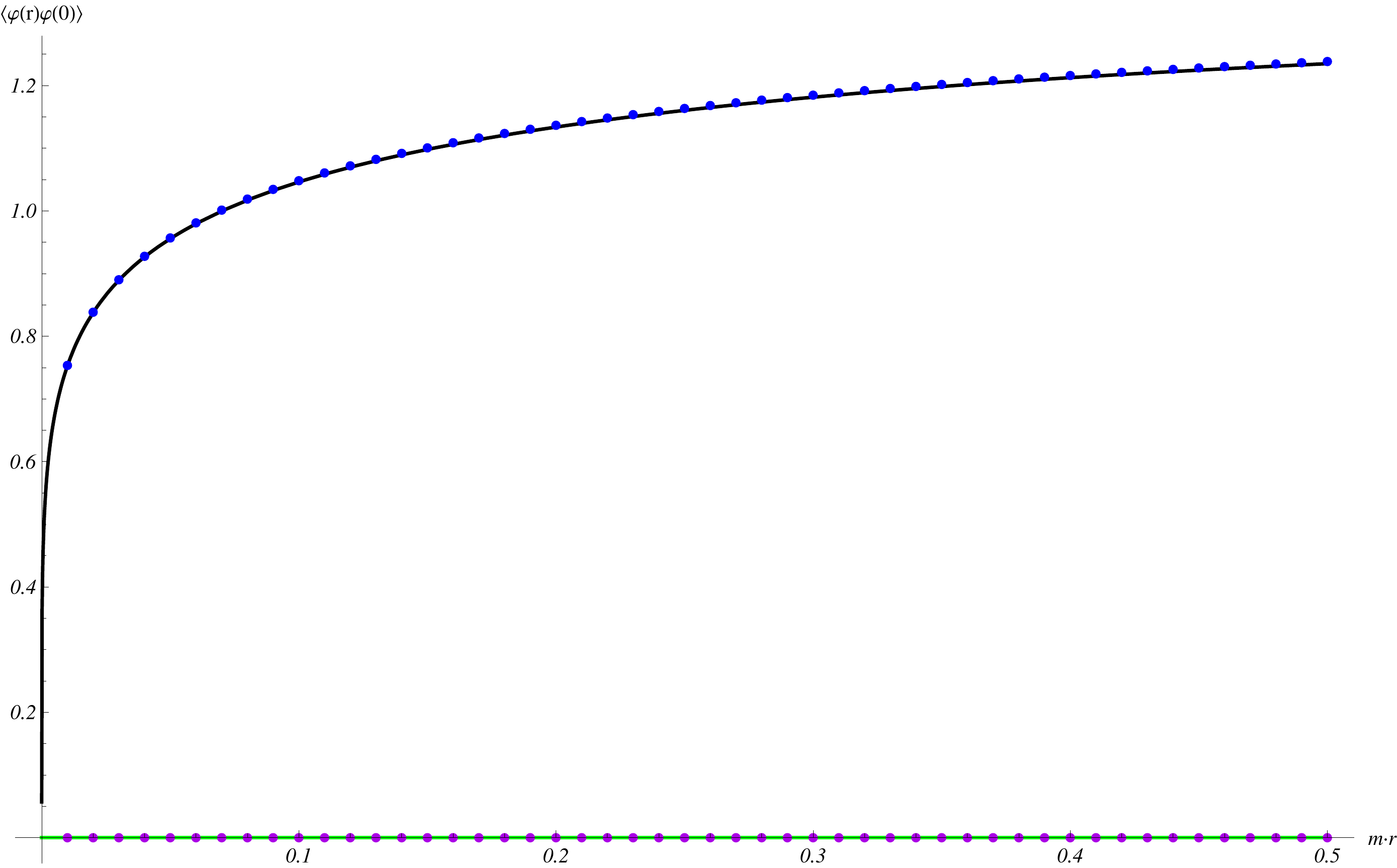}\caption{$\varphi\varphi$ correlation function: blue and purple dots show
the real and imaginary part calculated by form factor expansion up
to two particle terms, while black and green curves the first order
CFT perturbation results\label{fig:pertphipertphi}}
\end{figure}

Figure \ref{fig:pertphipertphi}, finally, exhibits the correlation
functions 
\begin{equation}
\langle\varphi(r)\varphi(0)\rangle\sim C_{\varphi\varphi}^{I}r^{2/5}+C_{\varphi\varphi}^{\varphi}\langle\varphi\rangle r^{1/5}\qquad\quad\langle\bar{\varphi}(r)\bar{\varphi}(0)\rangle\sim C_{\bar{\varphi}\bar{\varphi}}^{I}r^{2/5}+C_{\bar{\varphi}\bar{\varphi}}^{\bar{\varphi}}\langle\bar{\varphi}\rangle r^{1/5}
\end{equation}
where $C_{\varphi\varphi}^{I}=C_{\bar{\varphi}\bar{\varphi}}^{I}=-1$
and $C_{\varphi\varphi}^{\varphi}=C_{\bar{\varphi}\bar{\varphi}}^{\bar{\varphi}}=\frac{\alpha}{\beta}$.
In all of these formulas, the vacuum expectation values of defect
operators from section \ref{sub:Exact-vacuum-expectation} are used. 

\subsection{One particle form factors}

In this subsections we present the various particle form factors We
start with the one-particle form factors of the fields $\Phi_{\pm}$.
Data are collected from various Bethe-Yang quantization numbers and
volumes. 

\begin{figure}[H]
\begin{centering}
\includegraphics[width=0.45\textwidth]{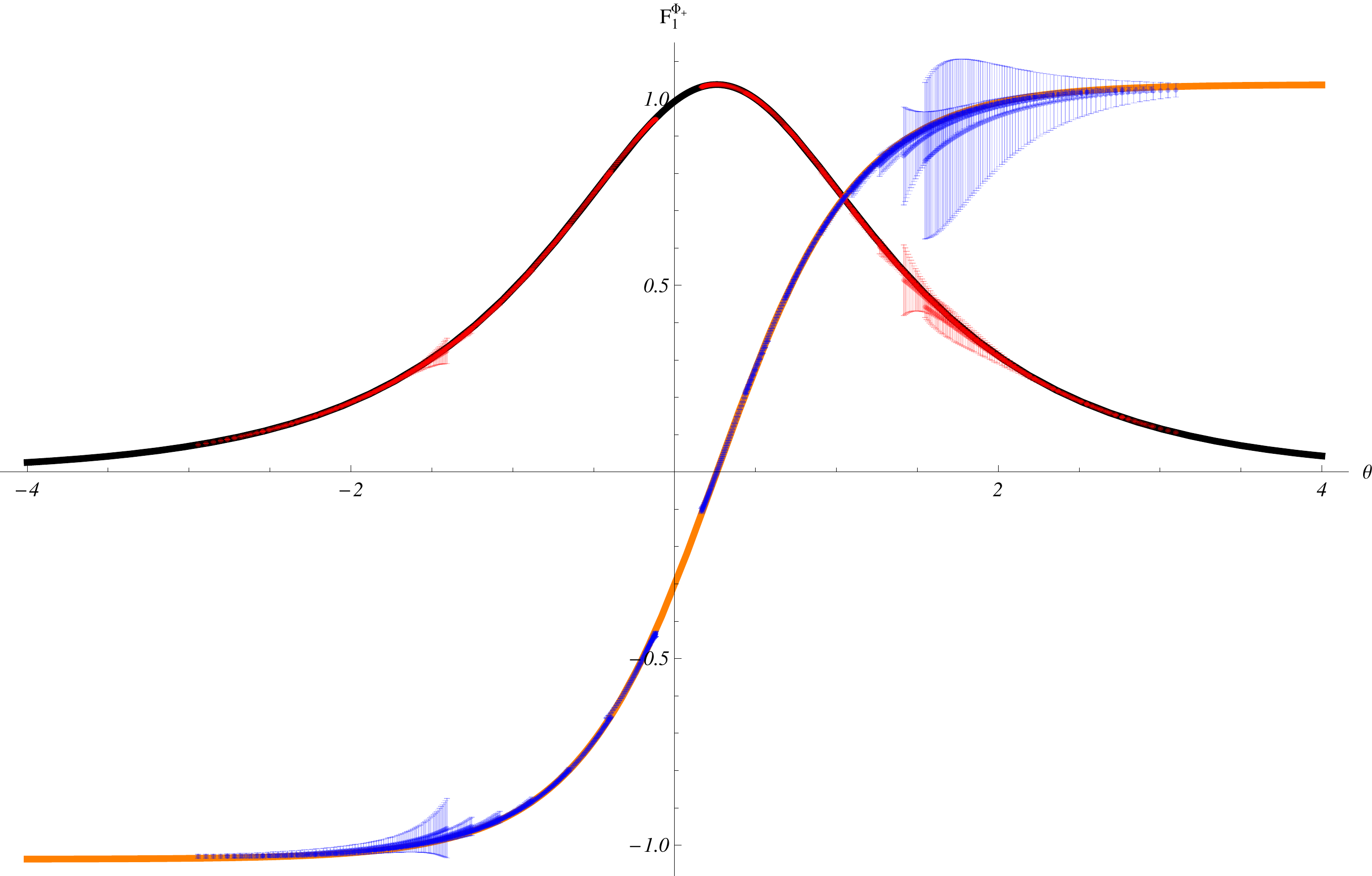}\includegraphics[width=0.45\textwidth]{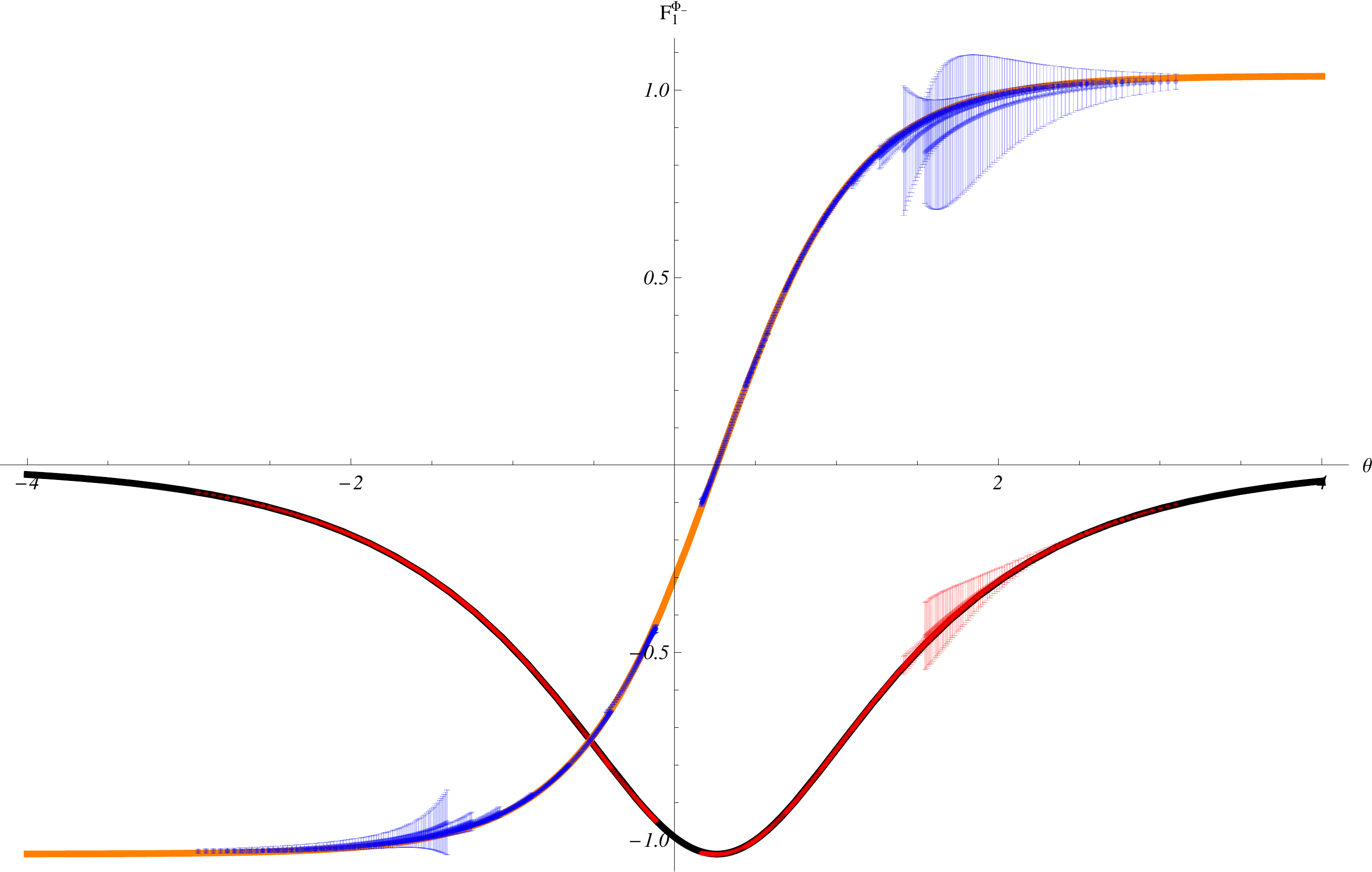}
\par\end{centering}

\label{fig:PhipPhimFF}\caption{Comparison between the extrapolated TCSA data (dots with confidence
bars) and the theoretical prediction (solid line) for the one-particle
form factor of the operators $\Phi_{+}$and $\Phi_{-}$.}
\end{figure}

\subsection{Multiparticle form factors}

In this subsection we present the comparison of data for multiparticle
form factors. 

\begin{figure}
\begin{centering}
\includegraphics[width=0.45\textwidth]{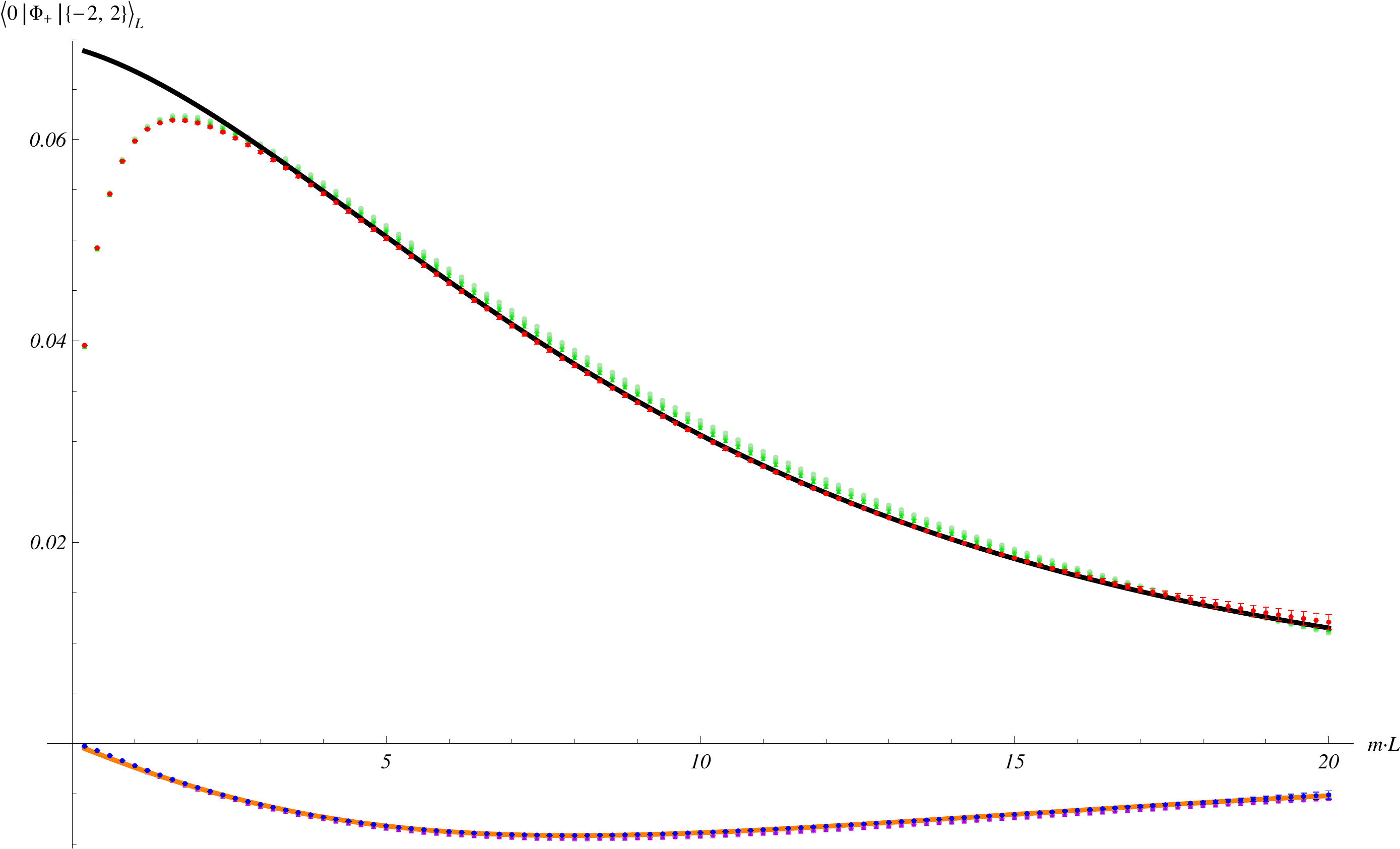}\includegraphics[width=0.45\textwidth]{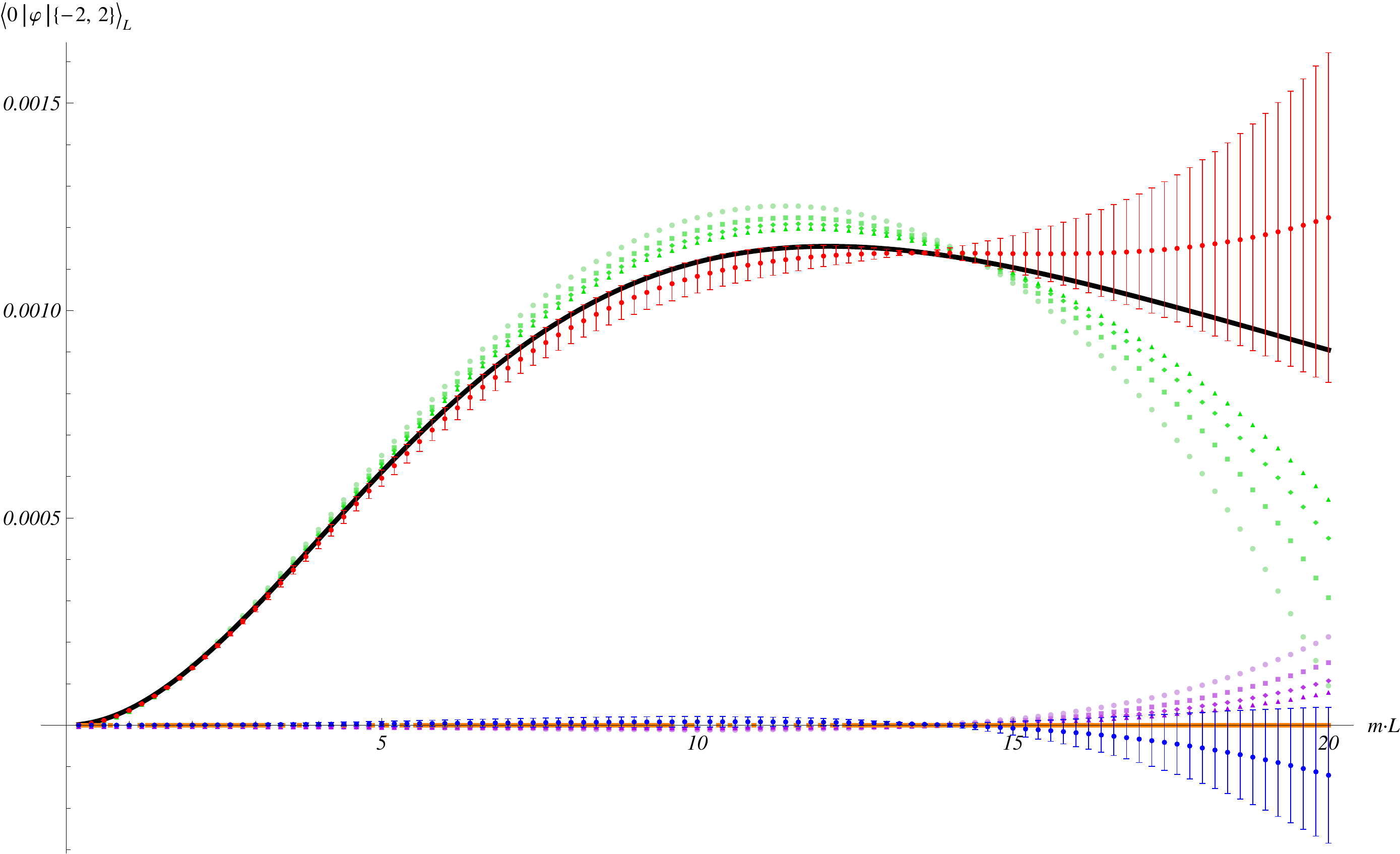}
\par\end{centering}

\caption{Left: two-particle form factor of the operator $\Phi_{+}$ on the
state labeled by quantum numbers $n_{1}=-2,\, n_{2}=2$. Right: two-particle
form factor of the operator $\varphi$ on the state $n_{1}=-2,\, n_{2}=2$.
The solid lines are computed from formula (\ref{eq:Finite_volume_FF}),
while the dots with confidence bars are obtained by extrapolated TCSA
data. (The consistent legend is described in the beginning of subsection
\ref{sub:Form-factors-and})}
\end{figure}
\begin{figure}
\begin{centering}
\includegraphics[width=0.45\textwidth]{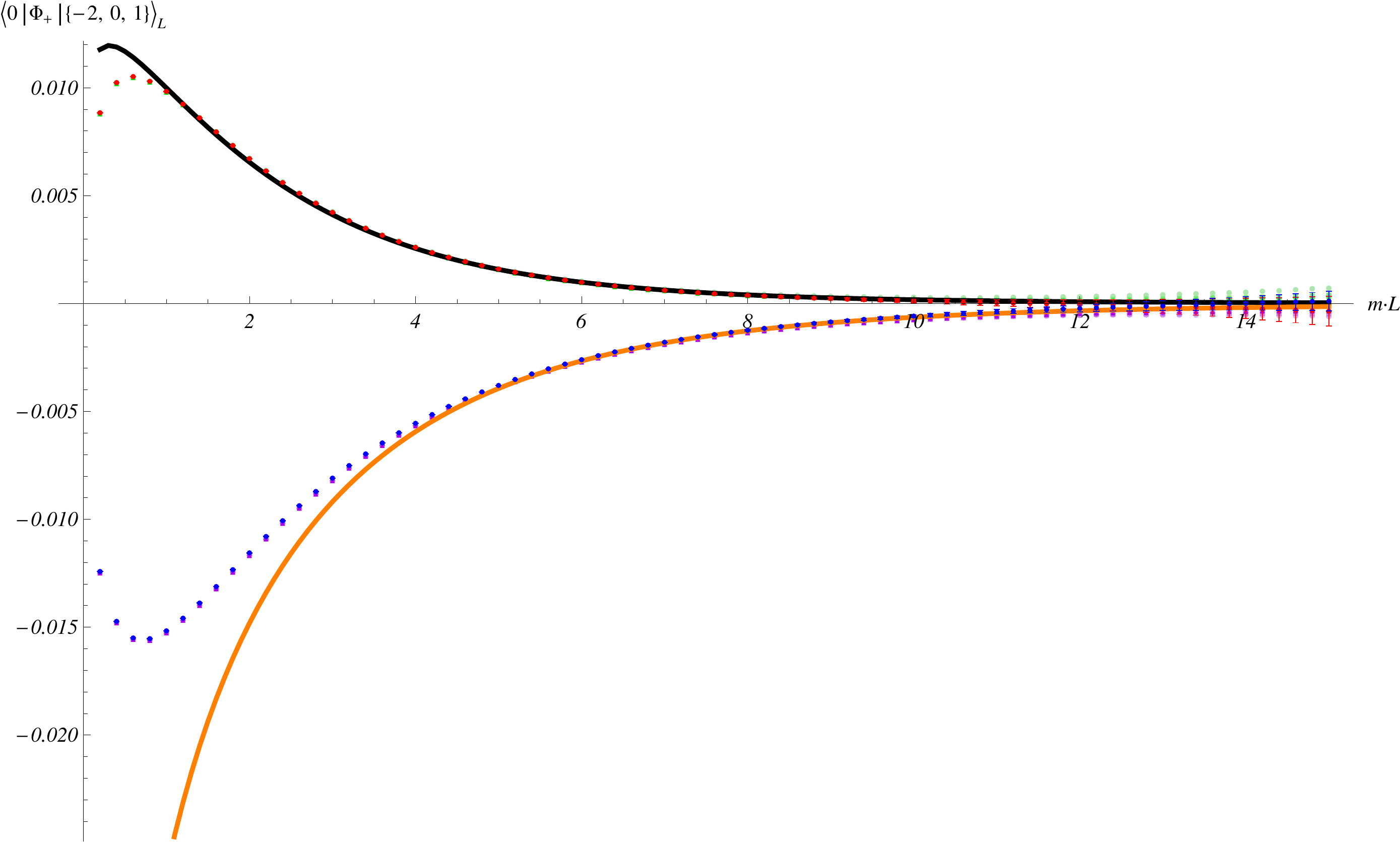}\includegraphics[width=0.45\textwidth]{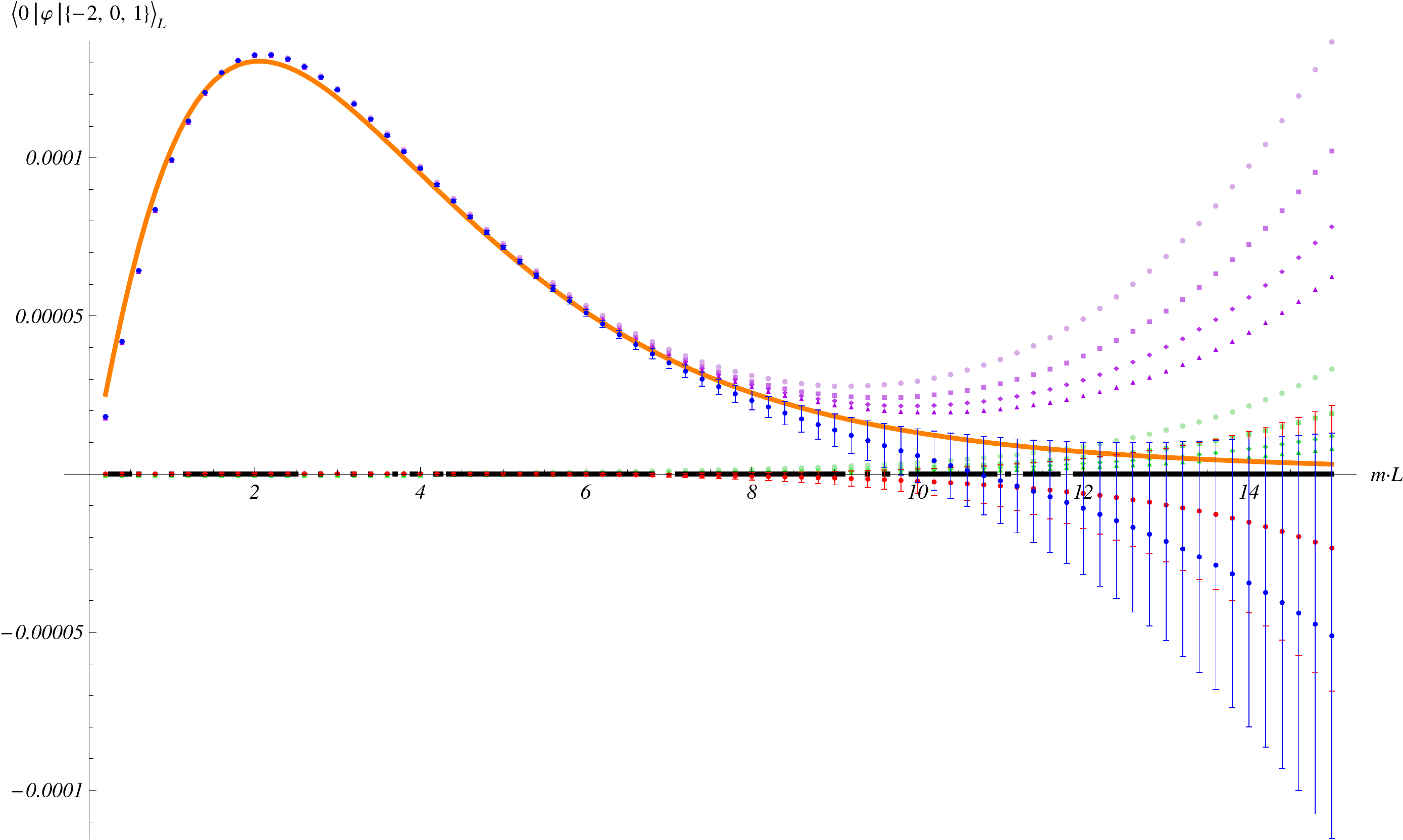}
\par\end{centering}

\caption{Left: three-particle form factor of the operator $\Phi_{+}$ on the
state labeled by quantum numbers $n_{1}=-2,\, n_{2}=0,\, n_{3}=1$.
Right: three-particle form factor of the operator $\varphi$ on the
state $n_{1}=-2,\, n_{2}=0,\, n_{3}=1$. The solid lines are computed
from formula (\ref{eq:Finite_volume_FF}), while the dots with confidence
bars are obtained by extrapolated TCSA data. (The consistent legend
is described in the beginning of subsection \ref{sub:Form-factors-and})}
\end{figure}
\begin{figure}
\begin{centering}
\includegraphics[width=0.45\textwidth]{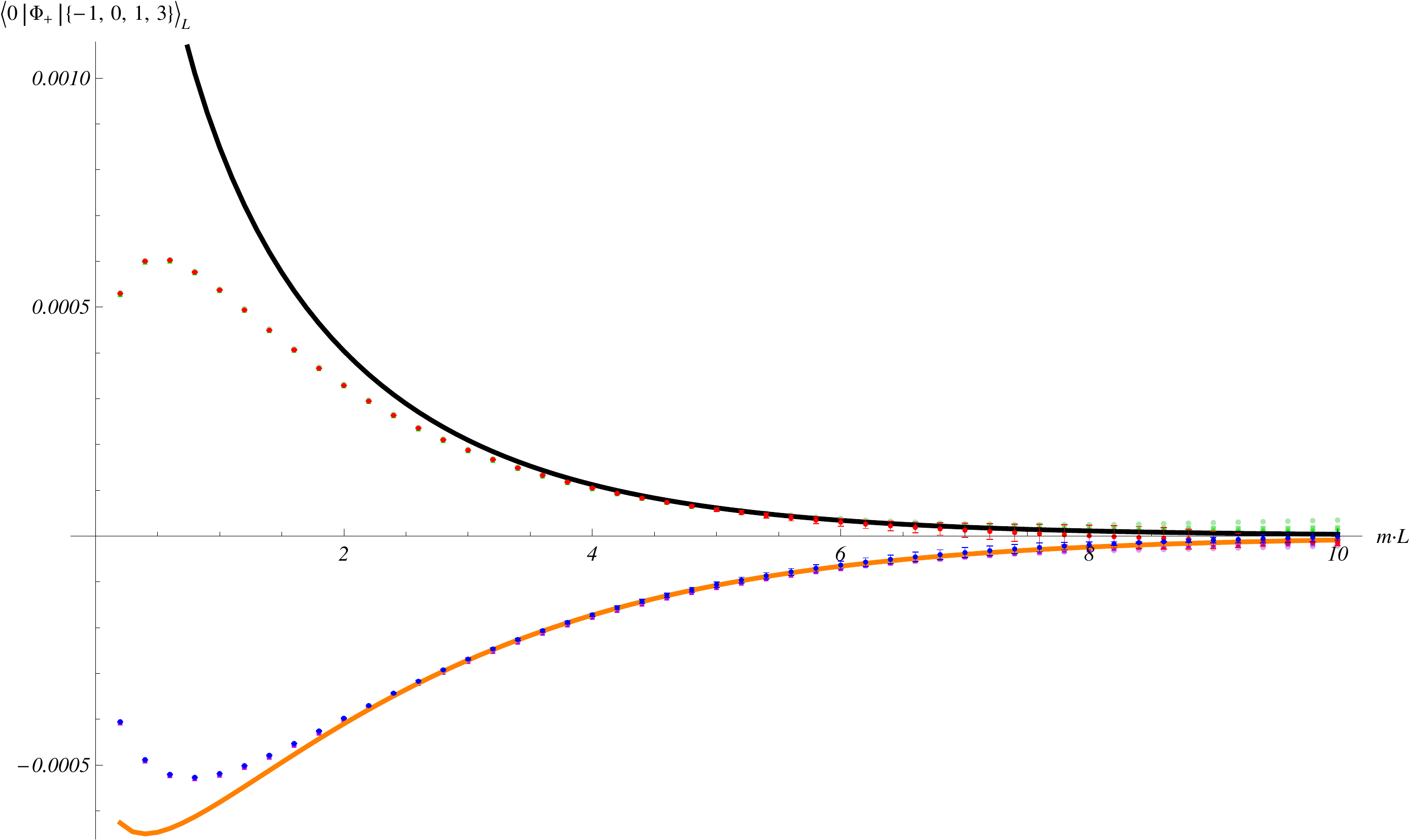}\includegraphics[width=0.45\textwidth]{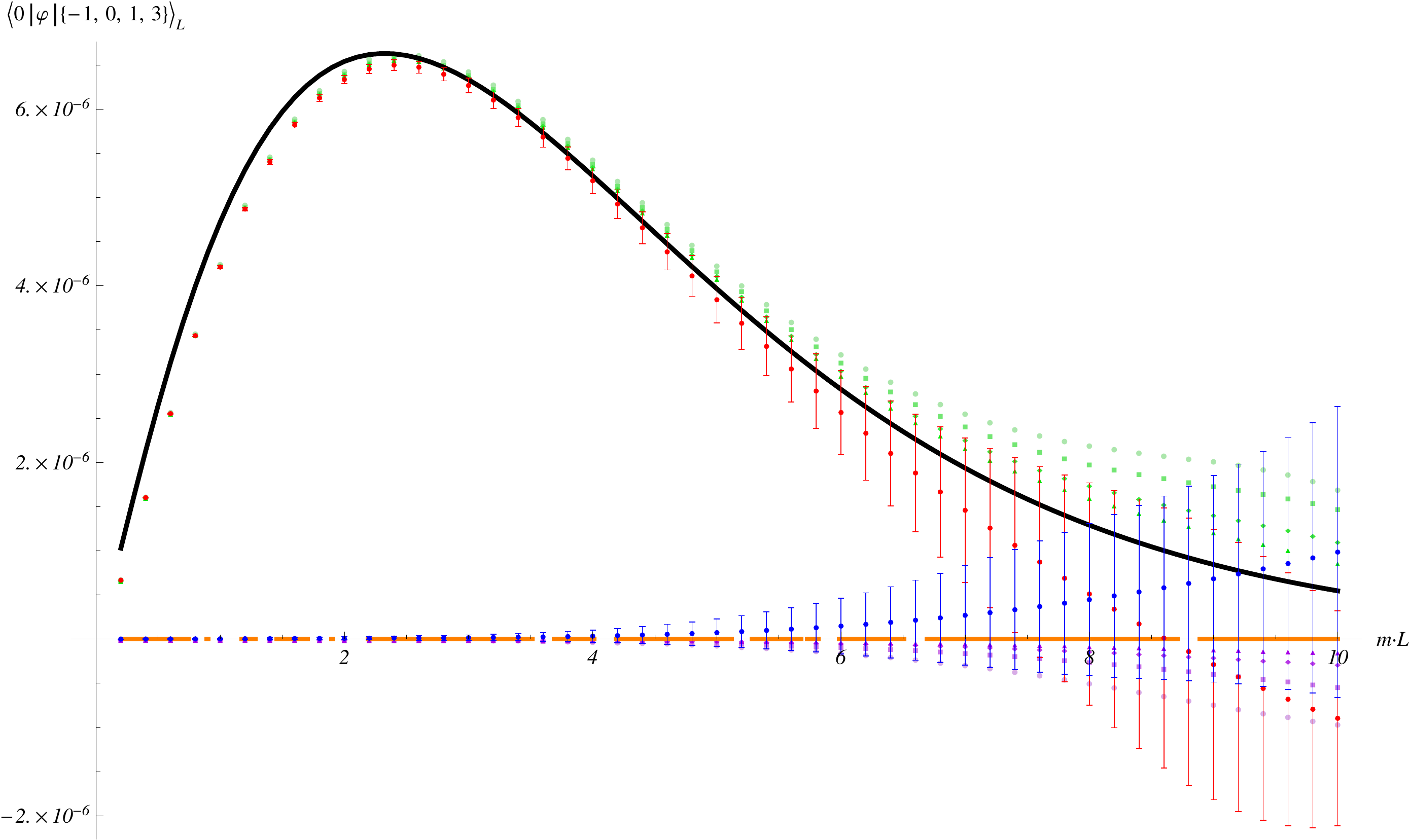}
\par\end{centering}

\caption{Left: four-particle form factor of the operator $\Phi_{+}$ on the
state labeled by quantum numbers $n_{1}=-1,\, n_{2}=0,\, n_{3}=1,\, n_{4}=3$.
Right: four-particle form factor of the operator $\varphi$ on the
state $n_{1}=-1,\, n_{2}=0,\, n_{3}=1,\, n_{4}=3$. The solid lines
are computed from formula (\ref{eq:Finite_volume_FF}), while the
dots with confidence bars are obtained by extrapolated TCSA data.
(The consistent legend is described in the beginning of subsection
\ref{sub:Form-factors-and})}
\end{figure}

\subsection{Diagonal form factors}

This subsection contain some data for diagonal form factors with various
particle numbers.

\begin{figure}[h]
\begin{centering}
\includegraphics[width=0.48\textwidth]{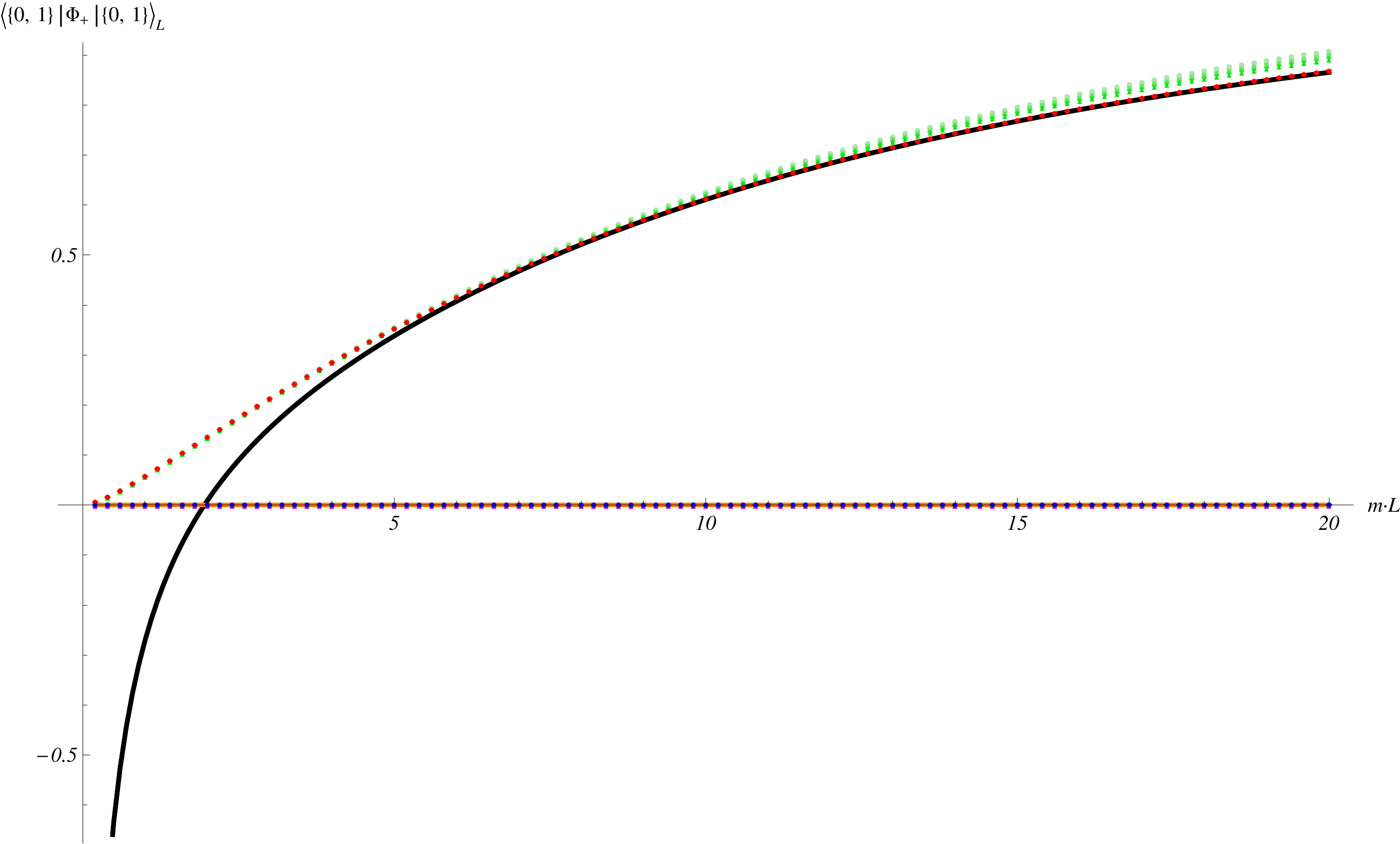}\includegraphics[width=0.48\textwidth]{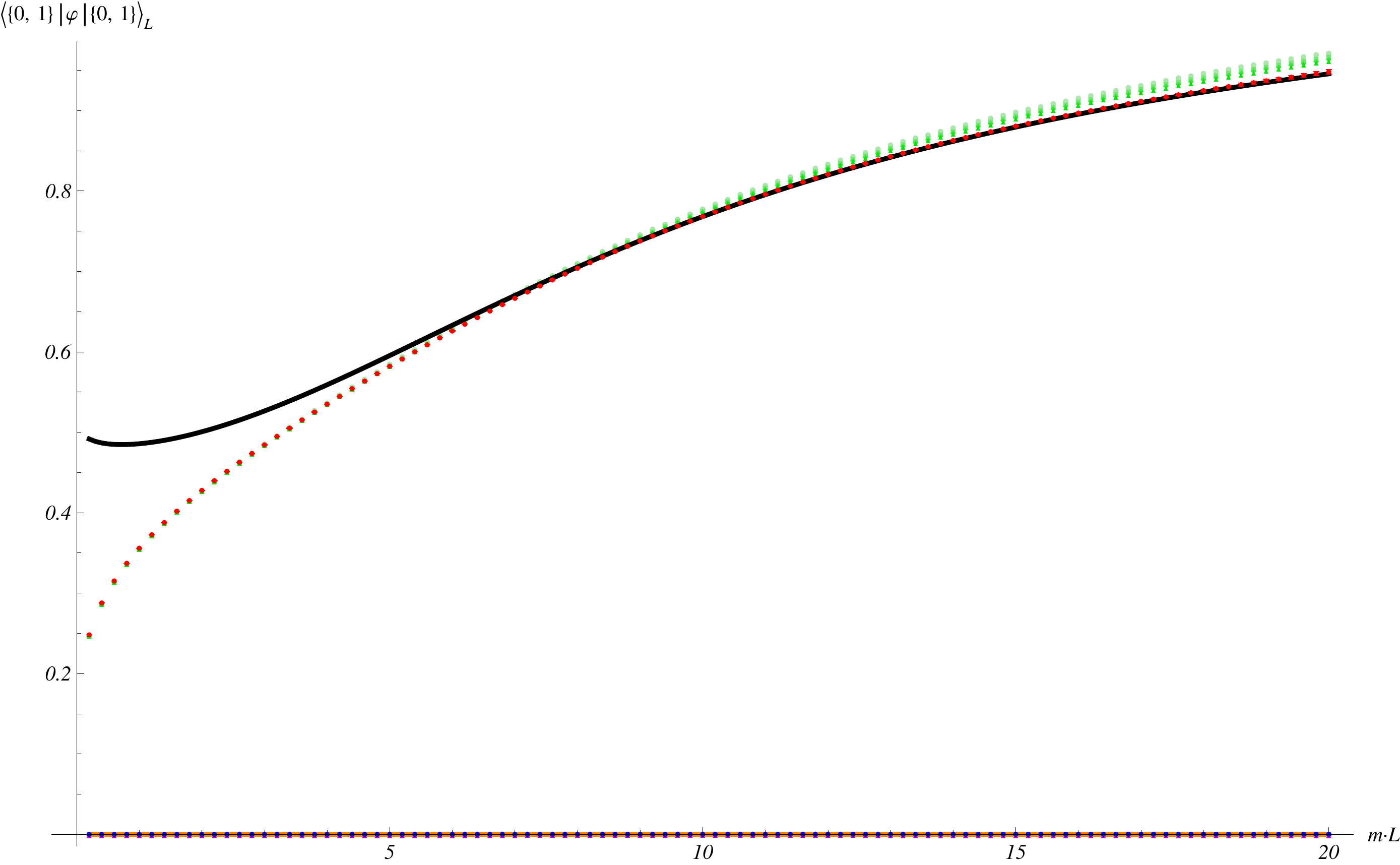}
\par\end{centering}

\caption{Left: two-particle diagonal form factor of the operator $\Phi_{+}$
on the state labeled by quantum numbers $n_{1}=0,\, n_{2}=1$. Right:
two-particle diagonal form factor of the operator $\varphi$ on the
state $n_{1}=0,\, n_{2}=1$. The solid lines are computed from formula
(\ref{eq:diaggenrulesaleur}), while the dots with confidence bars
are obtained by extrapolated TCSA data. (The consistent legend is
described in the beginning of subsection \ref{sub:Form-factors-and})}
\end{figure}

\begin{figure}[h]
\begin{centering}
\includegraphics[width=0.48\textwidth]{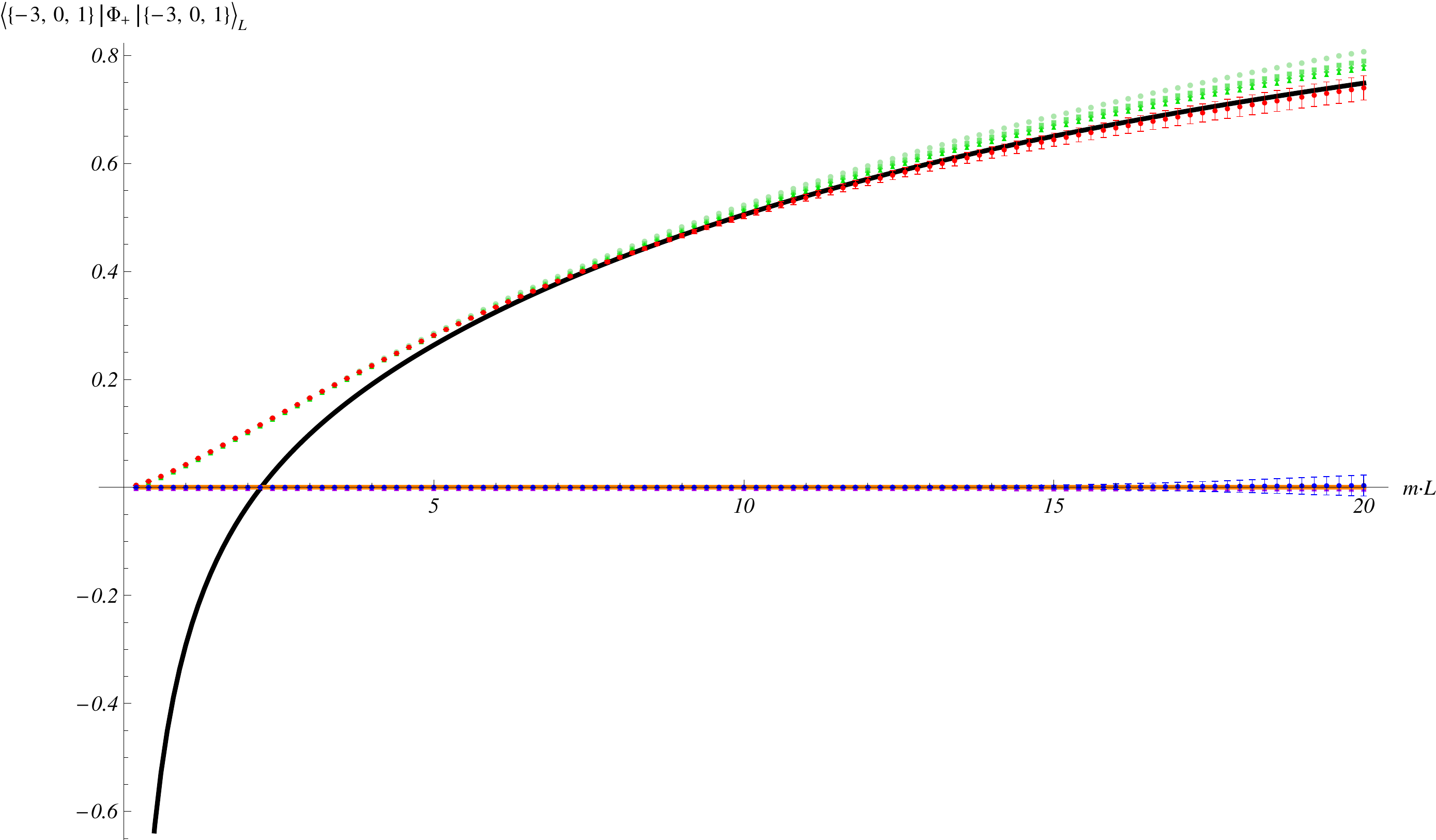}\includegraphics[width=0.48\textwidth]{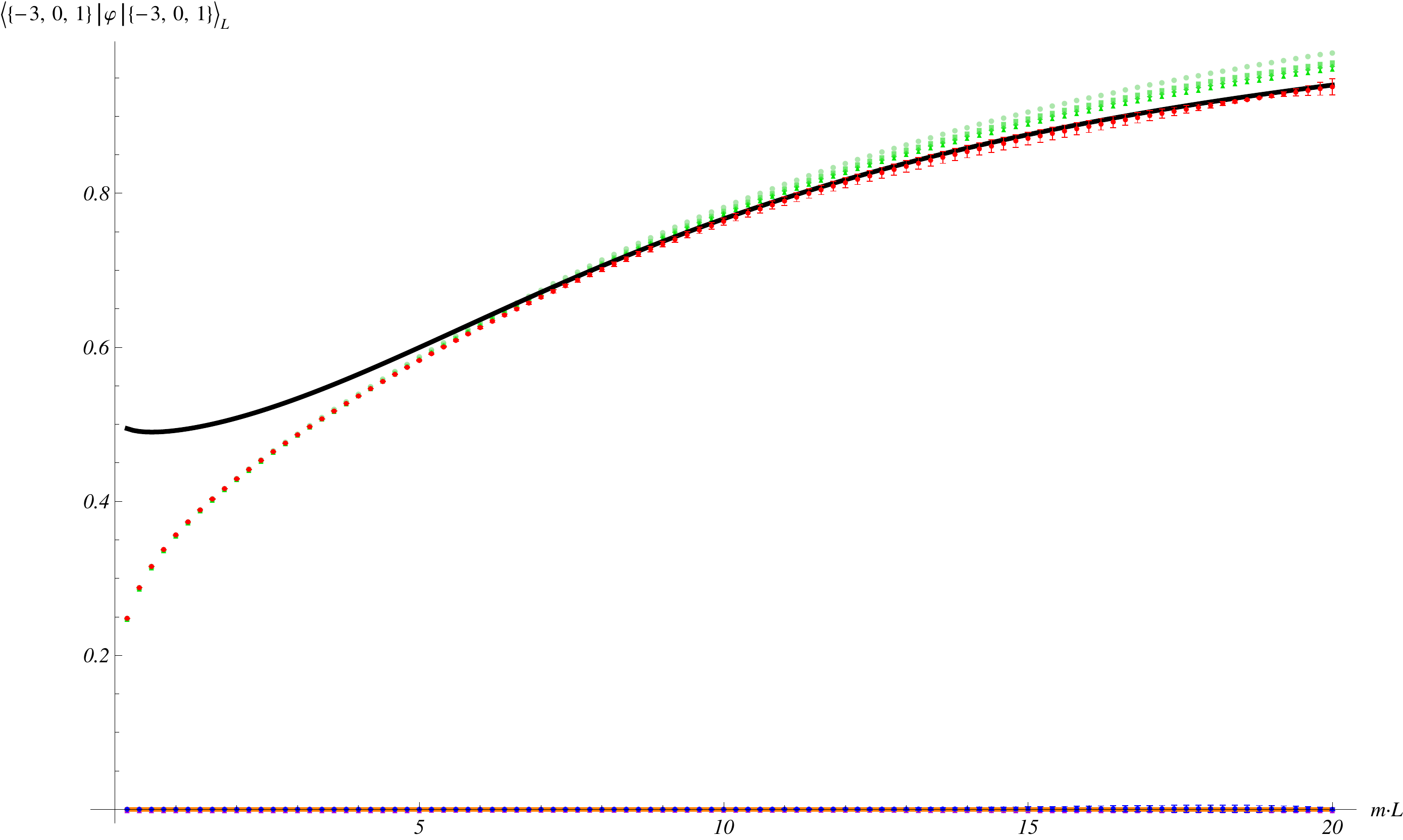}
\par\end{centering}

\caption{Left: three-particle diagonal form factor of the operator $\Phi_{+}$
on the state labeled by quantum numbers $n_{1}=-3,\, n_{2}=0,\, n_{3}=1$.
Right: three-particle diagonal form factor of the operator $\varphi$
on the state $n_{1}=-3,\, n_{2}=0,\, n_{3}=1$. The solid lines are
computed from formula (\ref{eq:diaggenrulesaleur}), while the dots
with confidence bars are obtained by extrapolated TCSA data. (The
consistent legend is described in the beginning of subsection \ref{sub:Form-factors-and})}
\end{figure}

\begin{figure}
\begin{centering}
\includegraphics[width=0.48\textwidth]{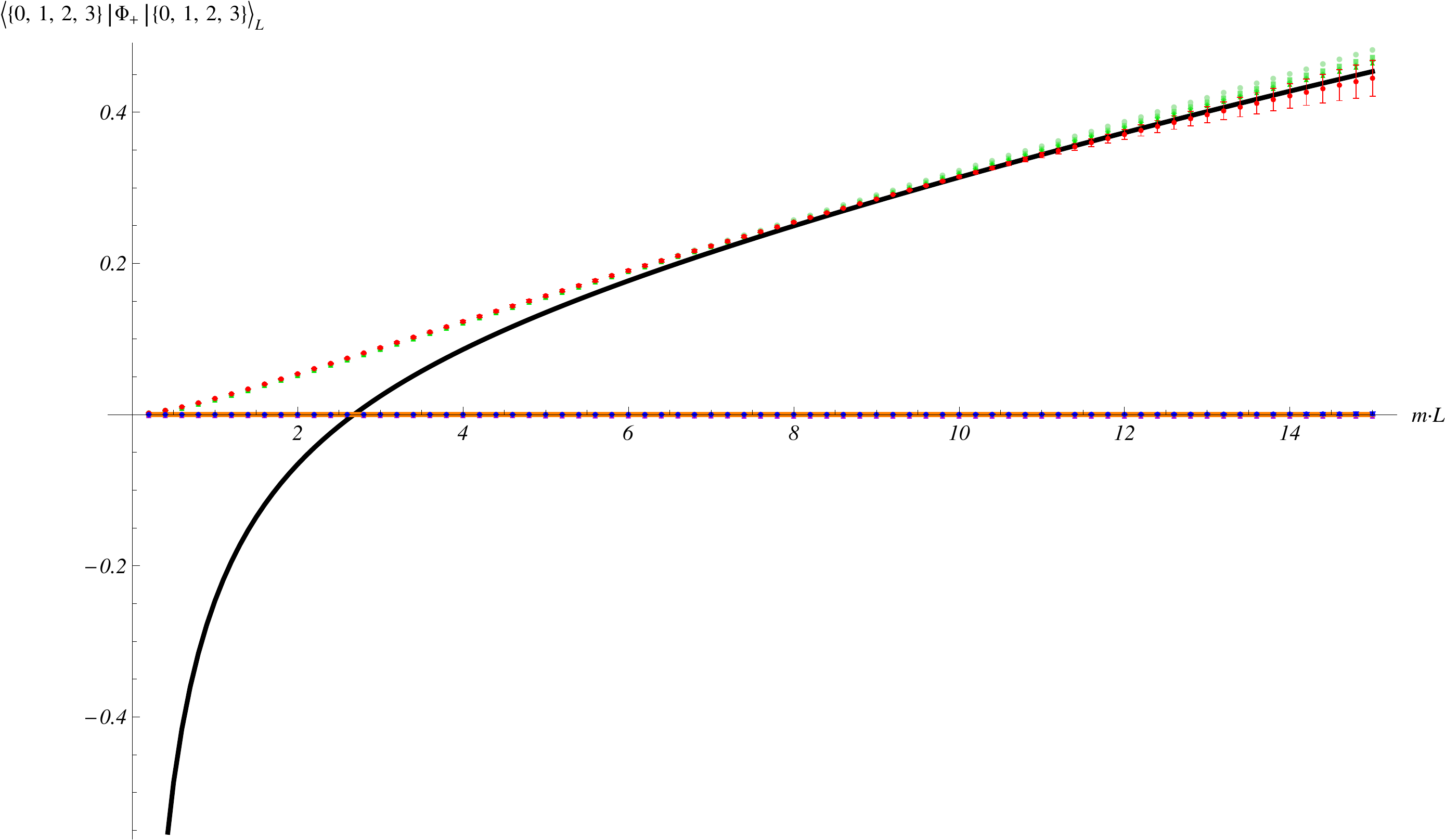}\includegraphics[width=0.48\textwidth]{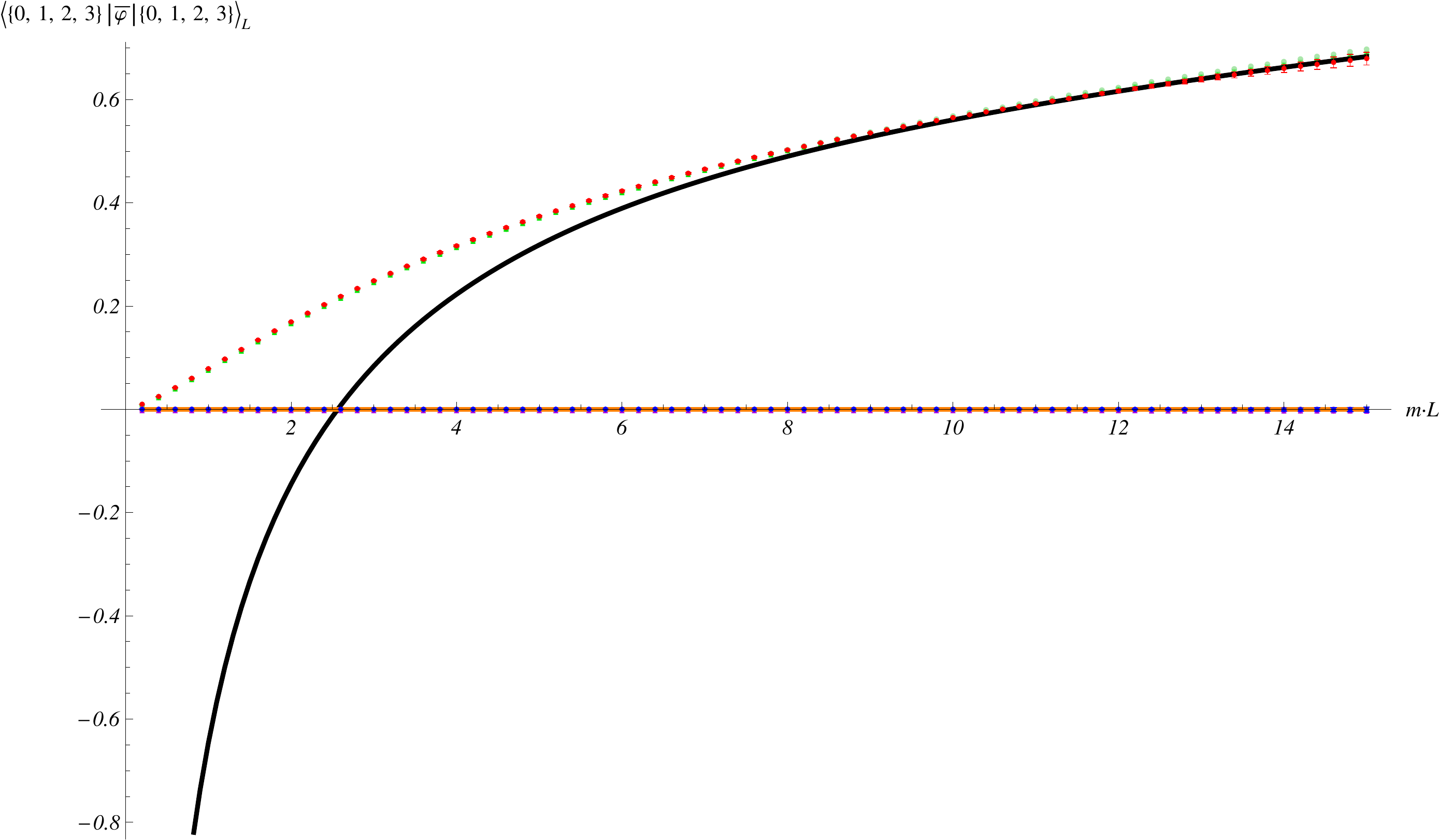}
\par\end{centering}

\caption{Left: four-particle diagonal form factor of the operator $\Phi_{+}$
on the state labeled by quantum numbers $n_{1}=0,\, n_{2}=1,\, n_{3}=2,\, n_{4}=3$.
Right: four-particle diagonal form factor of the operator $\bar{\varphi}$
on the state $n_{1}=0,\, n_{2}=1,\, n_{3}=2,\, n_{4}=3$. The solid
lines are computed from formula (\ref{eq:diaggenrulesaleur}), while
the dots with confidence bars are obtained by extrapolated TCSA data.
(The consistent legend is described in the beginning of subsection
\ref{sub:Form-factors-and})}
\end{figure}


\newpage

\begin{thebibliography}{10}

\bibitem{Avan:2012rb}
Jean Avan and Anastasia Doikou.
\newblock {The sine-Gordon model with integrable defects revisited}.
\newblock {\em JHEP}, 1211:008, 2012.

\bibitem{Bajnok:2004jd}
Z.~Bajnok and A.~George.
\newblock {From defects to boundaries}.
\newblock {\em Int.J.Mod.Phys.}, A21:1063--1078, 2006.

\bibitem{Bajnok:2007jg}
Z.~Bajnok and Zs. Simon.
\newblock {Solving topological defects via fusion}.
\newblock {\em Nucl.Phys.}, B802:307--329, 2008.

\bibitem{Bajnok:2009hp}
Zoltan Bajnok and Omar el~Deeb.
\newblock {Form factors in the presence of integrable defects}.
\newblock {\em Nucl.Phys.}, B832:500--519, 2010.

\bibitem{Bajnok:2013}
Zoltan Bajnok, Laszlo Hollo, and Gerard Watts.
\newblock {Defect scaling Lee-Yang model from the perturbed DCFT point of
  view}, July 2013.

\bibitem{Bowcock:2004my}
P.~Bowcock, Edward Corrigan, and C.~Zambon.
\newblock {Affine Toda field theories with defects}.
\newblock {\em JHEP}, 0401:056, 2004.

\bibitem{Bowcock:2005vs}
P.~Bowcock, Edward Corrigan, and C.~Zambon.
\newblock {Some aspects of jump-defects in the quantum sine-Gordon model}.
\newblock {\em JHEP}, 0508:023, 2005.

\bibitem{Corrigan:2010ph}
E.~Corrigan and C.~Zambon.
\newblock {Integrable defects in affine Toda field theory and infinite
  dimensional representations of quantum groups}.
\newblock {\em Nucl.Phys.}, B848:545--577, 2011.

\bibitem{Corrigan:2007gt}
Edward Corrigan and C.~Zambon.
\newblock {On purely transmitting defects in affine Toda field theory}.
\newblock {\em JHEP}, 0707:001, 2007.

\bibitem{Delfino:1994nr}
G.~Delfino, G.~Mussardo, and P.~Simonetti.
\newblock {Scattering theory and correlation functions in statistical models
  with a line of defect}.
\newblock {\em Nucl.Phys.}, B432:518--550, 1994.

\bibitem{Doikou:2012fc}
Anastasia Doikou and Nikos Karaiskos.
\newblock {Sigma models in the presence of dynamical point-like defects}.
\newblock {\em Nucl.Phys.}, B867:872--886, 2013.

\bibitem{Dorey:2000eh}
P.~Dorey, M.~Pillin, R.~Tateo, and G.M.T. Watts.
\newblock {One point functions in perturbed boundary conformal field theories}.
\newblock {\em Nucl.Phys.}, B594:625--659, 2001.

\bibitem{Feverati:2006ni}
Giovanni Feverati, Kevin Graham, Paul~A. Pearce, Gabor~Zs. Toth, and Gerard
  Watts.
\newblock {A Renormalisation group for TCSA}.
\newblock 2006.

\bibitem{Giokas:2011ix}
Philip Giokas and Gerard Watts.
\newblock {The renormalisation group for the truncated conformal space approach
  on the cylinder}.
\newblock 2011.

\bibitem{Konik:2007cb}
Robert~M. Konik and Yury Adamov.
\newblock {A Numerical Renormalization Group for Continuum One-Dimensional
  Systems}.
\newblock {\em Phys. Rev. Lett.}, 98:147205, 2007.

\bibitem{Kormos:2007qx}
M.~Kormos and G.~Takacs.
\newblock {Boundary form-factors in finite volume}.
\newblock {\em Nucl.Phys.}, B803:277--298, 2008.

\bibitem{Luscher:1985dn}
M.~Luscher.
\newblock {Volume Dependence of the Energy Spectrum in Massive Quantum Field
  Theories. 1. Stable Particle States}.
\newblock {\em Commun.Math.Phys.}, 104:177, 1986.

\bibitem{Luscher:1986pf}
M.~Luscher.
\newblock {Volume Dependence of the Energy Spectrum in Massive Quantum Field
  Theories. 2. Scattering States}.
\newblock {\em Commun.Math.Phys.}, 105:153--188, 1986.

\bibitem{Luscher:1990ux}
Martin Luscher.
\newblock {Two particle states on a torus and their relation to the scattering
  matrix}.
\newblock {\em Nucl.Phys.}, B354:531--578, 1991.

\bibitem{Maiani:1990ca}
L.~Maiani and M.~Testa.
\newblock {Final state interactions from Euclidean correlation functions}.
\newblock {\em Phys. Lett.}, B245:585--590, 1990.

\bibitem{Pozsgay:2008bf}
B.~Pozsgay.
\newblock {Luscher's mu-term and finite volume bootstrap principle for
  scattering states and form factors}.
\newblock {\em Nucl.Phys.}, B802:435--457, 2008.

\bibitem{Pozsgay:2013jua}
B.~Pozsgay.
\newblock {Form factor approach to diagonal finite volume matrix elements in
  Integrable QFT}.
\newblock 2013.

\bibitem{Pozsgay:2007kn}
B.~Pozsgay and G.~Takacs.
\newblock {Form-factors in finite volume I: Form-factor bootstrap and truncated
  conformal space}.
\newblock {\em Nucl.Phys.}, B788:167--208, 2008.

\bibitem{Pozsgay:2007gx}
B.~Pozsgay and G.~Takacs.
\newblock {Form factors in finite volume. II. Disconnected terms and finite
  temperature correlators}.
\newblock {\em Nucl.Phys.}, B788:209--251, 2008.

\bibitem{Pozsgay:2010xd}
Balazs Pozsgay.
\newblock {Mean values of local operators in highly excited Bethe states}.
\newblock {\em J.Stat.Mech.}, 1101:P01011, 2011.

\bibitem{Szecsenyi:2013gna}
I.M. Szecsenyi, G.~Takacs, and G.M.T. Watts.
\newblock {One-point functions in finite volume/temperature: a case study}.
\newblock {\em JHEP}, 1308:094, 2013.

\bibitem{Takacs:2011nb}
G.~Takacs.
\newblock {Determining matrix elements and resonance widths from finite volume:
  The Dangerous $\mu$-terms}.
\newblock {\em JHEP}, 1111:113, 2011.

\bibitem{Yurov:1989yu}
V.P. Yurov and A.B. Zamolodchikov.
\newblock {Truncated conformal space approach to scaling Lee-Yang model}.
\newblock {\em Int.J.Mod.Phys.}, A5:3221--3246, 1990.

\bibitem{Zamolodchikov:1989cf}
A.B. Zamolodchikov.
\newblock {Thermodynamic Bethe ansatz in relativistic models. scaling three
  state potts and Lee-Yang models}.
\newblock {\em Nucl.Phys.}, B342:695--720, 1990.

\bibitem{Zamolodchikov:1990bk}
A.B. Zamolodchikov.
\newblock {Two point correlation function in scaling Lee-Yang model}.
\newblock {\em Nucl.Phys.}, B348:619--641, 1991.

\end{thebibliography}
\end{document}